\documentclass{aa}  

\usepackage{graphicx}
\usepackage{longtable}
\usepackage{txfonts}
\usepackage{diagbox}
\begin{document}

   \title{Dispersal timescale of protoplanetary disks in the low-metallicity young cluster Dolidze~25}


   \author{M. G. Guarcello
          \inst{1}
          \and
          K. Biazzo\inst{2}
          \and
          J. J. Drake\inst{3}
          \and
          G. Micela\inst{1}
          \and
          L. Prisinzano\inst{1}
          \and
          S. Sciortino\inst{1}
          \and
          F. Damiani\inst{1}
          \and
          E. Flaccomio\inst{1}
          \and
          C. Neiner\inst{4}
          \and
          N. J. Wright\inst{5}
          }

   \institute{INAF - Osservatorio Astronomico di Palermo, Piazza del Parlamento 1, I-90134, Palermo, Italy\\
              \email{mario.guarcello@inaf.it}
              \and
              INAF—Osservatorio Astrofisico di Catania, Via Santa Sofia 78, I-95123 Catania, Italy \\
              \and
              Smithsonian Astrophysical Observatory, 60 Garden Street, Cambridge, MA 02138, USA \\
              \and
              LESIA, Paris Observatory, France \\
              \and
              Astrophysics Group, Keele University, Keele, ST5 5BG, UK
             }

   \date{}

 
  \abstract
   {The dispersal of protoplanetary disks sets the timescale available for planets to assemble, and thus it is one of the fundamental parameters in theories of planetary formation. Disk dispersal is determined by several properties of the central star, the disk itself, and the surrounding environment. In particular, the metallicity of disks may impact their evolution, even if to date controversial results exist: in low-metallicity clusters disks seem to rapidly disperse, while in the Magellanic Clouds some evidence supports the existence of accreting disks few tens of Myrs old.} 
   {In this paper we study the dispersal timescale of disks in Dolidze~25, the young cluster in proximity of the Sun with lowest metallicity, with the aim of understanding whether disk evolution is impacted by the low-metallicity of the cluster.}
   {We have analyzed $Chandra$/ACIS-I observations of the cluster and combined the resulting source catalog with existing optical and infrared catalogs of the region. We selected the disk-bearing population in a 1$^\circ$ circular region centered on Dolidze~25 from criteria based on infrared colors, and the disk-less population within a smaller central region among the X-ray sources with OIR counterpart. In both cases, criteria are applied to discard contaminating sources in the foreground/background. We have derived stellar parameters from isochrones fitted to color-magnitude diagrams.}
   {We derived a disk fraction of $\sim$34\% and a median age of the cluster of 1.2$\,$Myrs. To minimize the impact of incompleteness and spatial inhomogeneity of the list of members, we restricted this calculation to stars in a magnitude range where our selection of cluster members is fairly complete and by adopting different cuts in stellar masses. By comparing this estimate with existing estimates of the disk fraction of clusters younger than 10$\,$Myrs, our study suggests that the disk fraction of Dolidze~25 is lower than what is expected from its age alone.}
   {Even if our results are not conclusive given the intrinsic uncertainty on stellar ages estimated from isochrones fitting to color-magnitude diagrams, we suggest that disk evolution in Dolidze~25 may be impacted by the environment. Given the poor O star population and low stellar density of the cluster, it is more likely that disks dispersal timescale is dictated more by the low metallicity of the cluster rather than external photoevaporation or dynamical encounters.}


   \maketitle
%

\section{Introduction}

The dispersal of protoplanetary disks is a crucial topic in astronomy for its importance in setting the time available for the formation of planetary systems around young stars \citep[e.g.,][]{Helled2014prpl.conf.643H}. The timescale for disk dispersal has been observationally set by determining in clusters at different age the fraction of stars still hosting a protoplanetary disk \citep{HaischLL2001,HernandezHMG2007,Richert2018MNRAS.477.5191R}. These studies have found that most of the disks disperse in a few Myrs: starting from disk fractions as high as 60\%-80\% in very young clusters, the typical fraction in 5$\,$Myrs old clusters is about 20\%, and in $\geq$10$\,$Myrs old regions, such as the TW~Hydra, $\sigma$Ori, and NGC~7160 associations, primordial disks become exceedingly rare as attested by a very low incidence of less than 5\% \citep{Sicilia-Aguilar2006ApJ.638.897S,HernandezHMG2007}. However, these numbers must be interpreted as a general trend, since individual stars may retain their disks for longer times \citep[e.g.,][]{Armitage2003MNRAS.342.1139A}.\par

    \subsection{Disk evolution in different environments}
    \label{intro_env}

    The trend outlined before refer to disks whose evolution is not affected by the surrounding environment. In the past decades, several environmental feedback mechanisms that can potentially affect the dispersal of protoplanetary disks have been explored. \par
    
    Local stellar density is important since during the dynamical evolution of the parental cluster, stars can experience close encounters with other members during which the mutual gravitational interaction may affect the evolution of their disks. During these encounters, part of the disk material can be dispersed in the surrounding medium or even captured by the other star \citep{ClarkePringle1993,PfalznerUH2005,ThiesKGS2010}. The importance of close encounters is studied by simulating the dynamical relaxation of clusters with different stellar densities with the aim of estimating the rate of destructive encounters in a time interval comparable to disk lifetimes. For instance, \citet{ClarkePringle1993} found that a density of about 100 stars/pc$^3$ is required in order to have 1\% chance for potentially destructive close encounters for 100$\,$AU disks in 1$\,$Myr; \citet{SteinhausenPfalzner2014AA} found that only a negligible fraction of protoplanetary disks experience destructive close encounters in 2$\,$Myrs in clusters with a stellar density smaller than 3000 stars/pc$^3$; \citet{VinckeBP2015} found that in 5$\,$Myrs in clusters with a stellar density smaller than 90 stars /pc$^3$ no disks are shrink down to 10$\,$AU by close encounters, while about 10\%-17\% can be dispersed down to 100$\,$AU. \par
    
    By looking at the typical stellar density of known clusters in the Milky Way, it can be concluded that only the most extreme clusters such as the Arches may have a stellar density so high to result in a significant probability for destructive encounters \citep{OlczakKHP2012}. However, even in these cases, a more destructive feedback would be provided by externally induced photoevaporation. In this case disks are dispersed because of the incidence of energetic UV radiation \citep[e.g.,][]{JohnstoneHB1998} emitted by nearby massive stars. UV photons dissociate and ionize Hydrogen molecules and atoms, increasing the gas temperature up to more than one thousand degrees and driving a photo-evaporative wind away from the disk. Since the UV radiation is provided by massive stars, externally induced photoevaporation is expected to be important in clusters with at least a few thousand members that are expected to host massive stars \citep{Weidner2010MNRAS.401.275W}, and the effects of photoevaporation are more dramatic within a few parsecs from such massive stars. Direct observations of evaporating disks were obtained in the Trapezium in Orion \citep{ODellW1994,BallyOM2000,Fang2016ApJ.833L.16F}, Cygnus~OB2 \citep{WrightDDG2012,GuarcelloDWG2014}, NGC~2244 \citep{BalogRSM2006}, NGC~1977 \citep{Kim2016ApJ.826L.15K}, and Carina \citep{Mesa-Delgado2016ApJ.825L.16M}. Indirect evidence supporting a fast erosion of protoplanetary disks in proximity of massive stars was  obtained by observing a decline of the disk fraction close to massive stars or in regions with high local UV fields in massive clusters and associations such as: NGC~2244 \citep{BalogMRS2007}, NGC~6611 \citep{GuarcelloPMD2007,GuarcelloMDP2009,GuarcelloMPP2010}, and Pismis~24 \citep{FangBKH2012}. \citet{RichertFGK2015} instead found no evidence supporting a lower disk fraction near massive stars in the sample of massive clusters included in the MYStIX project \citep{Feigelson2013ApJS.209.26F}, suggesting that evidence supporting the external disks photoevaporation found by earlier studies was affected by selection effects. This has been refuted by careful later studies of e.g. NGC 6231 \citep{Damiani2016AA.596A.82D}, Cygnus OB2 \citep{GuarcelloDWA2016arXiv}, and Trumpler 14 and 16 \citep{ReiterParker2019MNRAS.486.4354R}.  \par
    
The metallicity of disks, which is typically assumed to be equal to that of their parental clusters, is also expected to play an important role in determining disk dispersion timescales by affecting the relative content of dusts, which regulates important disks properties such as opacity. The first, and so far only, observational confirmation of a fast erosion of disks selected from infrared photometry in low metallicity environments has been provided by \citet{YasuiKTS2009,YasuiKTS2010,Yasui2016AJ.151.115Y} and \citet{Yasui2016AJ.151.50Y}, who derived the disk fractions in six clusters in the outer Galaxy, characterized by [O/H]$\sim$-0.7$\,$dex and dust/gas ratio of $\sim$0.001. The clusters of their sample younger than 1$\,$Myr (Cloud2-N and -S, Sh2-209, and Sh2-208) have a disk fraction between 7$\pm$1\% and 27$\pm$7\%, much smaller than the typical disk fraction of  60\% - 80\% observed in clusters with similar age but solar metallicity. Similarly, the only cluster in their sample with an age of 2-3$\,$Myrs (Sh2-207) has a disk fraction of 5.1$\pm$4.6\%, while clusters with this age and solar metallicity have a disk fraction of 30\%-40\%. \par

    \citet{YasuiKTS2010} stated that a faster dispersal of protoplanetary disks in low-metallicity environments is unlikely to be a consequence of a more efficient dust aggregation process, which instead is expected to proceed slowly because of the low dust content. Alternatively, the authors considered it more likely that in low-metallicity disks, the ionization fraction is larger than in disks with higher metallicity, increasing their accretion rates. This hypothesis was supported by previous works \citep[e.g.,][]{Hartmann2006ApJ.648.484H,Hartmann2009apsf.book.H} claiming that accretion is mainly driven by magnetorotational instability \citep[MRI,][]{BalbusHawley1991}, whose efficiency increases with increasing disk ionization rate. Also \citet{GortiHollenbach2009ApJ.690.1539G} claimed that FUV photons penetrate more deeply in disks with small dust/gas ratio, and that the dispersal time of disks decreases with increasing dust opacity. In the last years, general consensus shifted toward a more important role of magnetically driven disk wind in removing mass and angular momentum from the disk, also driving mass accretion toward the inner disk \citep{Suzuki2010ApJ.718.1289S,BaiStones2013ApJ.769.76B,Simon2013ApJ.775.73S}. Since in this picture MRI is still responsible to trigger magnetohydrodynamic (MHD) turbulence that drive the vertical gas outflow \citep{Suzuki2010ApJ.718.1289S} and both mass accretion and mass loss rate are expected to increase with increasing the penetration depth of ionizing photons \citep{Bai2016ApJ.818.152B}, the low disk opacity in low metallicity can still be responsible for a faster disk dispersal than at solar metallicity. \par


    Some studies explored the possibility that low metallicity increases the effectiveness of photoevaporation in removing gas and small dust grains from protoplanetary disks. This is in line with the results obtained by \citet{GortiHollenbach2009ApJ.690.1539G}. Focussing on photoevaporation induced by X-ray photons emitted by the central stars \citep{ErcolanoDRC2008ApJ, ErcolanoCLarkeDrake2009ApJ.699.1639E}, \citet{ErcolanoClarke2010MNRAS.402.2735E} have found that the disk dispersal timescale due to photoevaporation (t$_{phot}$) increases with the disk metallicity following the relation t$_{phot}\propto$Z$^{0.52}$. Following these authors, the larger efficiency of photoevaporation in low-metallicity disks is due to smaller dust opacity. It is interesting to note that, according to \citet{ErcolanoClarke2010MNRAS.402.2735E}, disk dissipation timescales are instead expected to strongly decrease with increasing disks metallicity when disks are mainly dispersed by planet formation, since the larger metallicity results in a more efficient solid coagulation into planetesimals \citep{PollackHBL1996Icar,Hubickyj2005Icar.179.415H}. This is in line with the evidence that the incidence of Jovian planets around dwarf stars increases with the metallicity of the stars \citep[e.g.,][]{FischerValenti2005ApJ.622.1102F}. The larger efficiency of photoevaporation in low-metallicity has been more recently confirmed also by the simulations presented by \citet{Nakatani2018ApJ.857.57N,Nakatani2018ApJ.865.75N}. On the other hands, these models do not include the effects of metallicity over the stellar UV and X-ray emission. For instance, coronal X-ray radiation is dominated by line emission from highly-ionized atomic species, and thus it depends on the abundance of heavy elements \citep{Pizzolato2001AA.373.597P}.   \par
    
In the context of disk evolution in low-metallicity environments, it is worth mentioning the evidence of more intense accretion rates found in stars with disks in the Magellanic Clouds. Several works that focused on accretors in low-metallicity star-forming complexes of the Large Magellanic Clouds \citep{deMarchi2010ApJ.715.1D,deMarchi2017ApJ.846.110D,SpezziDPS2012,Biazzo2019ApJ.875.51B}, on average with Z=0.007 \citep{Maeder1999AA.346.459M}, and Small Magellanic Clouds \citep[e.g.,][]{deMarchi2011ApJ.740.11D} reported larger mass accretion rates compared with stars with similar masses of the Milky Way \citep[e.g.,][]{Beccari2015AA.574A.44B}. Both the higher accretion rates and longer accretion timescales in low-metallicity star-forming regions have been interpreted by the authors as a consequence of a less intense radiation pressure experienced by the inner disks when the dust content is smaller. These results can also be understood if disk photoevaporation discussed above is overwhelmed by other effects that can results in longer disk lifetimes and larger accretion rates: larger ionization inducing more intense mass accretion rates, lower disk opacity which results in a lower disk temperature, smaller viscosity and thus longer viscous timescale \citep{Durisen2007prpl.conf.607D}, or a slower formation of protoplanets due to the smaller concentration of solid bodies and thus less efficient grain aggregation process, slowing down the dispersal of protoplanetary disks \citep{DullemondDomink2005AA.434.971D}. \par

   
    \subsection{Dolidze~25}
    \label{intro_dol25}

   	\begin{figure}[]
	\centering	
	\includegraphics[width=8.5cm]{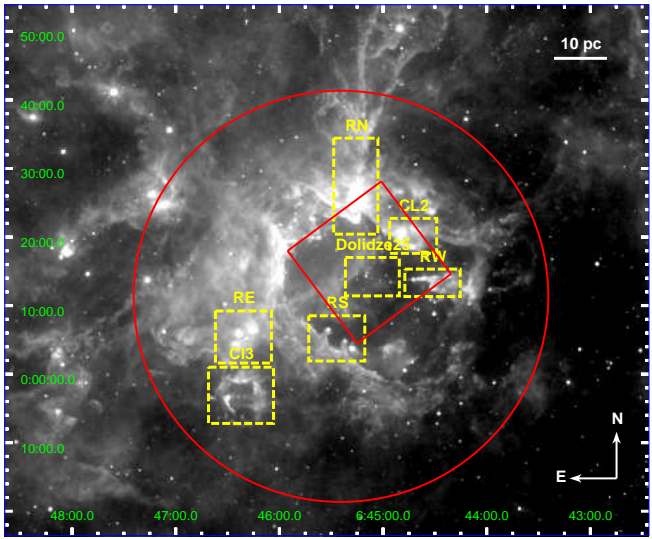}
	\caption{WISE 12$\,\mu$m image of the area surrounding Dolidze~25. The red circle marks the area where we selected stars with disks; the red box delimits the ACIS-I FoV; the dashed yellow boxes encompass the regions identified and studied by \citet{Puga2009AA.503.107P}; the segment in the upper right corner shows the angular extent corresponding to 10$\,$pc at the distance of Dolidze~25.}
	\label{field_img}
	\end{figure}

The young open cluster Dolidze~25 (aka C~0642+0.03 or OCL-537; l=212$^\circ $, b=-1.3$^\circ $) is one of the best targets to study the evolution and dispersal timescale of protoplanetary disks in a low-metallicity environment, being one of the known rare cases of Galactic low-metallicity environments. Cluster metallicity was determined for the first time by \citet{LennonDFG1990} on the basis of high-resolution spectroscopy of three OB stars that were found to be deficient in metals by a factor of $\sim$6, and later confirmed by \citet{Fitzsimmons1992MNRAS.259.489F} and \citet{Negueruela2015AA.584A.77N}. The latter authors have derived a metallicity -0.3 dex below solar for Silicon and of -0.5 dex below solar for Oxygen. Even if these values are not as low as those reported by \citet{LennonDFG1990} and are not fully inconsistent with the radial slope of the metallicity gradient in our Galaxy, when the observed data-points dispersions is taken into account \citep{Rolleston2000AA.363.537R,Esteban2013MNRAS.433.382E}, Dolidze~25 is confirmed as one of the young clusters with the lowest metallicity known in our Galaxy.  \par

    The determination of the main parameters of Dolidze~25, such as distance and age, is quite controversial. The first estimates were based on the few massive stars (about ten OB stars). \citet{MoffatVogt1975AAS.20.85M} determined a distance of 5.25$\,$kpc from UBVH$\alpha$  photometry; \citet{LennonDFG1990} placed the cluster at 3.6$\,$kpc from isochrones fit to the upper main sequence of the cluster; \citet{TurbideMoffat1993AJ.105.1831T} determined an age of $\sim$6$\,$Myrs and a distance of $\sim$5$\,$kpc from optical photometry. More recently, \citet{DelgadoDA2010} have analyzed $UBVRIJHK$ photometry of the central area of Dolidze~25, identifying 214 candidate cluster members and setting the cluster distance equal to 3.6$\,$kpc. These authors claimed that two distinct populations belong to the cluster: A younger pre-main sequence population 5$\,$Myrs old, and an older population with an age of 40$\,$Myrs. Following studies found no evidence for such an old cluster population. For instance, \citet{Negueruela2015AA.584A.77N} set an upper limit to the cluster age of $\sim$3$\,$Myrs by noting that none of the most massive stars of Dolidze~25 (the O6~V star S33 and the O7~V stars S15 and S17, following the nomenclature based on the WEBDA\footnote{https://webda.physics.muni.cz/} database) show evidence of any evolution off the main sequence, and adopted a distance of 4.5$\,$kpc from the trigonometric parallax distance of the HII region IRAS~06501+0143 in the proximity of the cluster. \citet{CusanoRAG2011} have analyzed VIMOS@VLT, 2MASS and Spitzer data of Dolidze~25. They have set a distance of 4$\,$kpc from the spectroscopic parallax of three OB members and an average age of 2$\,$Myrs. They also found evidence for a significant age spread and a sequential star formation process across the whole area. \citet{Kalari2015ApJ.800.113K} estimated an age between 2 and 3$\,$Myrs for cluster members in the center of Dolidze~25 selected from infrared photometry and the analysis of the H$\alpha$ line, and found no evidence of a more intense accretion in members with disks with respect to the coeval populations with Solar metallicity. In this paper, we adopt a distance to Dolidze~25 equal to 4.5$\pm$0.5$\,$kpc, estimated from the Gaia/EDR3 counterparts of the 10 OB stars included in \citet{Negueruela2015AA.584A.77N} catalog and with error in parallaxes smaller than 0.2$\,$mas. \par
    
    \begin{figure*}[]
	\centering	
	\includegraphics[width=8.5cm]{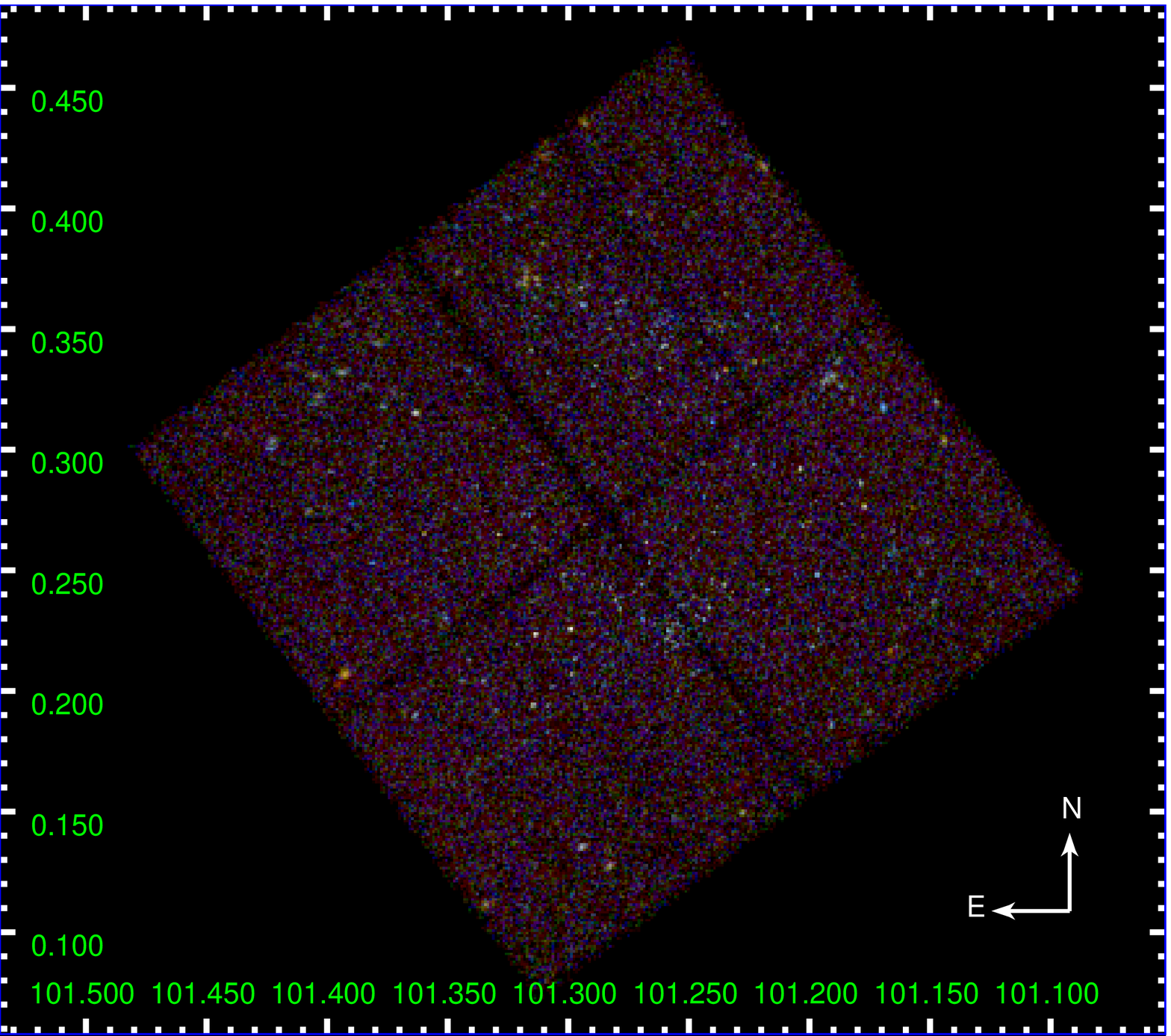}
	\includegraphics[width=8.5cm]{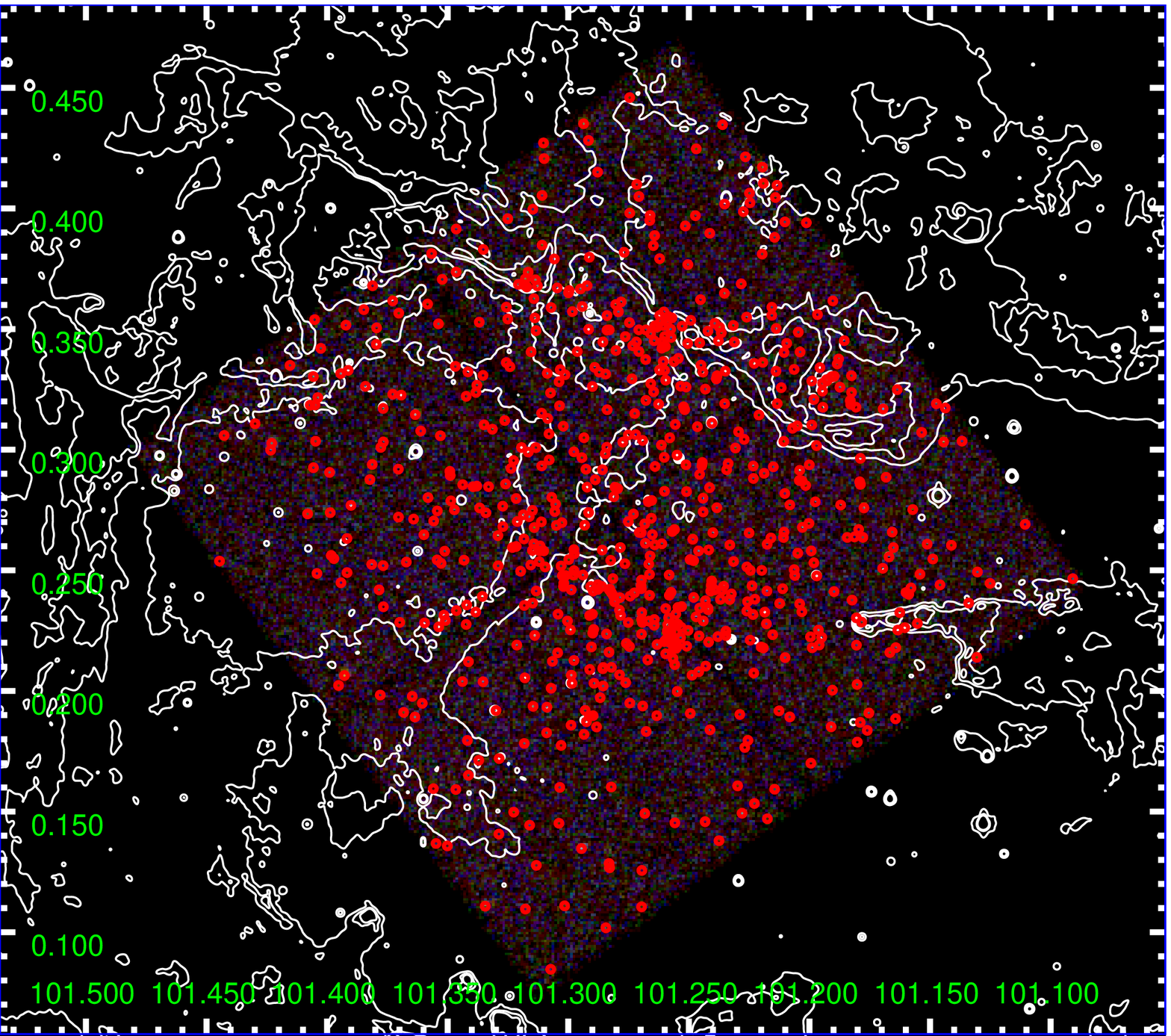}	
	\caption{Combined $Chandra$/ACIS-I Images of Dolidze~25. In the left panel, we show an RBG image, with events in the hard energy band marked in red, in green those in the ``medium'' band (1.21-1.99$\,$keV), and in blue those in the soft band. In the right panel, the white polygons mark the contours of the continuum emission at 8$\,\mu$m from Spitzer/IRAC, while the red circles mark the position of the validated X-ray sources.}
	\label{acis_img}
	\end{figure*}
    
Dolidze~25 is part of a vast star-forming complex classified as Sh2-284 by \citet{Sharpless1959ApJS.4.257S}. The most comprehensive determination to date of the pre-main sequence population of the entire area has been performed by \citet{Puga2009AA.503.107P} from the analysis of Spitzer observations. They selected a total of 155 Class~I and 183 Class~II objects, clustered in different regions of the complex: In the central cluster Dolidze~25; around the large HII cavity surrounding the central cluster, which is ionized by the stars S33, S15 and S17; and in the compact HII regions IRAS~06439-0000, IRAS~06446+0029 and IRAS~06454+0020. The presence of pillars and globules containing young stars and pointing toward the central cluster apparently supports the hypothesis of some level of triggered star formation across the area \citep{CusanoRAG2011}. Fig. \ref{field_img} shows the WISE \citep{Wright2010AJ.140.1868W} image at 12$\,\mu$m of the area surrounding Dolidze~25, with marked the large area where we searched for stars with disks, the field observed with $Chandra$/ACIS-I, and the regions identified and analyzed by \citet{Puga2009AA.503.107P}. In this paper we will adopt the nomenclature defined by the latter authors to indicate these regions. \par

\section{The multi-wavelength catalog}

In this section we describe the multi-wavelength catalog of the studied area, which includes archival optical and infrared data, together with an X-ray catalog built from specific observations performed with $Chandra$/ACIS-I.

\subsection{Chandra/ACIS-I observations}
\label{chandra_sec}

Dolidze~25 was observed with $Chandra$/ACIS-I on 1$\rm^{st}$ and 3$\rm^{rd}$ December 2013 (Obs.IDs: 14565 and 16543, respectively; P.I.: Guarcello). The two observations were co-pointed at R.A.=06:45:05.10 and Dec=+00:16:15.60, with exposure times of 76.67 and 68.44$\,$ksec and both with a roll angle of 53$^\circ$. We produced the Level 2 event files from the Level 1 files using the CIAO \citep{Fruscione2006SPIE.6270E.1VF} script {\it chandra\_repro}. We then combined the two event files by using the tool {\it merge\_obs}. Before merging the two event files, we registered the astrometry of the Obs.ID 16543 onto the 14565
through the following procedure: we first run {\it Wavdetect} \citep{Freeman2002ApJS.138.185F} to detect sources in the two images separately, considering only the 100 brightest detected sources; we then matched the two resulting catalogs with a closest-neighbor approach; and then we updated the astrometry of the event files using the CIAO tool {\it WCS\_update}. Exposures maps in three bands (broad: 0.5-7$\,$keV; soft: 0.5-1.2$\,$keV; hard: 2-7$\,$keV) were calculated using the standard CIAO tools {\it asphist}, {\it mkinstmap}, and {\it mkexpmap}. \par


    Source detection in the three energy bands was performed using both {\it Wavdetect} and the {\it PWDetect} \citep{Damiani1997ApJ.483.350D} detection algorithms. {\it Wavdetect} detected a total of 696 sources in the broad energy band (272 in the soft band, 420 in the hard band) adopting a threshold of 10$^{-4}$, while {\it PWDetect} detected 367 sources in the broad band (657 in the soft band, 191 in the hard band) adopting a threshold of $\sigma$=4.6. The resulting six catalogs were merged in a unique list containing 2105 candidate sources adopting a closest-neighbors approach and visually inspecting the photon distributions in the event files. Even if this list was clearly dominated by spurious detections, mainly in the soft band, we decided to temporarily keep it since we validated each candidate X-ray source with a rigorous approach. \par
    
 Photon extraction and sources validation were performed using the {\it IDL} software {\it ACIS Extract}\footnote{The ACIS Extract software package and User’s Guide are available online at \url{http://www.astro.psu.edu/xray/docs/TARA/ae_users_guide.html}.} \citep[AE,][]{BroosTFG2010}. AE performs photon extraction by defining for each source the PSF at 1.5$\,$keV, reducing the PSF size of crowded sources down to 40\%. The individual background regions are defined as an annulus centered on each source, with an inner radius equal to 1.1 times the 99\% of the PSF, and the outer radius set in order to encompass 100 background photons. For sources in crowded regions, AE constructs a background model that accounts for the contamination due to nearby bright sources. In this latter case, the background model is improved after multiple iterations and extractions. \par

    AE estimates the probability for each source of being a background fluctuation and it saves it in the parameter {\it prob\_no\_source} ($P_B$). Following most of the existing works on similar data analysis \citep[e.g.,][]{WrightDGA2014}, we considered as probable spurious sources those with $P_B>0.01$. We thus pruned our list by removing all isolated sources with $P_B>0.01$. If a group of crowded sources met the requirement $P_B>0.01$, we removed only the faintest source and then we repeated the photon extraction process for the remaining sources with the attempt of improving their $P_B$. After repeating the procedure 5 times, and after a visual inspection of those sources marked by AE as probable spurious detections due to the {\it hook-shaped} feature discovered in the {\it Chandra} PSF\footnote{http://cxc.harvard.edu/ciao/caveats/psf artifact.html}, we removed 1487 sources from the initial list, producing a final list of 618 confirmed sources. Fig. \ref{acis_img} shows an RGB $Chandra$/ACIS-I image of the combined event files and the positions of the validated X-ray sources, together with the contours of the diffuse emission at 8$\,\mu$m from {\it Spitzer}/IRAC observations. 

In each of the five iterations of the photons extraction process, we allowed AE to correct source positions. Following AE's guidelines, three position estimates were calculated for each source: the {\it mean data position} which is obtained from the centroid of the extracted events, and it is typically used for on-axis sources ($\rm\theta$<5$^\prime$); the {\it correlation position} which is calculated from the correlation between the PSF and the events distribution, and it is typically used for off-axis sources ($\rm\theta$>5$^\prime$); and the {\it maximum Likelihood position}, which is calculated from the maximum-Likelihood image of source neighborhood and it is typically used for crowded sources. The catalog of the X-ray sources, which is made available on-line, is described in the Appendix \ref{app_xraycat}. \par

\subsection{Optical and Infrared catalogs}  

We collected the optical and infrared data available on a circular area of 0.5 degrees radius (39.3$\,$pc at the distance of 4500$\,$pc) centered on Dolidze~25, in order to cover not only the cluster, but also a significant part of the surrounding parental cloud. Table \ref{catal_table} shows the list of the catalogs we used. Some of these catalogs were not directly necessary for the aim of this paper, but they were included nevertheless for future analysis of the stellar population of Dolidze~25 and Sh2-284. In Table \ref{catal_table}, for each catalog, we show the total number of sources falling in the selected region, together with the number of sources we retained after pruning away spurious sources, artifacts, and sources with not reliable photometry following the various explanatory manuals or the references listed in the last column. The criteria adopted to clean each catalog are summarized in the $criteria$ column. \par

\begin{figure*}[!h]
	\centering	
	\includegraphics[width=18cm]{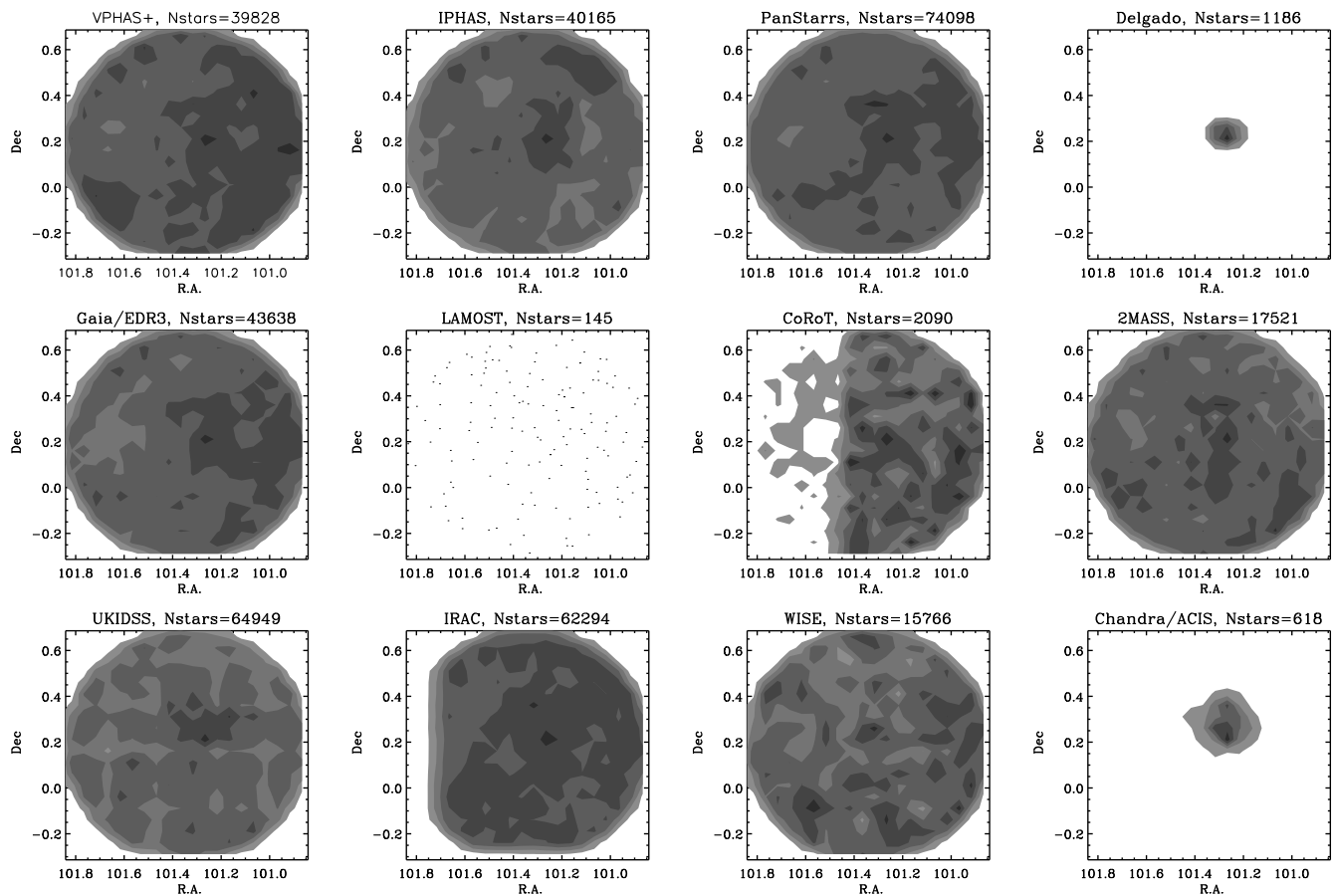}
	\caption{Spatial coverage of the sources with good photometry in the catalogs used in this work. A central overdensity, corresponding to the approximate location of Dolidze~25, is evident in all the catalogs.}
	\label{catalog_spadis_img}
	\end{figure*}

    The optical photometry is provided by the {\it Second Data Release} of the {\it VST Photometric H{$\alpha$} Survey of the Southern Galactic Plane and Bulge} \citep[VPHAS+,][]{Drew2014MNRAS.440.2036D}, the {\it Second Data Release of the INT/WFC Photometric H$\alpha$ Survey of the Northern Galactic Plane} \citep[IPHAS,][]{BarentsenFDG2014}, the {\it Panoramic Survey Telescope and Rapid Response System} \citep[Pan-STARRS,][]{Chambers2016arXiv161205560C}, the {\it Second} and {\it Early Third Data Release of the Gaia catalog} \citep{Gaia2016AA.595A.1G} providing parallaxes for 31531 sources and radial velocities for 205 sources, and the optical catalog published by \citet{DelgadoDA2010} based on observations taken with ALFOSC at the 2.6$\,$m Nordic Optical Telescope (NOT), which covers only a small $7^\prime\times8^\prime$ central area. Infrared photometry is instead provided by the {\it Two Micron All-Sky Survey} (2MASS) {\it Point Source Catalog} \citep{CutriSDB2003} and the tenth data release of the {\it UKIRT InfraRed Deep Sky Surveys} \citep[UKIDSS,][]{LawrenceWAE2007} in the $JHK$ bands, together with the catalog obtained from observations with Spitzer/IRAC during the Cycle 4 (2005 March 28, Program ID: 3340, P.I.: Neiner), presented in \citet{Puga2009AA.503.107P}, and the {\it AllWISE Source Catalog} \citep{Wright2010AJ.140.1868W}. We also included: 2090 optical light curves obtained from the {\it Convection, Rotation and Planetary Transits} satellite \citep[\emph{CoRoT},][]{BaglinABD2006ESASP} taken with a cadence of 32$\,$s or 512$\,$s; and the spectral classification of 145 stars obtained from the {\it Fourth Data Release of the Large sky Area Multi-Object Spectroscopic Telescope} (LAMOST) based on observations of the 4$\,$m telescope located at the Xinglong Observatory northeast of Beijing \citep[China,][]{Luo2015arXiv150501570L}. Fig. \ref{catalog_spadis_img} shows the spatial coverage of the catalogs included in the multi-band catalog. Most of them have a rather uniform distribution, with a clear overdensity of sources in the center of the field, roughly corresponding to the cavity cleared by Dolidze~25, and westward of the central cavity. The distribution of the UKIDSS sources is easily explained by the fact that the adopted pruning criteria removed most of the sources along the CCD edges. 

\begin{table*}
\caption{Optical and infrared catalogs used in this work}             
\label{catal_table}      
\centering                          
\begin{tabular}{c c c c c c}        
\hline\hline                 
Catalog & Bands & Initial & Selected & Criteria & References \\    
\hline                        
VPHAS+/DR2 & $ugriH\alpha$ & 79470 & 39828 & $fPrimary$=1 & \citet{Drew2014MNRAS.440.2036D,Drew2016yCat.2341.0D} \\
       &                                         &        &        & Detected in $\geq$2 bands &   \\
\hline
IPHAS/DR2  & $riH\alpha$   & 64300  & 40165  & $mergedClass$=1,2 & \citet{BarentsenFDG2014} \\
       &                 &       &               & Detected in $\geq$2 bands & \\
       &                 &       &               & visual inspection & \\       
\hline
Pan-STARRS/DR1 & $grizy$ & 77335 & 74098 & $Qual$=4 & \citet{Chambers2016arXiv161205560C} \\
& & & & Detected in $\geq$2 bands &   \\
& & & & visual inspection & \\
\hline
Gaia/DR2, EDR3 & $G$,$B_P$,$R_P$  & 43919 & 43919 & & \citet{Gaia2016AA.595A.1G} \\
\hline
Delgado  & $UBVRIJHK$ & 1673 & 1186 & Detected in $\geq$2 bands & \citet{DelgadoDA2010} \\
\hline
CoRoT    &            & 2090 & 2090 &       & \citet{Debosscher2009AA.506.519D} \\
& & & & & \citet{Affer2012MNRAS.424.11A} \\
& & & & & \citet{Carone2012AA.538A.112C} \\
& & & & & \citet{Guenther2012AA.543A.125G} \\
& & & & & \citet{Sebastian2012AA.541A.34S} \\
& & & & & \citet{COROT2014yCat.102028C} \\
\hline
LAMOST/DR4 & 370-900$\,$nm & 145 & 145 & &\citet{Luo2016yCat.5149.0L} \\
\hline
UKIDSS/DR10 & $JHK$ & 114024 & 64949 & $priorsec$=0 & \citet{LawrenceWAE2007} \\
& & & & $merged\_class$=-1 & \\
& & & & Detected in $\geq$2 bands& \\
\hline
2MASS/PSC & $JHK$ & 18422 & 17521 & $ph\_qual\neq$F,E,U & \citet{CutriSDB2003} \\
& & & & $rd\_flg\neq$6 & \\
& & & & $cc\_flg\neq$p,d,s,b & \\
\hline
Spitzer/IRAC & [3.6],[4.5],[5.8],[8.0] & 62336 & 62294 & visual inspection & \citet{Puga2009AA.503.107P} \\
\hline                                   
WISE  & [3.4],[4.6],[12],[22] & 15774 & 15766 & $ccf\neq$D,P,H,0 & \citet{Cutri2012yCat.2311.0C} \\
& & & & $qph\neq$C,X,U & \\
\hline
\hline
\end{tabular}
\end{table*}

\subsection{The merged catalog}
\label{mergedcat_sect}

The optical, infrared, and X-ray catalogs were merged in a multi-wavelength catalog with the procedure described in details in Appendix \ref{AppA}. The catalog contains 101722 entries. In particular, among the 618 X-ray sources, 463 are matched with at least one optical-infrared (OIR) counterpart. Considering the multiple coincidences between X-ray, optical, and infrared sources, the catalog contains a total of 593 X+OIR sources, with an expected contamination by spurious coincidences of about 10\%.

\section{Selection of stars with disks}
\label{disks_sec}

\begin{figure}[]
	\centering	
	\includegraphics[width=9cm]{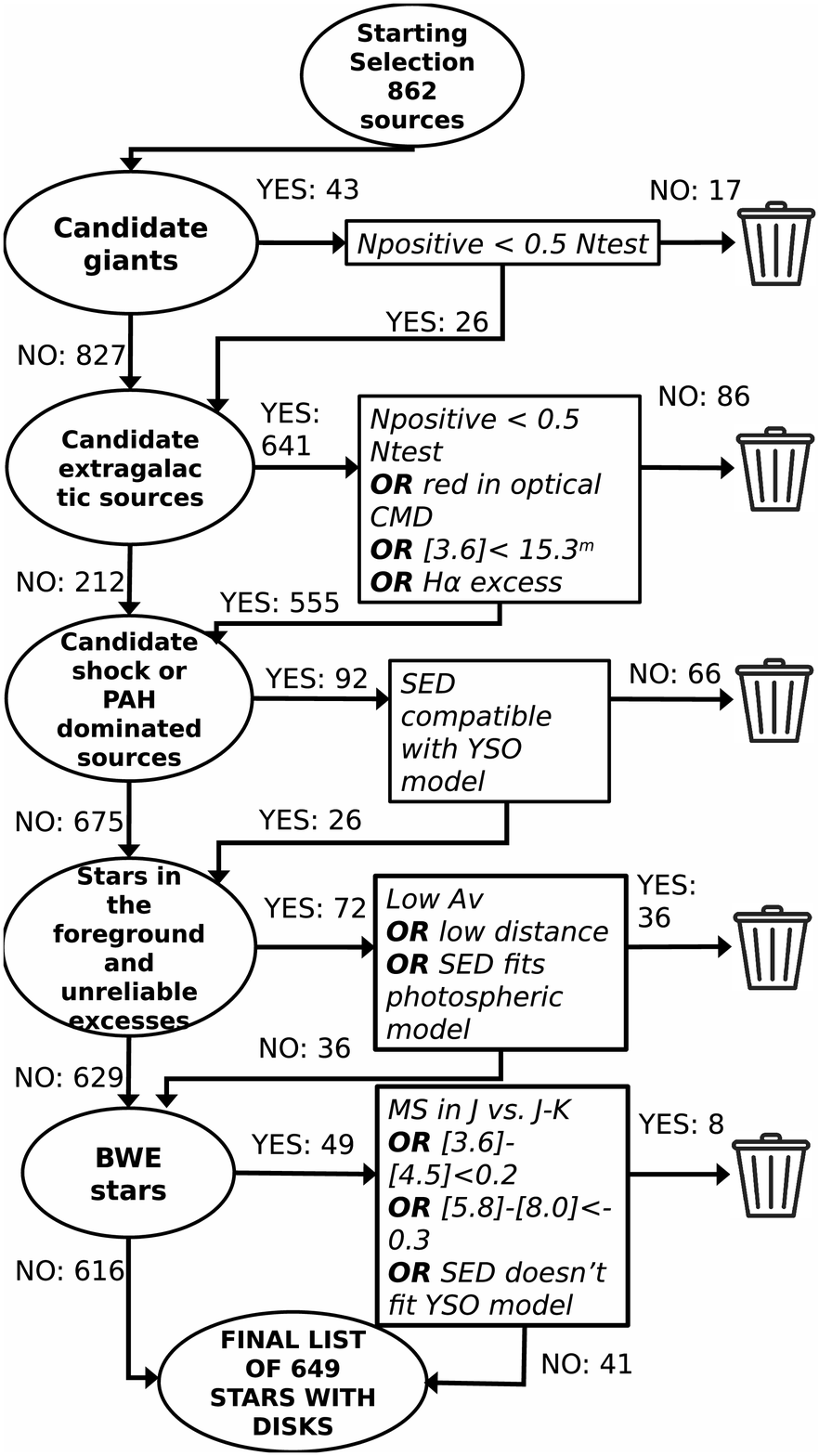}
	\caption{Summary flowchart of the procedure adopted to select stars with disks and prune candidate contaminants from the list. In each step represented by an oval, the list is pruned of a given class of contaminants. For any selection of contaminants, sources are verified with the corresponding test shown in the boxes. A detailed description of the tests is provided in the text.}
	\label{disksel_img}
	\end{figure}

Stars with disks were selected by adopting criteria based on 2MASS, UKIDSS, IRAC and WISE photometry. However, these methods potentially select also various types of contaminants, e.g. extragalactic sources, giants with circumstellar dust, PAH-contaminated sources, foreground stars. In order to obtain an inclusive selection of stars with disks removing all possible contaminants, we first selected all stars meeting at least one of the criteria defined to select stars with disks, and then we pruned the list applying different tests, each one aimed at selecting specific classes of contaminants (see Fig. \ref{disksel_img}).

\subsection{Initial list of candidate stars with disks}
\label{diskssel_sec}

The preliminary list of candidate stars with disks was produced by selecting all sources satisfying at least one of the criteria defined by \citet{GutermuthMMA2009ApJ,GuarcelloDWD2013ApJ} and \citet{KoenigLeisawitz2014ApJ.791.131K}: 

\begin{enumerate}
\item from the IRAC [3.6]-[4.5] vs. [4.5]-[5.8] diagram, sources with: \par
     $\bullet$ [3.6]-[4.5]$>$0.7 AND [4.5]-[5.8]$>$0.7;
\item from the IRAC [3.6]-[5.8] vs. [4.5]-[8.0] diagram, sources with:\par
    $\bullet$ [4.5]-[8.0]$>$0.5 AND \par
    $\bullet$ [3.6]-[5.8]$>$0.35 AND \par
    $\bullet$ [3.6]-[4.5]$\leq$0.5+0.14$\times$([4.5]-[8.0]-0.5);
\item from the IRAC [3.6]-[4.5] vs. [5.8]-[8.0] diagram, sources with: \par
    $\bullet$ [3.6]-[4.5]$>$0.2 AND [5.8]-[8.0]$>$0.3;
\item from the IRAC [4.5]-[5.8] vs. [5.8]-[8.0] diagram, sources with:\par
    $\bullet$ -0.1$\leq$[4.5]-[5.8]$<$1.4 AND [5.8]-[8.0]$>$0.2;
\item from the WISE [3.4]-[4.6] vs. [4.6]-[12] diagram, sources with:\par
    $\bullet$ 2$\leq$[4.6]-[12]$<$4.5 AND \par
    $\bullet$ [3.4]-[4.6]$>$2.2-0.42$\times$([4.6]-[12]) AND \par
    $\bullet$ [3.4]-[4.6]$>$0.46$\times$([4.6]-[12])-0.9;
\item from the WISE [3.4]-[4.6] vs. [4.6]-[12] diagram, sources with: \par
    $\bullet$ [3.4]-[4.6]$>$0.25 AND \par
    $\bullet$ [3.4]-[4.6]$<$0.9$\times$([4.6]-[12])-0.25 AND \par
    $\bullet$ [3.4]-[4.6]$>$-1.5$\times$([4.6]-[12])+2.1 AND \par
    $\bullet$ [3.4]-[4.6]$>$0.46$\times$([4.6]-[12])-0.9 AND \par
    $\bullet$ [4.6]-[12]$<$4.5;
\item from the J-H vs. [3.4]-[4.6] diagram, sources with: \par 
    $\bullet$ H-K$>$0 AND \par
    $\bullet$ H-K$>$-1.76$\times$([3.4]-[4.6])+0.9 AND \par
    $\bullet$ H-K$<$(0.55/0.16)$\times$([3.4]-[4.6])-0.85 AND \par
    $\bullet$ [3.4]$\leq$13.
\end{enumerate}

These criteria were applied only to sources with errors in the relevant magnitudes smaller than 0.1$^m$ and colors smaller than $0.15^m$. The loci defined by these criteria are shown in Fig. \ref{diagrams3_img}, and Appendix \ref{AppE} Figs. \ref{diagrams4_img}, and \ref{diagrams5_img}. The resulting preliminary list of candidate stars with disks counts 862 sources. This list was then pruned removing different classes of contaminants, as explained below.

\subsection{Candidate giants with circumstellar dust}
\label{giant_sect}

Evolved giants with circumstellar dust can have intrinsic infrared red colors that can mimic the Spectral Energy Distribution (SED) typical of stars with disks. In order to account for this contamination, we have used the PARSEC isochrones\footnote{http://stev.oapd.inaf.it/cmd} \citep{BressanMGS2012MNRAS} in order to define in some color-magnitude diagrams the expected loci where giants with circumstellar dust suffering zero extinction can be found. We then projected these loci along the specific extinction vectors. We defined these loci after testing different compositions of the circumstellar dust using the available options on the PARSEC web interface. After inspecting all the possible criteria, we used those resulting in independent selections: 

\begin{itemize}
\item stars in the $r$ vs. $r-i$ Pan-STARRS diagram brighter than the line drawn projecting the point [-0.1,11] along the extinction vector; 
\item stars in the $J$ vs. $J-K$ diagram brighter than the line drawn projecting the point [0,11] along the extinction vector;
\item stars in the [4.5] vs. [4.5]-[8.0] diagram with [4.5]-[8.0]$<$0;
\item stars in the [3.4] vs. [3.4]-[4.6] diagram brighter than the line [3.4]=1.56$\times$([3.4]-[4.6])+9.31
\item stars in the [3.4]-[4.6] vs. [12]-[22] diagram with: \par 
$\bullet$ [12]-[22]$<$1.9 AND \par
$\bullet$ [3.4]-[4.6]$<$([12]-[22])-0.45 \par
(criterion defined by \citealp{KoenigLeisawitz2014ApJ.791.131K}).
\item SED analysis (see below)
\end{itemize}

These loci are shown in the Figs. \ref{diagrams3_img}, \ref{diagrams4_img}, \ref{diagrams5_img}, \ref{diagrams1_img}. For each star, the tests are considered as ``positive'', i.e. suggesting the star being a contaminant, if the star falls in the defined locus of giant stars. The SED test was applied to the candidate stars with disks meeting at least one of the other criteria adopted to select candidate giants. In this test, we used the Python SED fitter tool $sedfitter$\footnote{https://sedfitter.readthedocs.io/en/stable/} developed by \citet{Robitaille2017AA.600A.11R}. The tool allowed us to fit the observed SEDs with synthetic ones produced from an extensive set of models of Young Stellar Objects (YSOs) with different properties of the central star, disk, envelope, bipolar cavities, and surrounding medium. We added to the available convolved filters the WISE, Pan-STARRS and SDSS filters using the Python codes $mkfilter.py$ and $filtermanage.py$ publicly available\footnote{https://github.com/mpound/YSOproject}. We considered this test as ``positive'' when the observed SED did not fit that of any YSO model\footnote{With the inclusion of SED fitting as a test, we attempt to break the degeneracy between highly extinguished background giants with circumstellar dust and stars with disks showing infrared excesses at longest wavelength.}. \par

    For each candidate star with disk selected as possible background giant, we thus counted the number $N_{test}$ of tests that was possible to perform and the number $N_{positive}$ of positive tests. As shown in Fig. \ref{disksel_img}, we removed from the list of stars with disks as candidate giants stars for which $N_{positive}\geq$0.5$N_{test}$.

\subsection{Candidate extragalactic sources}
\label{galaxy_sect}

Galaxies of different type (e.g. AGN, PAH galaxies, etc...) have infrared colors in the IRAC and WISE bands similar to those of stars with disks. However they can be discriminated from YSOs thanks to their typical blue optical colors and faint magnitudes. We discarded candidate galaxies from the sample of stars with disks in two steps. First, we adopted a similar approach to the one used to select giants with circumstellar dust, by defining tests aimed at selecting candidate galaxies and counting for each star with disk the number of positive over the total number of tests. These tests are defined adopting criteria introduced by other authors \citep{GutermuthMMA2009ApJ} or by plotting in infrared diagrams the extragalactic sources included in existing surveys \citep{Stern2005,Treister2006ApJ.640.603T,Rafferty2011ApJ.742.3R,KoenigLeisawitz2014ApJ.791.131K}:

\begin{itemize}
\item candidate PAH galaxies from the IRAC [4.5]-[5.8] vs. [5.8]-[8.0] diagram, with the criteria: \par
$\bullet$ [4.5]-[5.8]$<$1.05$\times$([5.8]-[8.0]-1)/1.2 AND \par
$\bullet$ [4.5]-[5.8]$<$1.05 AND [5.8]-[8.0]$>$1;
\item PAH galaxies from the IRAC diagram [3.6]-[5.8] vs. [4.5]-[8.0], with the criteria: \par
$\bullet$ [3.6]-[5.8]$<$1.5$\times$([4.5]-[8.0]-1)/2 AND \par
$\bullet$ [3.6]-[5.8]$<$1.5 AND [4.5]-[8.0]$>$1 AND \par
$\bullet$ [4.5]$>$11.5;
\item candidate AGN from the IRAC diagram [4.5] vs. [4.5]-[8.0], with the criteria: \par
$\bullet$ [4.5]-[8.0]$>$0.5 AND \par
$\bullet$ [4.5]$>$13.5+([4.5]-[8.0]-2.3)/0.4 AND \par
$\bullet$ [4.5]$>$13.5;
\item candidate AGN from the IRAC diagram [4.5] vs. [4.5]-[8.0], with the criteria: \par
$\bullet$ [4.5]-[8.0]$>$0.2 AND \par
$\bullet$ ([4.5]$>$14.5-([4.5]-[8.0]-1.2)/0.3 or [4.5]$>$14.5);
\item candidate galaxies from the WISE diagram [3.4]-[4.6] vs. [4.6]-[12] with the criteria: \par
$\bullet$ [4.6]-[12]$>$2.3 AND \par
$\bullet$ [3.4]-[4.6]$>$0.2 AND \par
$\bullet$ [3.4]-[4.6]$<$1.0 AND \par
$\bullet$ [3.4]-[4.6] $<$ 0.46$\times$([4.6]-[12])-0.78 AND \par
$\bullet$ [3.4]$>$13.0.
\end{itemize}

These loci are shown in the Figs. \ref{diagrams4_img} and \ref{diagrams5_img}. We then marked as a candidate extragalactic contaminant each source for which $N_{positive}\geq$0.5$N_{test}$.\par

    We then took advantage of the expected blue optical colors and faint magnitudes of extragalactic sources to re-classify as stars with disks some sources marked as possible galaxies. We first used the catalogs published by \citet{Brescia2015MNRAS.450.3893B} and \citet{Usatov2018JAD.24.3U} to define a locus populated by extragalactic sources in the following diagrams: $r$ vs. $r-i$, $g$ vs. $g-r$, and $r$ vs. $g-r$ (both VPHAS and Pan-STARRS, see Fig. \ref{diagrams1_img}). We then repeated the adopted strategy, by calculating for each source the ratio $N_{positive}/N_{test}$, where here a test is positive when the given star falls in these loci of extragalactic sources. We then reclassified as stars with disks those sources for which $N_{positive}<$0.5$N_{test}$. This procedure should in principle also help us to avoid discarding genuine stars with disks with blue optical colors due to accretion and/or scattering (discussed in Sect. \ref{bwe_sect}), which are expected to be more blue in $g-r$ than in $r-i$. \par
    
We also re-classified as stars with disks candidate extragalactic sources with large excess in $r-H\alpha$, typical of accreting stars with disks. To select these stars, we used the IPHAS and VPHAS $r-i$ vs. $r-H\alpha$ diagrams (see Fig. \ref{diagrams2_img}). In the former diagram we selected those sources with $r-H\alpha$ larger than the colors of the ZAMS (Zero Age Main Sequence) locus with EW$\rm_{H\alpha}$=-40$\AA$ and E$\rm_{B-V}$=1 defined by \citet{BarentsenVDG2011}, while in the latter diagram we used an ad-hoc lower limit for $r-H\alpha$. We also re-selected as stars with disks candidate galaxies with [3.6]<15.3$^m$. This limit was chosen by plotting in the [3.6] vs. [3-6]-[4.5] diagram the sources from the extragalactic catalogs compiled by \citet{Treister2006ApJ.640.603T} and \citet{Rafferty2011ApJ.742.3R}.

\subsection{Candidate shock- or PAH- dominated sources}
\label{pah_sect}

    Another class of possible contaminants are sources whose photometry in the [5.8] and [8.0] bands is contaminated by nebular PAH emission or unresolved knots of shock emission. We followed the prescription presented in \citet{GutermuthMMA2009ApJ} to select candidate contaminants of these two classes, and discarded those whose SED did not fit any YSO model (3 candidate unresolved knots of shock emission and 63 PAH-contaminated sources). The typical loci populated by PAH-contaminated sources and unresolved shocks in the [3.6]-[4.5] vs. [4.5]-[5.8] diagram are shown in Fig. \ref{diagrams5_img}

\subsection{Foreground stars and unreliable excesses}
\label{fore_sect}

    YSOs lying in the foreground of Dolidze~25 can contaminate our list of members with disks. Even if Gaia/EDR3 parallaxes are not useful to identify low-mass stars associated with Dolidze~25 and the Sh2-284 cloud because of their large distances, they can still be useful to select and discard stars in the foreground. However, since distances obtained by simply inverting Gaia parallaxes are not fully reliable for distances larger than 1$\,$kpc \citep[e.g.,][]{Bailer-Jones2018AJ.156.58B}, we also adopted a photometric criterion. We thus selected and removed as candidate foreground objects stars with disks with a parallax error smaller than 0.2$\,$mas, distance from Gaia parallaxes smaller than 2.5$\,$kpc, and the color $i$-$z$ from Pan-STARRS smaller than 0.25$^m$. \par
    
We also defined the following criteria to select objects with ``unreliable excesses'':
    
\begin{itemize}
\item Stars bluer than the expected pre-main sequence locus in the $i$ vs. $i-z$ (Fig. \ref{diagrams1_img}) or $J$ vs. $J-K$ (Fig. \ref{diagrams3_img}) diagrams;
\item stars with blue colors in the $i-z$ vs. $z-y$ diagram (Fig. \ref{diagrams2_img});
\item stars with colors in the $g-r$ vs. $r-i$  (Fig. \ref{diagrams2_img}) or $J-H$ vs. $H-K$  (Fig. \ref{diagrams3_img}) diagrams typical of low-extinction sources;
\item stars lying in the branch populated by low-extinction M stars in the $r-i$ vs. $i-J$ diagram (Fig. \ref{diagrams3_img});
\item stars with [3.6]-[4.5]$<$0.15 and [5.8]-[8.0]$<$1.2 in the [3.6]-[4.5] vs. [5.8]-[8.0] diagram (Fig. \ref{diagrams4_img});
\item stars with [4.5]-[5.8]$<$-0.3 in the [3.6]-[4.5] vs. [4.5]-[5.8] diagram (Fig. \ref{diagrams5_img}).
\end{itemize}

Objects selected according to one of the above conditions are removed from the list of stars with disks if their SEDs are compatible with photospheric models with an extinction A$\rm_V$between 0$^m$ and 100$^m$, or if they are not compatible with any YSO model, or if low extinction or distance are suggested by other diagrams. 

\subsection{Blue stars with excesses}
\label{bwe_sect}

Candidate stars with disks populating the expected locus of foreground main sequence stars in optical color-magnitude diagrams are not necessarily contaminants (stars in the foreground or galaxies, which, however, have already been removed from the list at this step), but they can also be genuine stars with disks, that we retain in our list of disks classifying them as BWE \citep[Blue With Excesses,][]{GuarcelloDMP2010} stars with disks. \par

 In fact, the optical colors of stars with ongoing accretion and with thick disks can be affected by the emission from accretion hot spots heated by the accreting material and light scattered along the line of sight by the dust in the disks. In the paradigm of magnetospheric accretion \citep{MuzerolleHC1998}, the accreting material funneled by the magnetic field falls onto the star at free-fall velocities of a few hundreds km/sec. The energy released by the accretion shock heats the surrounding stellar atmosphere up to more than 10000$\,$K (accretion hot spot), emitting soft X-rays, UV, and optical radiation at short wavelengths \citep{CalvetGullbring1998}. Besides, micron-size dust grains in protoplanetary disks can scatter part of the optical stellar emission along the line of sight. Since short-wavelength optical photons are more efficiently scattered than long-wavelength ones, the scattered light modifies the optical SED of stars with disks making optical colors bluer than photospheric values \citep[e.g.,][]{GuarcelloDMP2010}. \par
 
    We have selected the candidate stars with disks with optical colors bluer than the expected pre-main sequence locus in the following diagrams: $r$ vs. $r-i$, and $g$ vs. $g-r$, $r$ vs. $g-r$ (both VPHAS and Pan-STARRS, see Fig. \ref{diagrams1_img}), and discarded from our list of stars with disks the sources that: 

\begin{itemize}
\item are bluer than the expected pre-main sequence locus in the $J$ vs. $J-K$ diagram (e.g., they have $J-K$ colors typical of foreground objects);
\item OR have [3.6]-[4.5]$<$0.2;
\item OR have [5.8]-[8.0]$<$-0.3;
\item OR whose SED does not fit any YSO model.
\end{itemize}

\subsection{Final list of stars with disks}

\begin{figure*}[]
	\centering	
	\includegraphics[width=9cm]{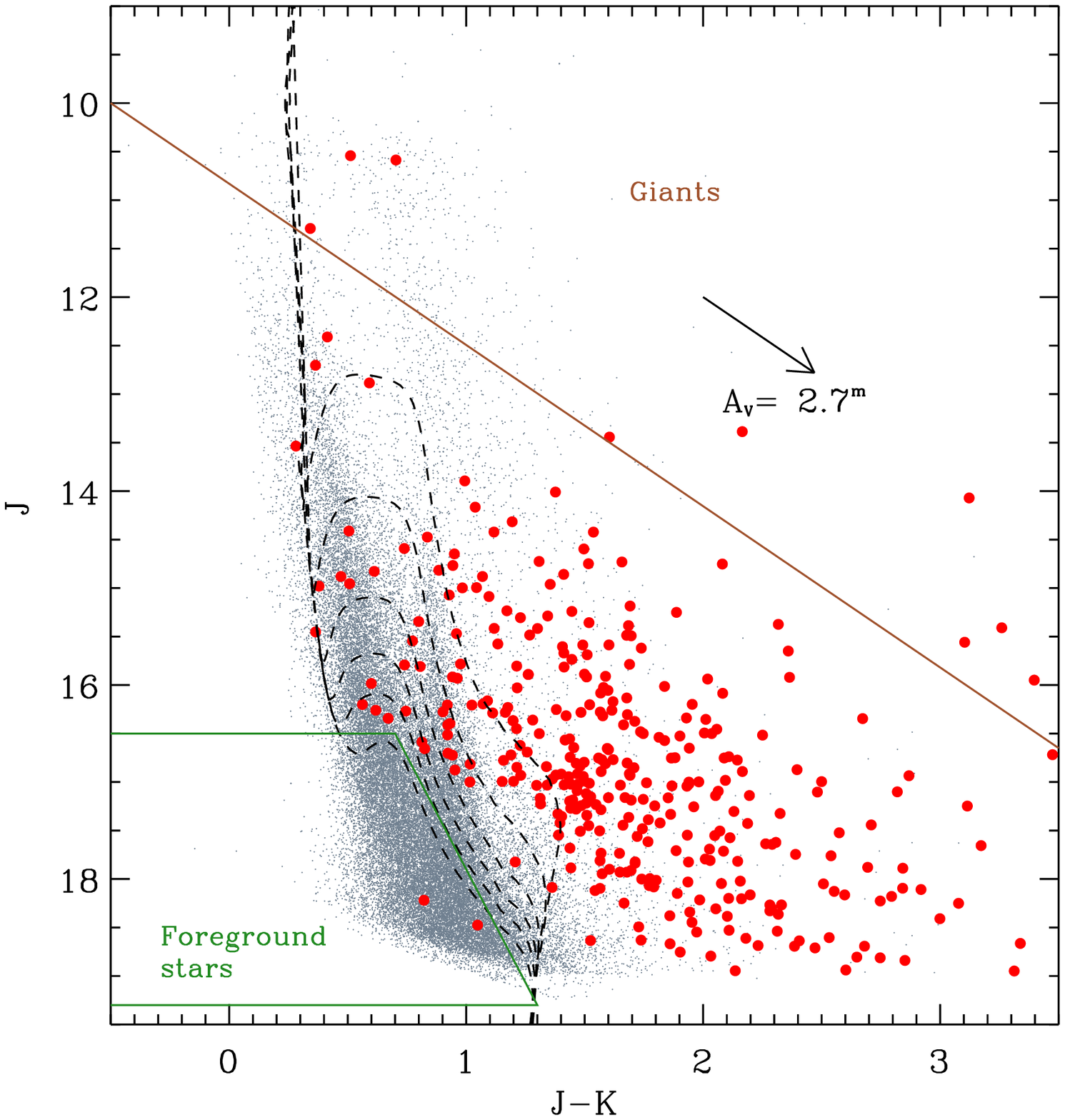}
	\includegraphics[width=9cm]{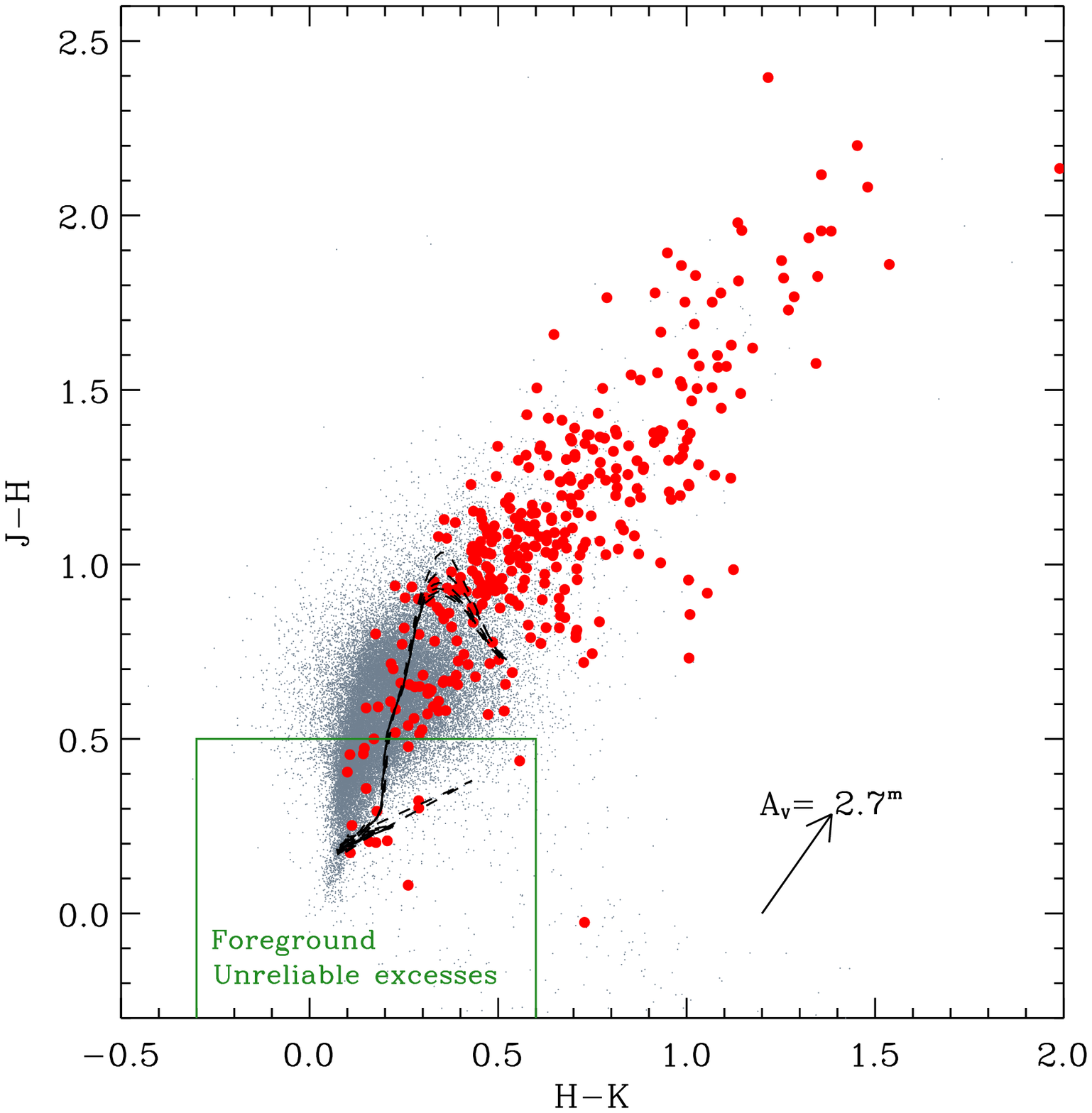}
	\includegraphics[width=9cm]{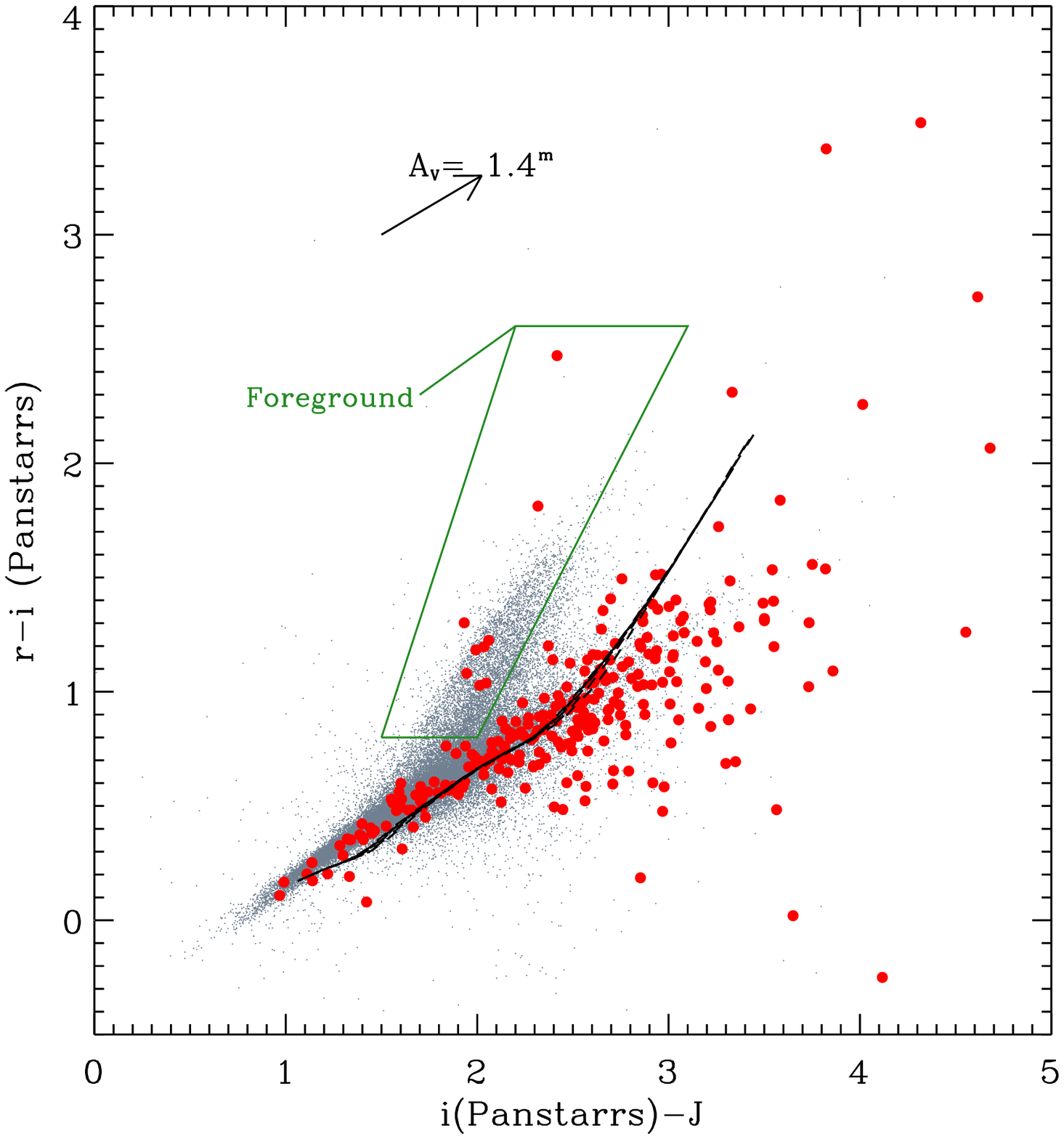}
	\includegraphics[width=9cm]{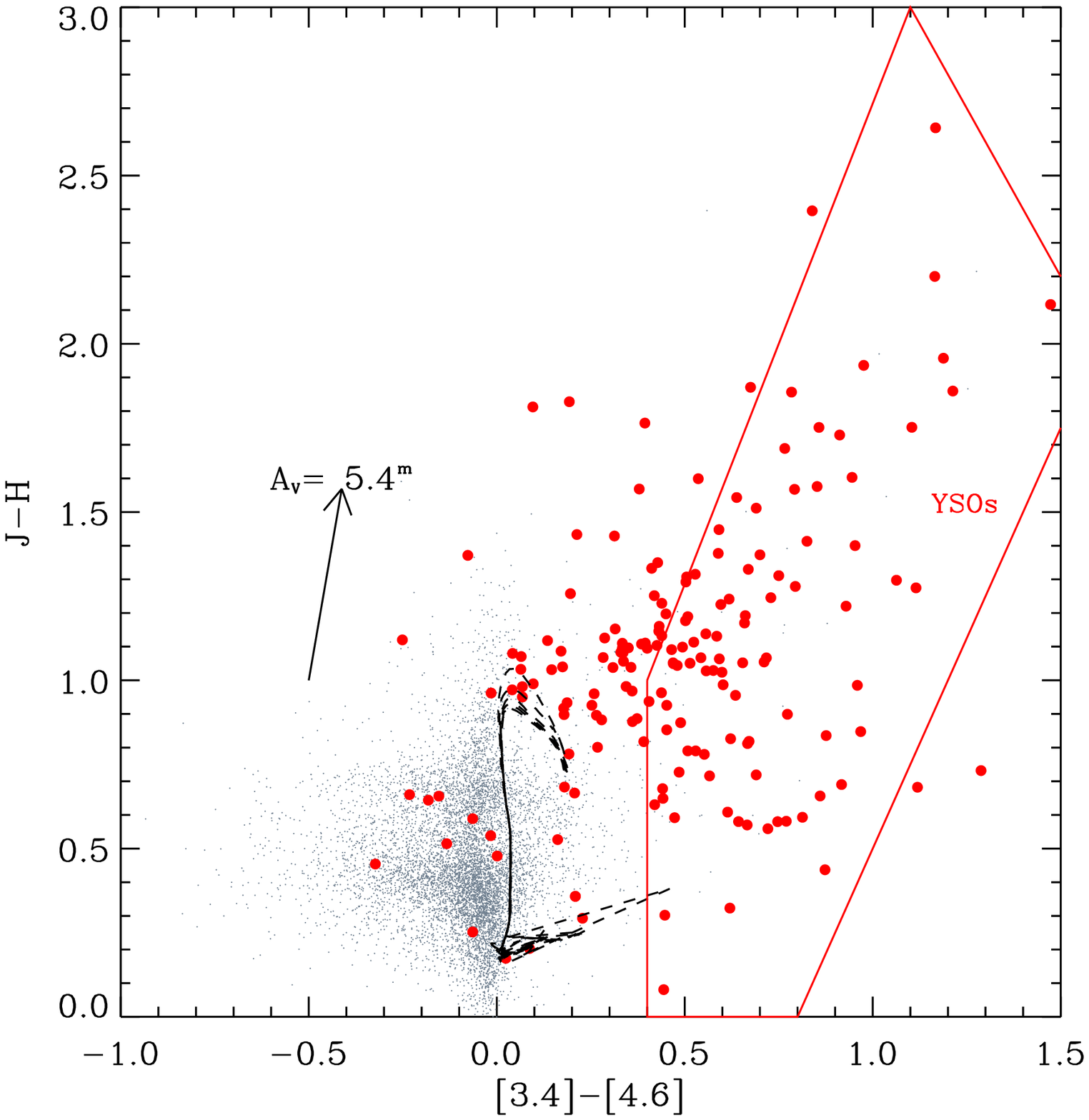}
	\caption{A subset of the infrared and optical-infrared diagrams of all sources falling in the studied field meeting the criteria of good photometry (e.g.: error in magnitude smaller than 0.1$^m$ and in color smaller than 0.15$^m$). The dashed lines show the isochrones for ages 0.5$\,$Myrs, 1.5$\,$Myrs, 3$\,$Myrs, 5$\,$Myrs, 8$\,$Myrs and 10$\,$Myrs, and with metallicity Z=0.004 \citep{DelgadoDA2010} from the PARSEC models, plotted adopting a distance of 4.5$\,$kpc and A$\rm_V$=2.7$^m$. Red dots mark the selected stars with disks retained in the final list. We also show the loci defined to select stars with disks and contaminants, delimited by red and green lines. In particular, in these diagrams we show the loci expected to be populated by giants, stars with unreliable excesses, foreground stars, and Young Stellar Objects (YSOs) with disks. All used diagrams are shown in Appendix \ref{AppE}}
	\label{diagrams3_img}
	\end{figure*}

\begin{figure*}[]
	\centering	
	\includegraphics[width=12cm]{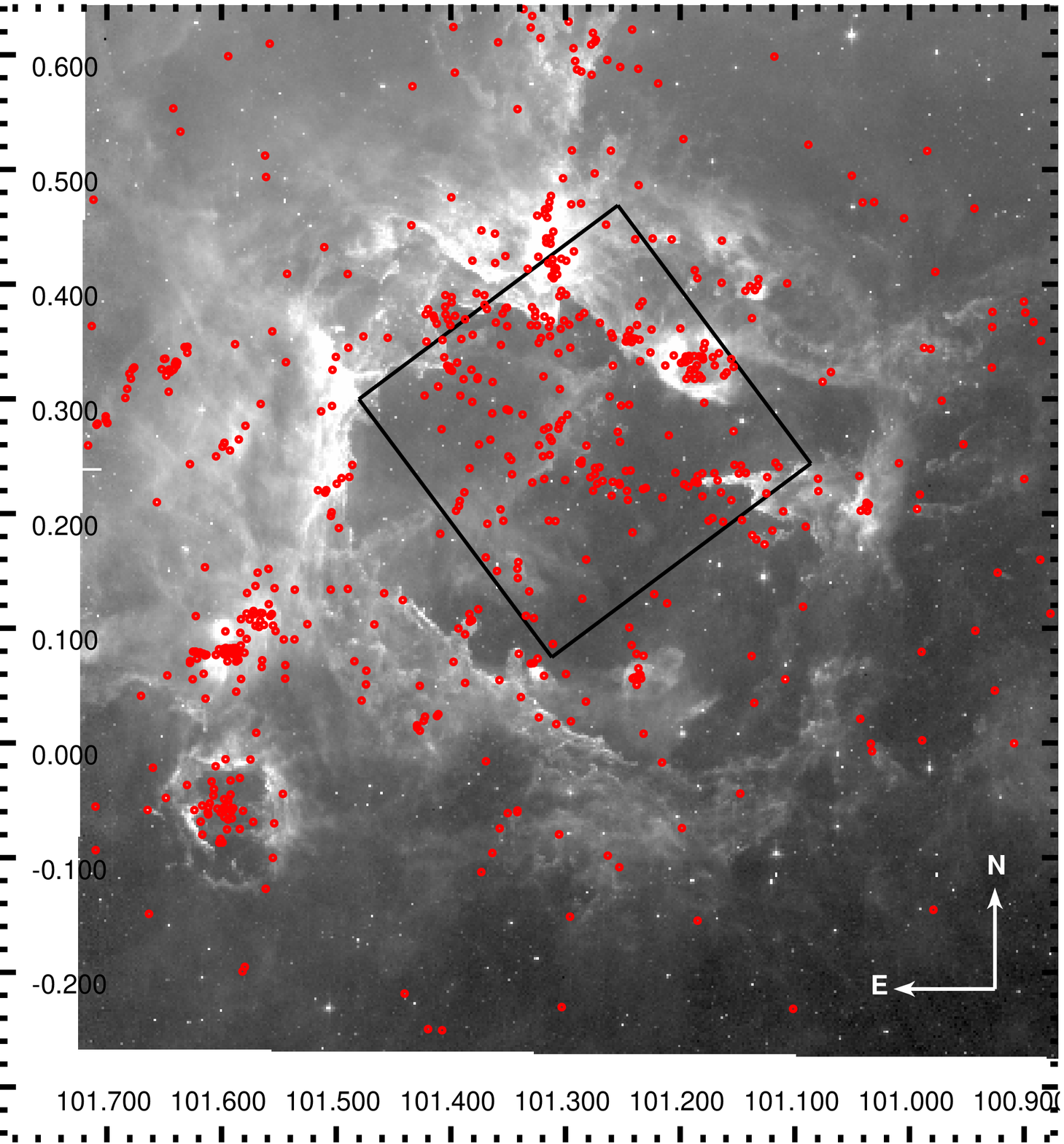}
	\caption{Spitzer/IRAC image in the [8.0] band of Dolidze~25 and the surrounding Sh2-284 complex, with marked the positions of selected stars with disks (red circles). The black square delimits the field observed with $Chandra$.}
	\label{spadis_img}
	\end{figure*}

After the pruning process, the list of stars with disks counts 659 stars. Figs. \ref{diagrams3_img}, \ref{diagrams5_img}, \ref{diagrams1_img}, and \ref{diagrams2_img}, \ref{diagrams4_img} show the color-color and color-magnitude diagrams of all stars with good photometry falling in the studied field, selected stars with disks, and the loci used to define all the adopted tests. Fig. \ref{spadis_img} shows the spatial distribution of the stars with disks in Dolidze~25 and the surrounding Sh2-284 cloud. Compared with the selection of stars with disks made by \citet{Puga2009AA.503.107P}, we have selected about a factor two more objects (659 vs. 329 sources). \par

\section{Candidate young stars without disks}
\label{class3_sect}

Intense magnetic activity of pre-main sequence stars produces an enhanced X-ray emission with respect to older stars \citep{Montmerle1996}. Young stars in star forming regions can thus be selected and separated from other sources falling in the same field of view by requiring detection in X-rays observations. As explained in Sect. \ref{chandra_sec}, the central cavity of Sh2-284 populated by Dolidze~25 stars was observed with $Chandra$/ACIS-I, and we detected and validated 618 X-ray sources. Among these sources, 486 have matched at least one optical-infrared star in our multi-wavelength catalog (considering the multiple coincidences, our catalog contains 542 X-ray sources with OIR counterparts, see Sect. \ref{mergedcat_sect}).\par

    \begin{figure}[]
	\centering	
	\includegraphics[width=9cm]{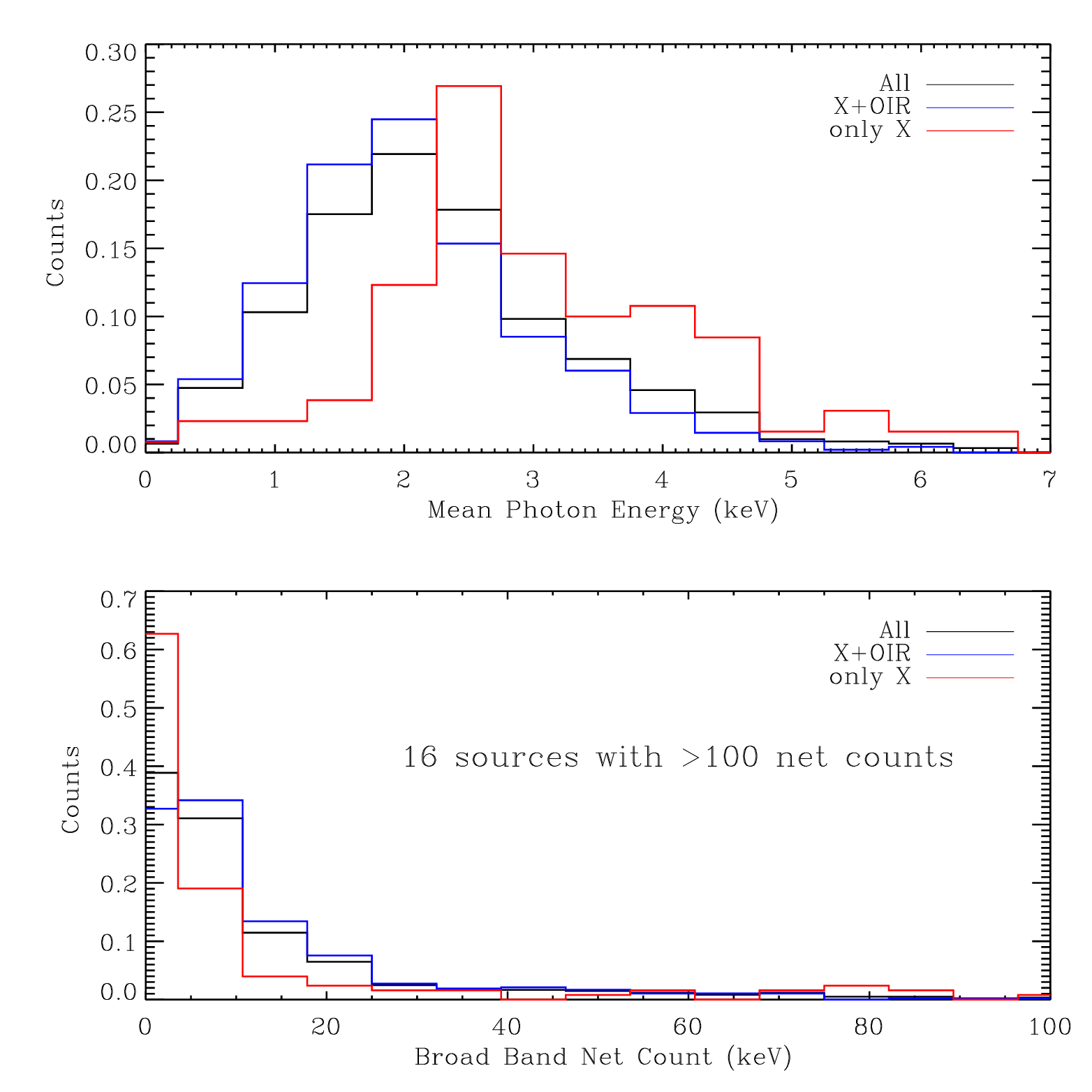}
	\caption{Distribution of the mean photon energy and net-counts in the broad energy band for the validated X-ray sources, and shown separately for the X-ray sources with and without OIR counterpart.}
	\label{xprop_img}
	\end{figure}
    
     Although the sample of X+OIR sources is expected to be dominated by young stars in the area, it can still contain a significant number of sources not associated with Dolidze~25, such as magnetically active stars not in the pre-main sequence phase, that must be identified and removed from the list of candidate members of the cluster. First, we considered the 131 X-ray sources that have not matched any OIR source. These sources can be either truly members of Dolidze~25 with very high extinction, false-negative produced in the catalogs-merging procedure, or extragalactic sources. Fig. \ref{xprop_img} shows the distribution of the mean photon energy and the net counts in the broad energy band of all the validated X-ray sources, and separately those of the X-ray sources with and without an OIR counterpart. The distribution of the mean photon energy of all X-ray sources and those with OIR counterparts peaks at $\sim$2.2$\,$keV. The distribution of the mean photon energy of the X-ray sources without counterpart has two peaks, with one at $\sim$2.5$\,$keV and one at about 4$\,$keV. This may suggest that this sample contains both stars and extragalactic sources. However, the typical net counts of these sources is $\leq$10 photons, which precludes further investigation. For this reason, since we can not distinguish between these two possibilities, we removed the X-ray sources without OIR counterpart from the list of disk-less members. The exception to this is a group of 16 X-ray sources without OIR counterparts, with a net count of more than 50 photons, and a mean photon energy between 1.2 and 3.2$\,$keV, that we retained in our list. We will attempt to constrain their nature by analyzing their X-ray spectra in a forthcoming work. \par 
     
Then, we selected and removed the X+OIR sources that are likely in the foreground or in the background. In order to search for extragalactic contaminants among the X-ray sources with OIR counterpart, we used the IRAC and WISE color-color and color-magnitude diagrams where we defined the typical loci of extragalactic sources (Figs. \ref{diagrams4_img} and \ref{diagrams5_img}). Only 6 X+OIR sources not classified as disk-bearing members populate these loci. Among these sources, two fall in the loci of extragalactic sources in more than half of the diagrams where they can be plotted, and thus they were removed from the list of candidate disk-less young stars. In order to select candidate X+OIR sources in the foreground, we have first removed 43 X+OIR sources with errors in the Gaia parallaxes smaller than 0.2$\,$mas and distances from parallax smaller than 2.5$\,$kpc. We also removed other 20 X+OIR sources with large parallax errors and falling the ``foreground'' or ``BWE'' loci in most of the diagrams where they can be plotted. After this pruning, we compiled a list of 379 candidate {\it unique} X-ray sources with at least one OIR counterpart, candidate for being disk-less young stars of Dolidze~25. As expected, the removed X+OIR sources have an almost uniform spatial distribution.

  \begin{figure*}[]
	\centering	
	\includegraphics[width=9cm]{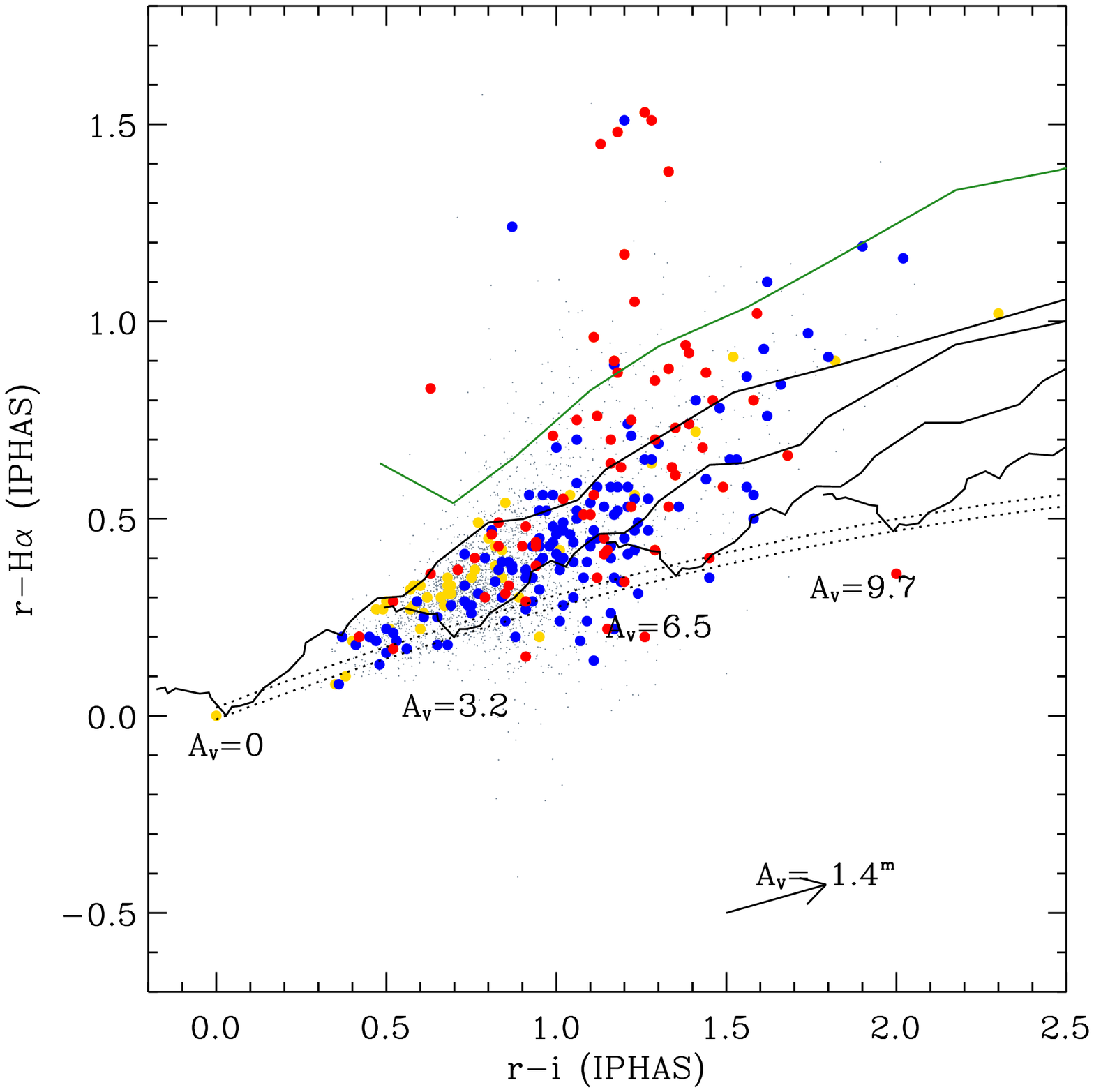}
	\includegraphics[width=9cm]{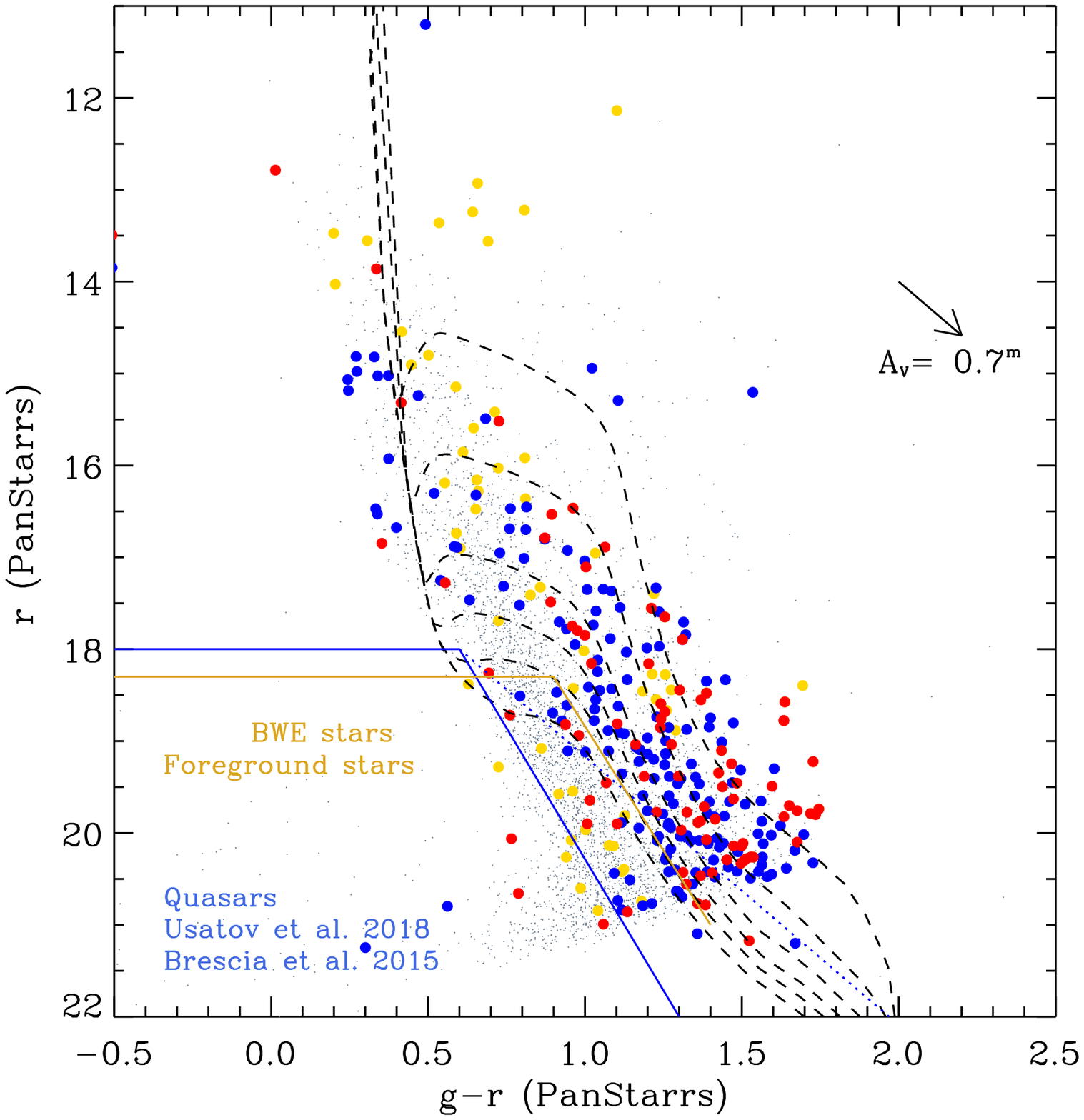}
	\includegraphics[width=9cm]{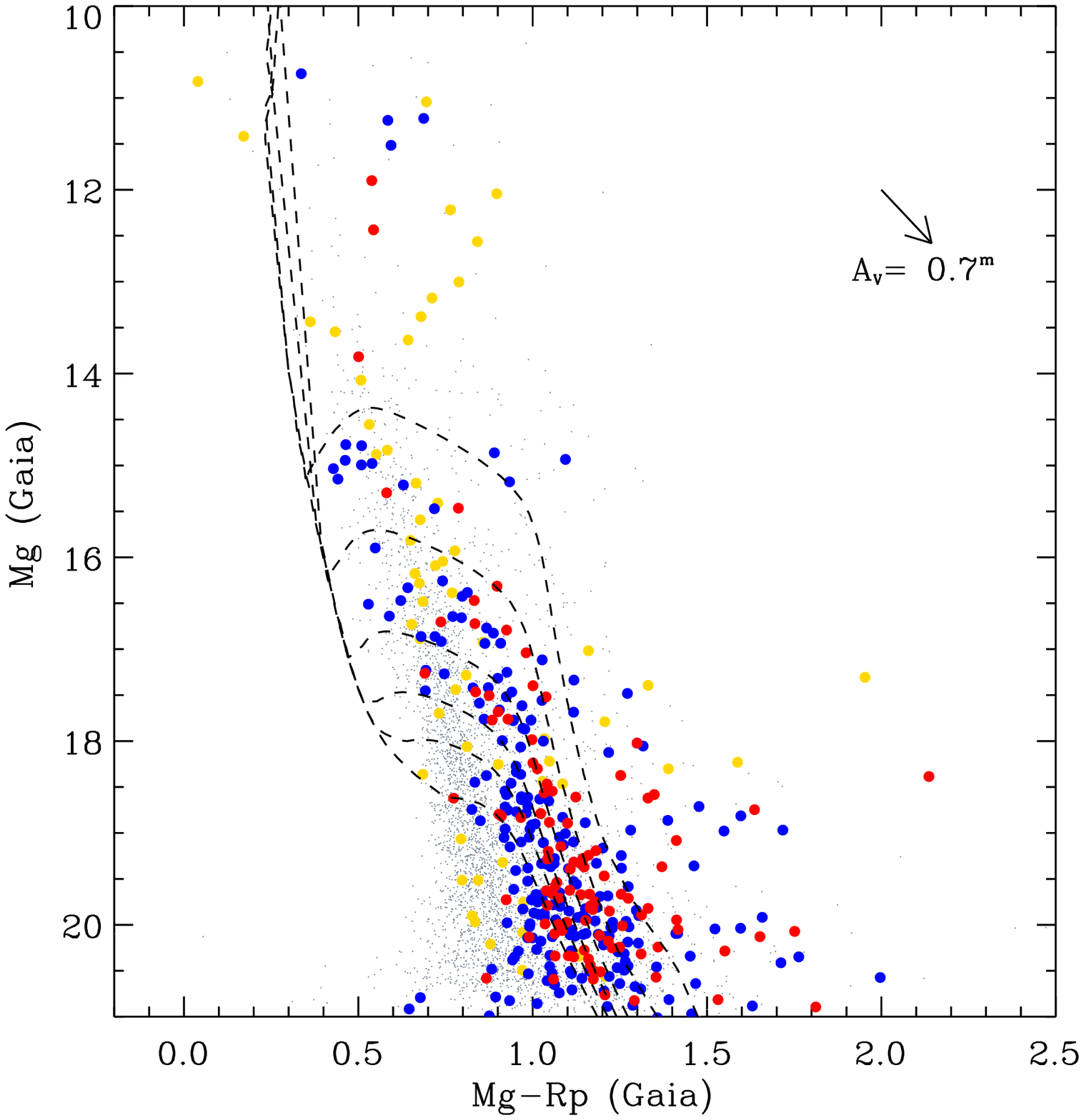}
	\includegraphics[width=9cm]{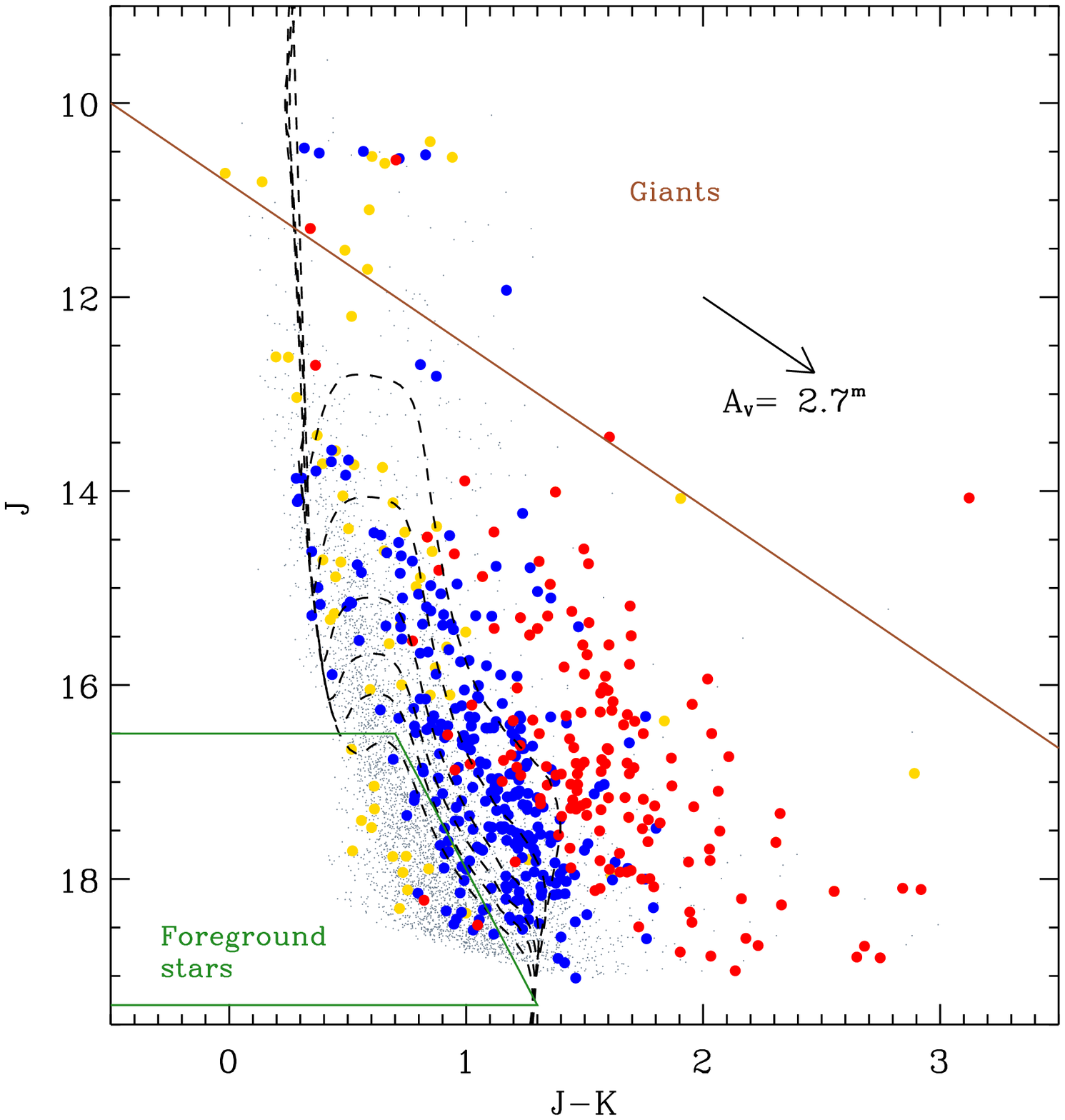}
	\caption{Diagrams of all sources falling in the ACIS FoV meeting the criteria of good photometry (gray points). Figure layout and symbols are as in Fig. \ref{diagrams3_img}. The large dots mark: candidate stars with disks inside the ACIS FoV (red), candidate young stars without disks (blue), X+OIR sources in the foreground or background (yellow).}
	\label{diagrams6_img}
	\end{figure*}

  \begin{figure}[]
	\centering	
    \includegraphics[width=9cm]{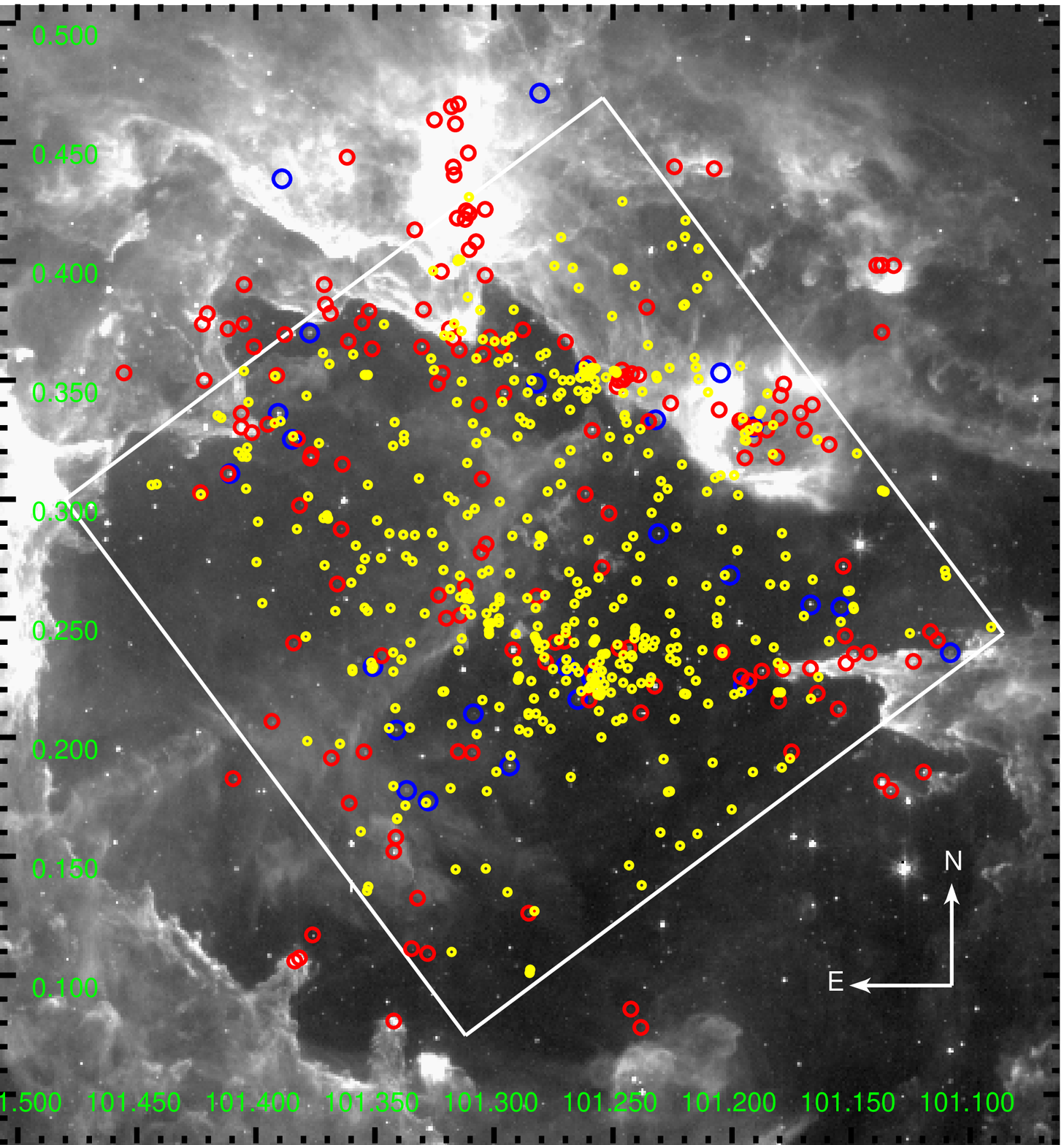}
	\caption{IRAC image in the [8.0] band of the field around Dolidze~25, together with the position of candidate stars with disks (red), without disks (yellow) and spectroscopic members (blue). We also indicate with a square the limits of the $Chandra$/ACIS-I field.}
	\label{class3_spadis}
	\end{figure}

	Fig. \ref{diagrams6_img} shows a selection of optical and infrared diagrams of all stars in the ACIS FoV matching the criteria for good photometry, the candidate stars with disks (inside the ACIS FoV), the members without disks and the X+OIR sources removed from the list of disk-less stars. In the optical color-magnitude diagrams, the pre-main sequence locus at the distance and extinction of Dolidze~25 is well defined by the selected candidate members of the cluster. Two candidate members without disks show a significant excess in H$\alpha$ despite not being included in the list of stars with disks, and would need to be further investigated to discern their nature. The X+OIR sources discarded from the list of members clearly show colors of foreground and background sources. Fig. \ref{class3_spadis} shows the spatial distribution of both disk-less and disk-bearing stars inside the cavity hosting Dolidze~25. With respect to the regions identified by \citet{Puga2009AA.503.107P}, we have identified a new group of members rich of disk-less stars along the cavity between the RN and Cl2 regions.

%

\section{Comparison with members from the literature and final selection}  
\label{othermemb_sect}

The list of members compiled by \citet{DelgadoDA2010} over a small area at the center of Dolidze~25 includes: 29 stars with IR excesses, 102 main sequence stars, and 103 pre-main sequence stars. Among these stars, we retrieved in our list of members: 8 of their stars with infrared excesses (3 disk-less and 5 with disks), 9 of their main sequence members (all as disk-less members), and 20 of their pre-main sequence stars (17 disk-less and 3 with disks). After checking the positions in the various optical and infrared diagrams of the stars selected by \citet{DelgadoDA2010} and not by us, we added only one star from their lists of members while the other stars do not lie with the other cluster members in all color-color and color-magnitude diagrams and thus were not included in our list of members. \par

    \citet{Puga2009AA.503.107P} selected 155 class~I and 183 class~II sources in the area around Dolidze~25 from specific Spitzer/IRAC observations of Sh2-284. Our list of disk-bearing members has in common with their list 116 class~I and 153 class~II sources. We discarded 17 of their members as probable contaminants. With the exception of three stars, that we added in our list, the few stars in the \citet{Puga2009AA.503.107P} list not included in our list do not meet the criteria we defined for good photometry in the IRAC bands. \par
    
\citet{Negueruela2015AA.584A.77N} studied optical spectra of the brightest stars in the field of Dolidze~25. Since the criteria that we defined to select and discard candidate giant stars with circumstellar dust (Sect. \ref{giant_sect}) and foreground stars (Sect. \ref{class3_sect}) are not designed for intermediate massive members of Dolidze~25, these stars have been automatically discarded from our list of members. We thus analyzed separately the stars included in the list of \citet{Negueruela2015AA.584A.77N}, including in our list of members 11 of these OB stars. According to Gaia data, we confirm that stars S9 and HD~48691 (from parallaxes) and HD~48807 (from proper motion) are likely stars in the foreground. We also confirm the IR and H$\alpha$ excesses of the stars S24 and SS57, classified as Herbig Be stars by \citet{Negueruela2015AA.584A.77N}. Considering among these stars those with errors in parallax smaller than 0.2$\,$mas, the median distance of these stars is 4.5$\pm$0.5$\,$kpc. As explained in Sect. \ref{intro_dol25}, this is the distance value that we adopted to plot the isochrones in all the color-magnitude diagrams shown in the paper. \par
    
    \citet{CusanoRAG2011} performed a spectroscopic and photometric analysis of 23 pre-main sequence objects in the center of Dolidze~25. Among the 6 stars identified by \citet{CusanoRAG2011} as disk-less members, two are detected in X-rays and classified as disk-less members also in our work. Since the spectroscopic evidence supporting the membership to Dolidze~25 of the remaining four stars are solid, we changed their status into disk-less members. We also selected as stars with disks 12 of the 17 stars in the list of class~II objects of \citet{CusanoRAG2011}. Among the remaining 5 stars, three do not match our criteria for good photometry in the diagrams we used to select stars with disks, and two were discarded as likely background giant contaminants (respectively an F0V and F2V star, according to the spectral classification provided by \citealt{CusanoRAG2011}). We re-classified these five stars as stars with disks of Dolidze~25. We also classify as class~II objects two other stars classified as class~I objects by \citet{Kalari2015ApJ.800.113K}.  \par

Our final list of confirmed young stars associated with Dolidze~25 and Sh2-284 counts 667 stars with disks, 424 stars without disks, and 10 spectroscopically confirmed massive and intermediately massive stars. The final catalog of the candidate young stars in Dolidze~25 and the surrounding area, available on-line, is described in the Appendix \ref{app_members}. 

\section{The disk fraction in Dolidze~25}
\label{df_sect}

In this section we calculate and analyze the disk fraction of Dolidze~25, where we selected both the disk-bearing and disk-less stellar population. A simple visual inspection of Fig. \ref{class3_spadis} shows that the spatial distribution of selected members inside the ACIS FoV is not homogeneous. The candidate members are apparently separated in two main populations: the main cluster inside the cavity and a population in the north along the bright rim of the cavity, mainly composed of the {\it cluster 2}, the bright rimmed cloud $RN$, and a population of disk-less sources between these two groups. An accurate analysis of the disk fraction in Dolidze~25 needs to account for any possible difference between the properties of these two populations, such as stellar age and mass content. \par

    \subsection{Analysis of the spatial distribution of members}
    \label{mst_sect}
    
    A detailed analysis of the subclustering in Dolidze~25, which would require also high quality astrometric and stellar dynamic data, is beyond the scope of this paper. However, we want to verify whether suitable statistical tests support the existence of two separated groups of cluster members. To this aim, we calculated the Minimum Spanning Tree (MST) of the cluster \citep{BarrowBS1985}. The MST consists in the unique set of branches connecting all points in a 2-D scatter plot with the minimum total length and not producing closed loops. We calculated the MST of the selected members using the {\it Analyses of Phyogenetics and Evolution} ($ape$) $R$ package \citep{Paradis10.1093/bioinformatics/bty633}. 
	
    \begin{figure}[]
	\centering	
	\includegraphics[width=9cm]{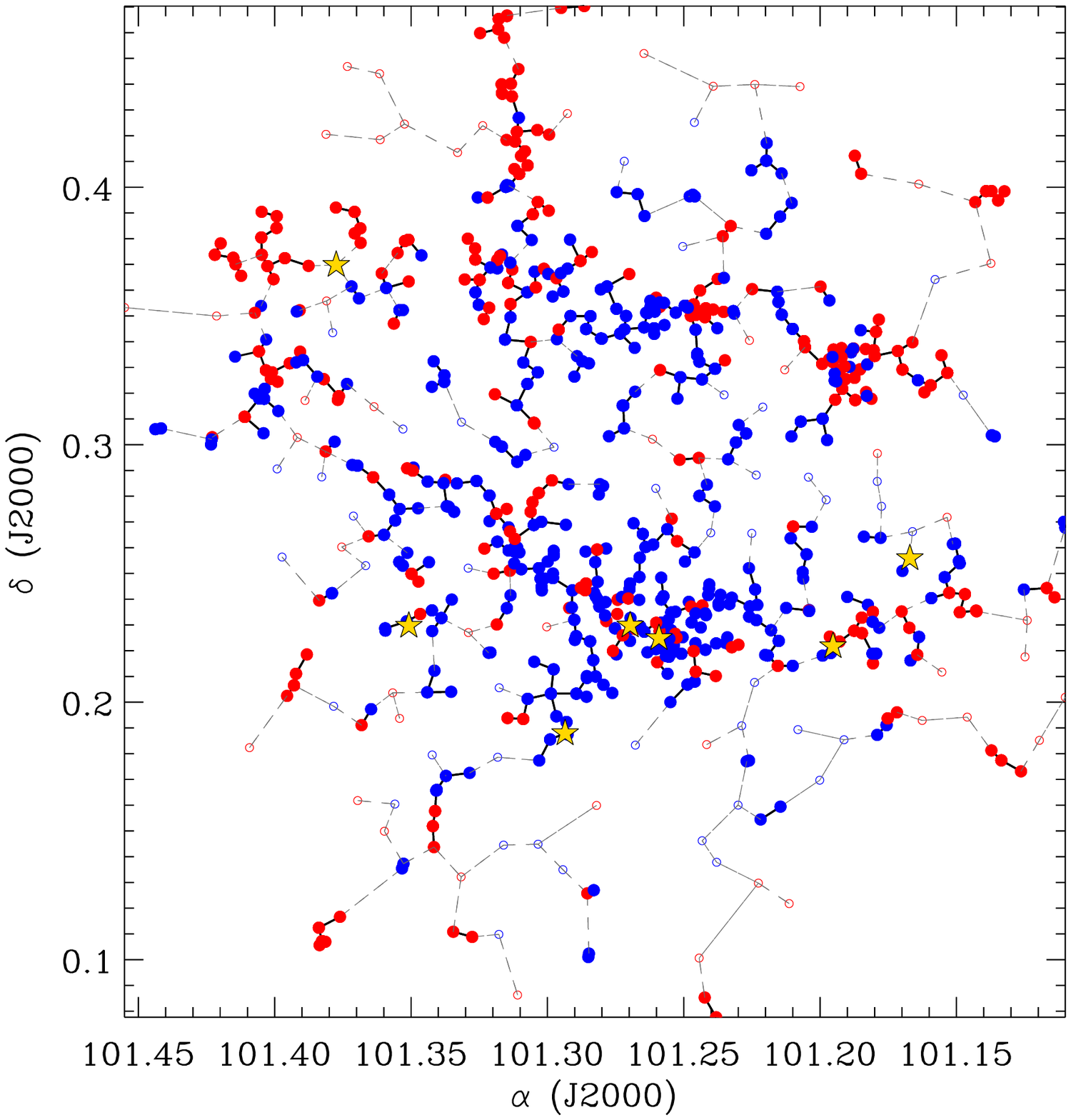}
	\caption{The Minimum Spanning Tree built on all the members selected inside the ACIS FoV. Members without disks are marked with blue dots, those with disks with red dots. The yellow star symbols mark the position of the most massive members. Filled dots mark the position of clustered members, e.g. closer than 34$^{\prime \prime}$.3 to at least one other member (see the text).}
	\label{mst_figure}
    \end{figure}

	Fig. \ref{mst_figure} shows the MST built on all the candidate members found in the ACIS FoV. In order to discern between clustered and non clustered members, we also estimated the {\it critical branch length} as the branch length at which the cumulative distribution of the branch lengths changes slope \citep{GutermuthMMA2009ApJ}. This calculation resulted in a critical branch length of 34$^{\prime \prime}$.3, corresponding to a projected distance of 0.75$\pm$0.08$\,$pc at a distance of 4.5$\,$kpc. We then considered as ``$clustered$'' all members separated from the closest member by a distance smaller than the critical branch length, and called ``$sparse$'' the remaining stars. In Fig. \ref{mst_figure}, ``clustered'' stars are marked with a filled dots, and the ``sparse'' population with empty dots. The MST confirms the presence of two main stellar groups: a group inside the cavity, corresponding to the main cluster, and an elongated northern group lying along the front of the cavity (called hereafter the $central$ and $northern$ populations, respectively). Several other small groups are identified, but the identification of these groups as real subclusters is beyond the scope of our work. We have also applied the method introduced by \citet{Allison2009MNRAS.395.1449A} to explore the presence of mass segregation in the cluster, and concluded that our data do not support this possibility.
	
	\subsection{Extinction across the field}
	\label{ext_sect}
	
	To estimate a reliable disk fraction of Dolidze~25 and compare it with that of other clusters, we need to quantify individual stellar parameters. We need to estimate the median age of cluster members to compare disk dispersal timescale in Dolidze~25 with other young clusters, while stellar masses are necessary to account for the incompleteness of our selection. Stellar parameters were evaluated by placing cluster members in derreddened color-color and color-magnitude diagrams, after evaluating individual extinctions. Individual extinctions can in principle be estimated by calculating in color-color diagrams the displacement along the extinction vector of stars from a representative isochrone drawn assuming zero extinction. \par
	
        The reliability of the estimate of individual extinctions may depend on the particular diagram that is used. A better estimate is in fact possible if the adopted isochrone is sufficiently regular in order to avoid unrealistic discontinuities in the distribution of the resulting extinctions, and if the slope of the isochrone is sufficiently different from that of the extinction vector, in order to avoid that photometric uncertainties would result in too large extinction uncertainties.   \par

   \begin{figure}[]
	\centering	
	\includegraphics[width=9cm]{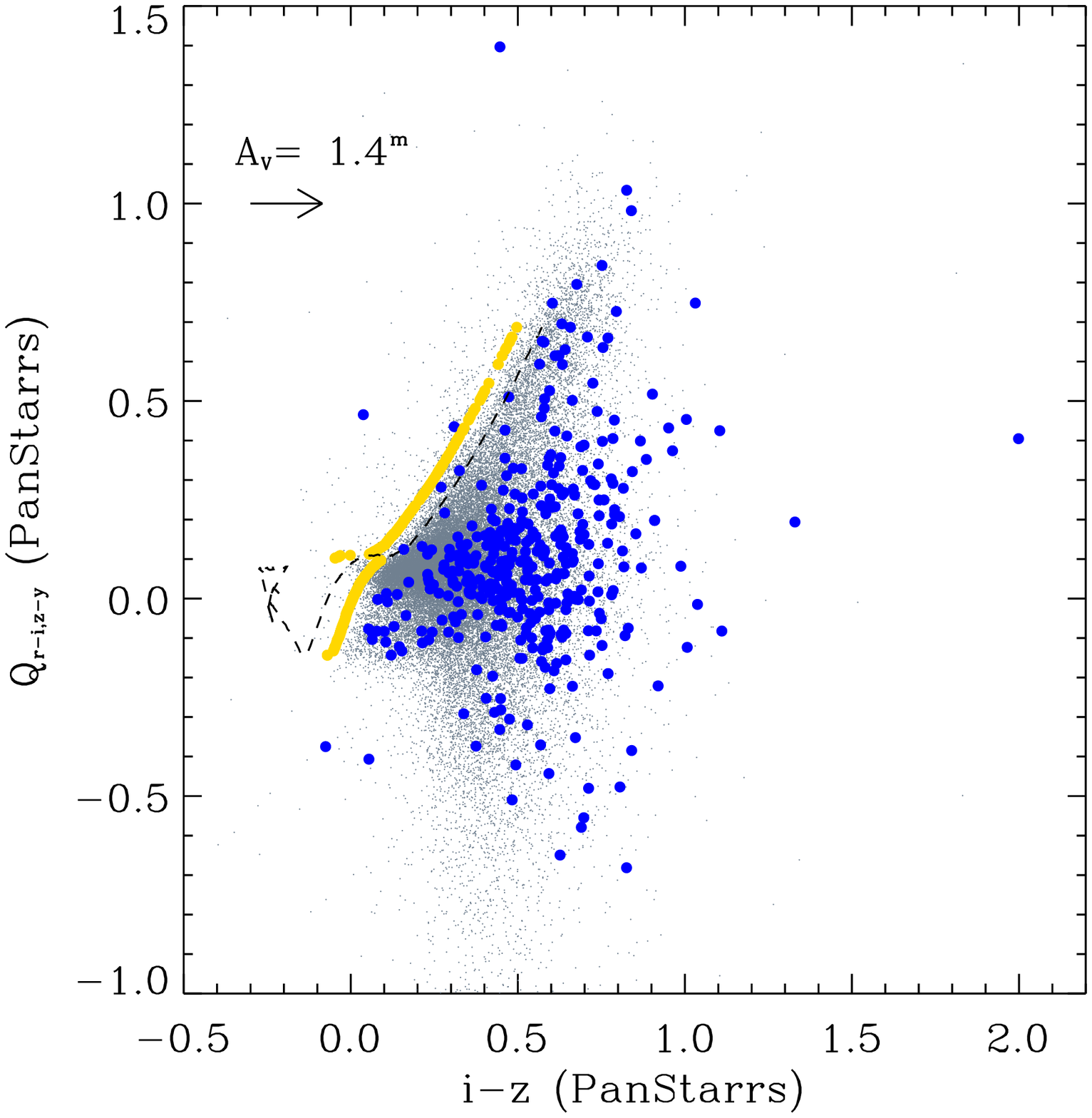}	
	\caption{Pan-STARRS $Q_{rizy}$ vs. $i-z$ diagram of all sources in the studied field with errors in colors smaller than 0.15$^m$. The blue dots mark the observed colors of the candidate members of Dolidze~25, while the yellow dots mark their positions after dereddening their colors. The dashed lines mark the zero-extinction low-metallicity PARSEC isochrones for the pre-main sequence phase assuming an age of 1.5$\,$Myrs.}
	\label{avdiag_figure}
    \end{figure}

	After several tests, we adopted the Pan-STARRS diagram $Q_{rizy}$ vs. $i-z$ (Fig. \ref{avdiag_figure}) to estimate individual stellar extinctions. $\rm Q_{rizy}$ is the reddening-free color index $(r-i)-(i-z)\times E_{r-i}/E_{z-y}$ \citep[similar to those defined by][]{Damiani2006} where the ratio $E_{r-i}/E_{z-y}$ in the Pan-STARRS bands is equal to 2.214 according to the reddening law of \citet{CardelliCM1989} and \citet{Donnell1994ApJ} (in Appendix \ref{AppB} we summarize the extinction coefficients adopted in all the bands used in this work). Since the optical colors of disk-bearing stars may not be fully representative of their stellar properties due to the emission from accretion hot-spots, light scattered by the disk and the partial occultation of the stars by their disks (see Sect. \ref{bwe_sect}), we will calculate individual extinction for both members with and without disks, but we will use only the estimate from the latter to derive the extinction map and median extinction of the cluster. As evident in the right panel of Fig. \ref{avdiag_figure}, the PARSEC isochrones in the $Q_{rizy}$ vs. $i-z$ diagram, considering only the PMS stage, present an horizontal segment at about $\rm Q_{rizy}$=0.1, which results in an unrealistic discontinuity in the distribution of the resulting extinctions. We thus averaged the $i-z$ values of the isochrones, increasing by 0.07$^m$ those for points with $\rm Q_{rizy}>0.1$ and decreasing by the same amount those with $\rm Q_{rizy}<0.1$. Moreover, we restricted the calculation of extinction from $Q_{rizy}$ vs. $i-z$ only for stars with -0.15$^m$$\rm \leq Q_{rizy} \leq$0.7$^m$, which is the range covered by the adopted isochrone. \par

\begin{figure*}[]
	\centering	
	\includegraphics[width=9cm]{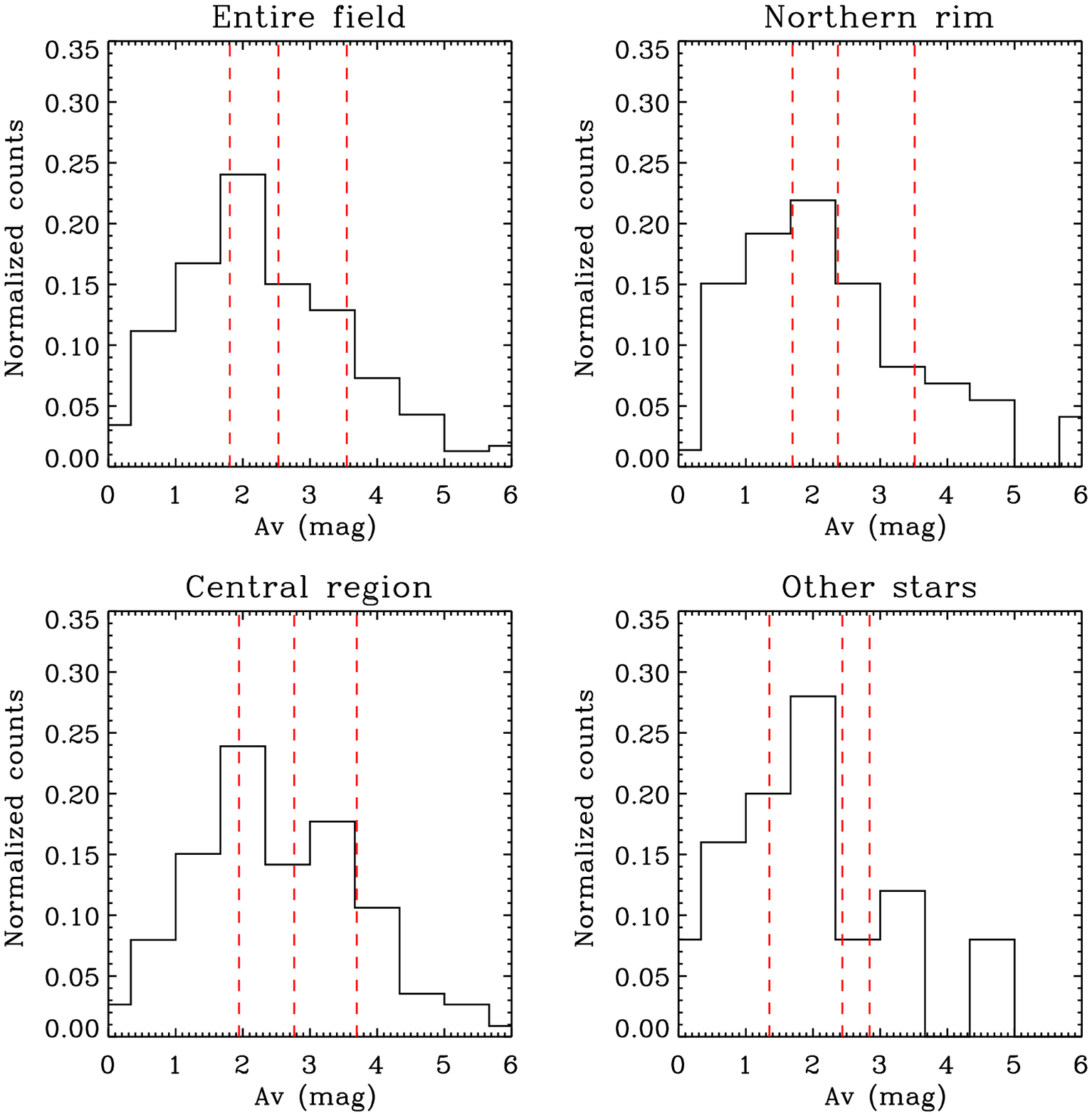}
	\includegraphics[width=9cm]{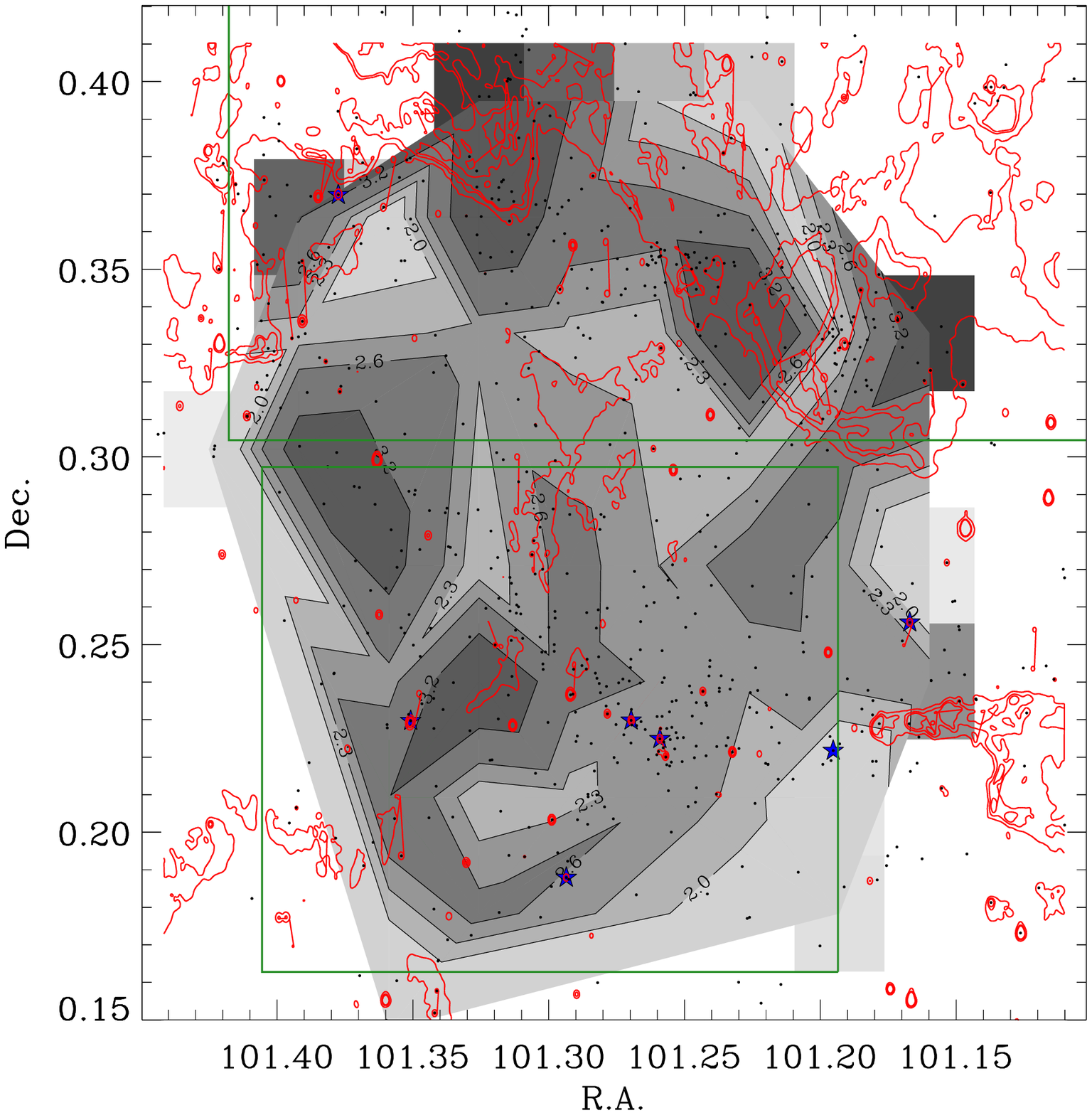}	
	\caption{Left: Distribution of the individual extinctions of all disk-less members, and separately for sources in the $central$ cluster, $northern$ group, and in the ``sparse'' population. The vertical lines mark the median and the 25\% and 75\% quantiles of each distribution. Right: Extinction map in the ACIS field. The red lines mark the emission level at $8.0\,\mu$m from IRAC images. The green boxes roughly delimit the central and northern groups.}
	\label{av_figure}
    \end{figure*}

We computed the individual extinctions for 241 candidate disk-less members. The resulting A$_V$ distribution has the 25\%, 50\%, and 75\% quantiles equal to 1.7$^m$, 2.3$^m$, 3.2$^m$, respectively. Fig. \ref{av_figure} shows the distributions of individual extinctions of all disk-less members, and separately those of the stars in the $central$ group, the $northern$ group, and the ``sparse'' population. The right panel also shows the resulting extinction map across the ACIS field, plotted together with the contours marking the emission levels at $8.0\,\mu$m in IRAC images, which help to visualize the distribution of nebular dust emission. Despite the largest values of extinctions being observed where the dust emission is more intense, such as in the northern part of the field, the A$\rm_V$ distributions of the $central$ and $northern$ groups are quite similar, with differences within 1 magnitude. 

	\subsection{Mass and age of cluster members}   
	\label{massage_sect}
 
    Masses and ages of candidate members were estimated by interpolating their positions in selected derreddened color-magnitude diagrams on a grid computed using the 0.5-10$\,$Myrs low-metallicity PARSEC isochrones. Details on how individual masses and ages were calculated are provided in the Appendix \ref{AppC}. Fig. \ref{agedist_figure} shows the distribution of individual stellar ages calculated for both disk-less and disk-bearing candidate members (the latter only within the ACIS field). The age distribution shows a clear peak at about log(age)$\sim$6.0 [Myrs], with a median age equal to log(age)$\rm _{median}$=6.2 [Myrs], with a standard deviation of 0.3 [Myrs].
 
    \begin{figure}[]
	\centering	
	\includegraphics[width=8cm]{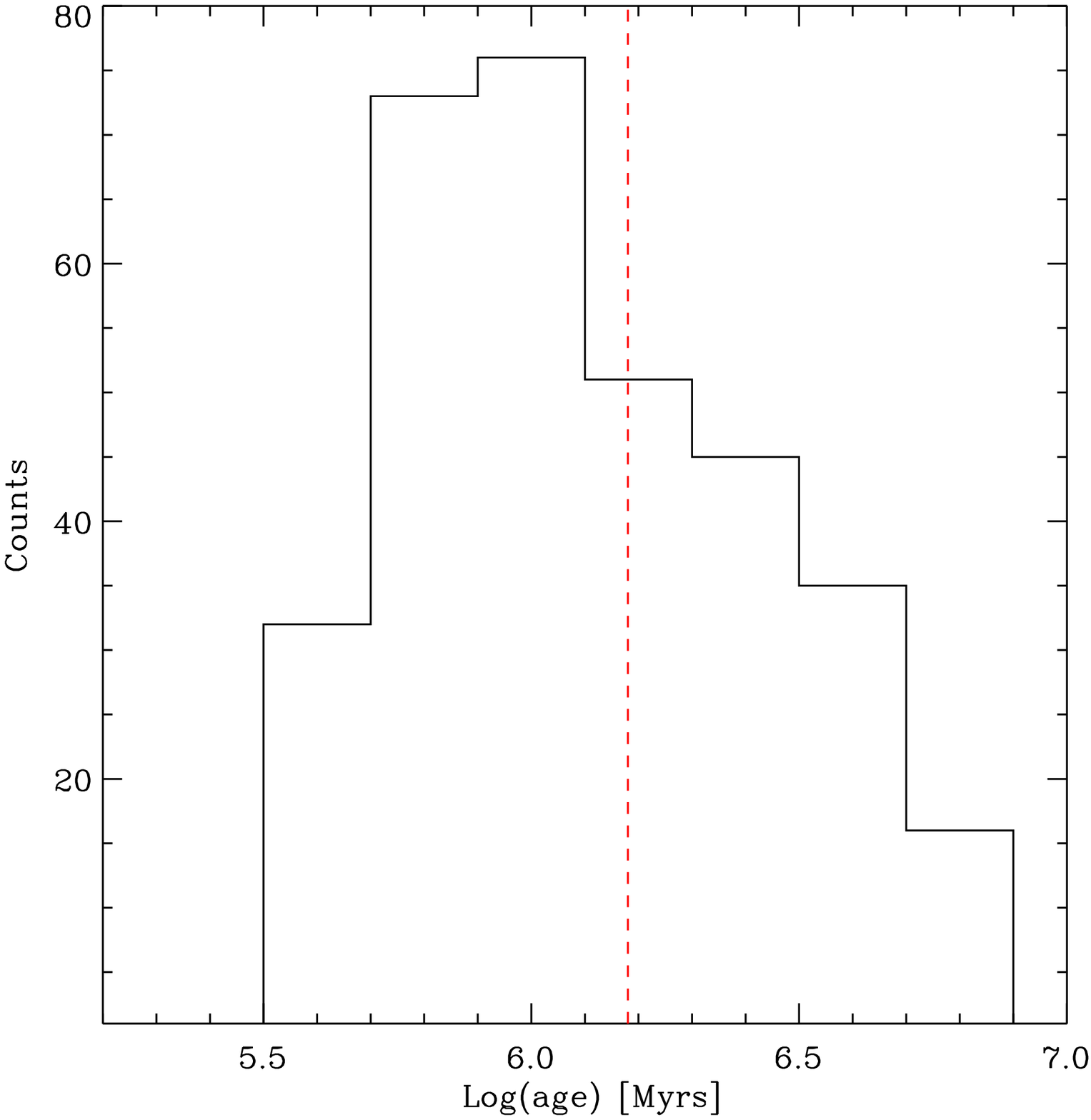}
	\caption{Distribution of individual stellar ages for candidate members within the ACIS field. The vertical lines mark the median value of the distribution. The distribution peaks at about log(age)$\sim$6.0 [Myrs], with a median age equal to log(age)$\rm _{median}$=6.2 [Myrs].}
	\label{agedist_figure}
    \end{figure}

Our estimate of the median age of Dolidze~25 is smaller than the estimate presented by \citet{DelgadoDA2010}, who selected two populations of candidate members from optical and infrared photometry, the youngest with a median age of log(age)=6.7$\pm$0.2 [Myrs] coexisting with a population older than 40$\,$Myrs, and by \citet{TurbideMoffat1993AJ.105.1831T} who estimated an age of 6$\,$Myrs fitting the upper main sequence to 12 candidate bright members. These differences can be understood as a consequence of the fact that our study is the first where disk-less members (for which age estimation from color-magnitude diagrams is more reliable) are selected down to the low-mass stars regime. Besides, the existence of an old population suggested by \citet{DelgadoDA2010} was not confirmed by other authors. Our estimate is instead more similar to what found by \citet{Negueruela2015AA.584A.77N}, who set an upper limit to the cluster age of 3$\,$Myrs from the photometric analysis of the most massive stars in the clusters (one O6V and two O7V stars), by \citet{CusanoRAG2011}, who estimated an age between 1 and 2$\,$Myrs from the photometric analysis of clusters members selected from spectroscopy, and by \citet{Kalari2015ApJ.800.113K} who estimated an age between 2 and 3$\,$Myrs for cluster members in the center of Dolidze~25 selected from infrared photometry and the analysis of the H$\alpha$ line.

    

\begin{table}[]
\caption{Stellar masses and ages in the central and northern field}             
\label{mass_table}      
\centering                          
\begin{tabular}{c c c c}        
\hline\hline                 
Field & Q10            & Q50             & log(median age) \\    
      & M$\rm_{\odot}$ & M$\rm_{\odot}$  & [Myrs]          \\
\hline                        
Central  & 0.5  & 0.9  & 6.2 \\                
Northern & 0.6  & 1.0  & 6.2 \\     
Sparse   & 0.5  & 1.2  & 6.3 \\           
\hline
\hline
\multicolumn{4}{l}{Q10 and Q50 are the 10\% and 50\% quantile, respectively.} \\ 
\end{tabular}
\end{table}
    
    In order to compare the photometric depth of our selections in the central and northern part of the cavity, Table \ref{mass_table} shows the 10\% and 50\% quantiles of the mass distributions for the ``central'', ``northern'', and ``sparse'' stellar populations. The average distributions of the ``central'' and ``northern'' populations show similar quantiles, suggesting that any possible difference in the average value of disk fraction may not be a consequence of a different depth of the list of members. \par

    \begin{figure}[]
	\centering	
	\includegraphics[width=8cm]{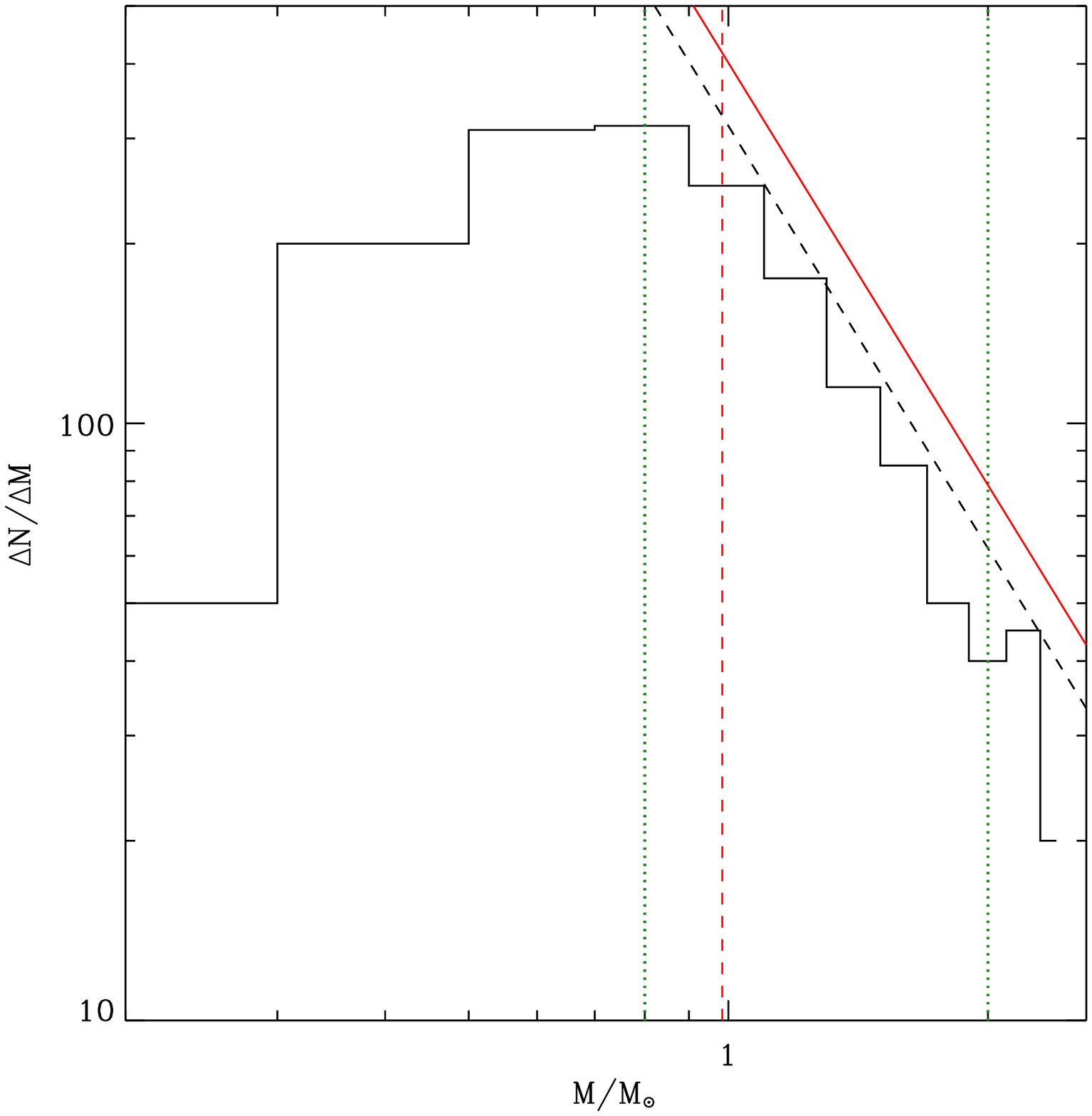}
	\caption{Mass function of both disk-less and disk-bearing candidate members of Dolidze~25 (the latter considered only if inside the ACIS field). The vertical red lines mark the median mass value. The dashed black line is obtained from a linear fit in the log-log space on the mass distribution between 0.8 and 2$\,\rm M_{\odot}$ (values marked with the green dotted lines). The red line shows the normalized Salpeter-Kroupa IMF with $\alpha$=2.35 \citep{KroupaWeidner2003ApJ.598.1076K}.}
	\label{imf_figure}
    \end{figure}
 
    Fig. \ref{imf_figure} shows the comparison between the mass distribution of all cluster members inside the ACIS FoV with the slope of the Salpeter-Kroupa Initial Mass Function (IMF), i.e. $\alpha$=2.35 \citep{KroupaWeidner2003ApJ.598.1076K}. The resulting slope, restricting the linear interpolation in the mass range 0.8-2$\,\rm M_{\odot}$, is consistent with the $\alpha$=2.35 value. Thus, even if the low metallicity of the cluster could have affected its IMF, we do not find any clues supporting a deviation of the IMF in Dolidze~25 with respect to the Salpeter-Kroupa slope.  \par
    

	\subsection{Completeness}
	\label{complete_sect}

	Estimating the completeness of our list of members, compiled by combining the outcome of several selection criteria all different from each other, is an almost impossible task. We can still try to assess the fraction of real cluster members per magnitude bin that we missed to select by comparing the catalog of sources inside the field observed with ACIS with that falling in a control field. \par
        
        In order to select the control field, we noted that the region in the south and south-west do not show any prominent nebular emission at 12$\,\mu$m (see Fig. \ref{field_img}). We thus selected an annular region between 22.8$^\prime$ and 30.6$^\prime$ from the center of the studied field and with $\delta \leq -0.0876678$, encompassing an area almost equal to that of the ACIS FoV. When we derive the magnitude distribution in a given bands, sources falling in this field and having the error in the particular band smaller than 0.1$^m$ form the ``control'' sample. We then selected all sources falling in the ACIS FoV not selected as members, and with the error in the particular band smaller than 0.1$^m$. These sources are the ``ACIS FoV'' sample. The two samples should contain the same foreground population, while the background population at the faint end of the magnitude distributions is expected to be richer in the ``control'' field due to the lower extinction with respect to the ``ACIS FoV'' sample. The latter sample should also contain any member of Dolidze~25 we missed to select, for instance because of the intense variability in IR \citep{Morales-CalderonSHG2011} and X-ray \citep{StassunBF2007ApJ} bands typical of pre-main sequence stars. The ``members'' sample, instead, contains all selected members with error in the given magnitude smaller than 0.1$^m$. \par
        
    \begin{figure*}[]
	\centering	
	\includegraphics[width=8cm]{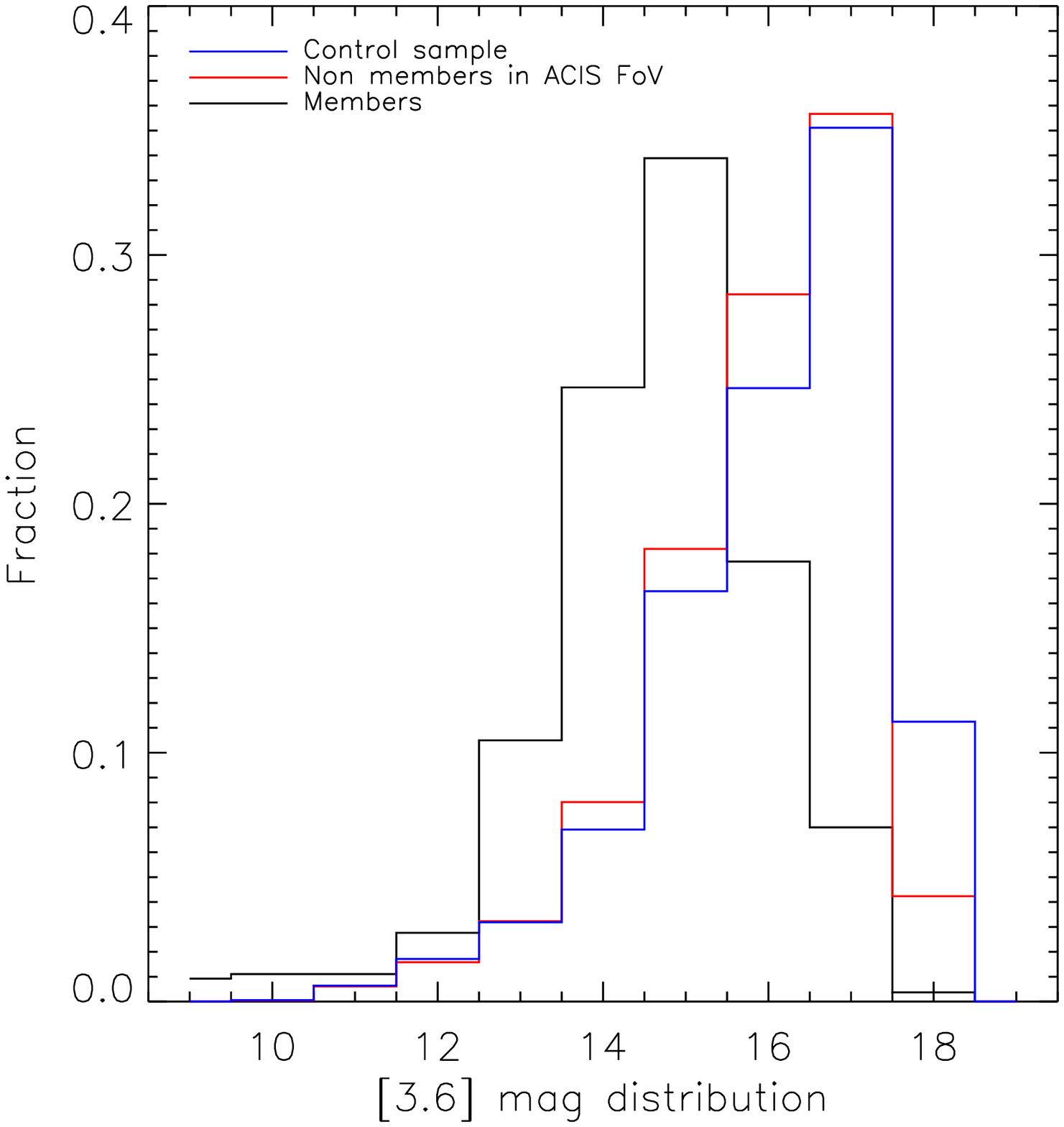}
	\includegraphics[width=8cm]{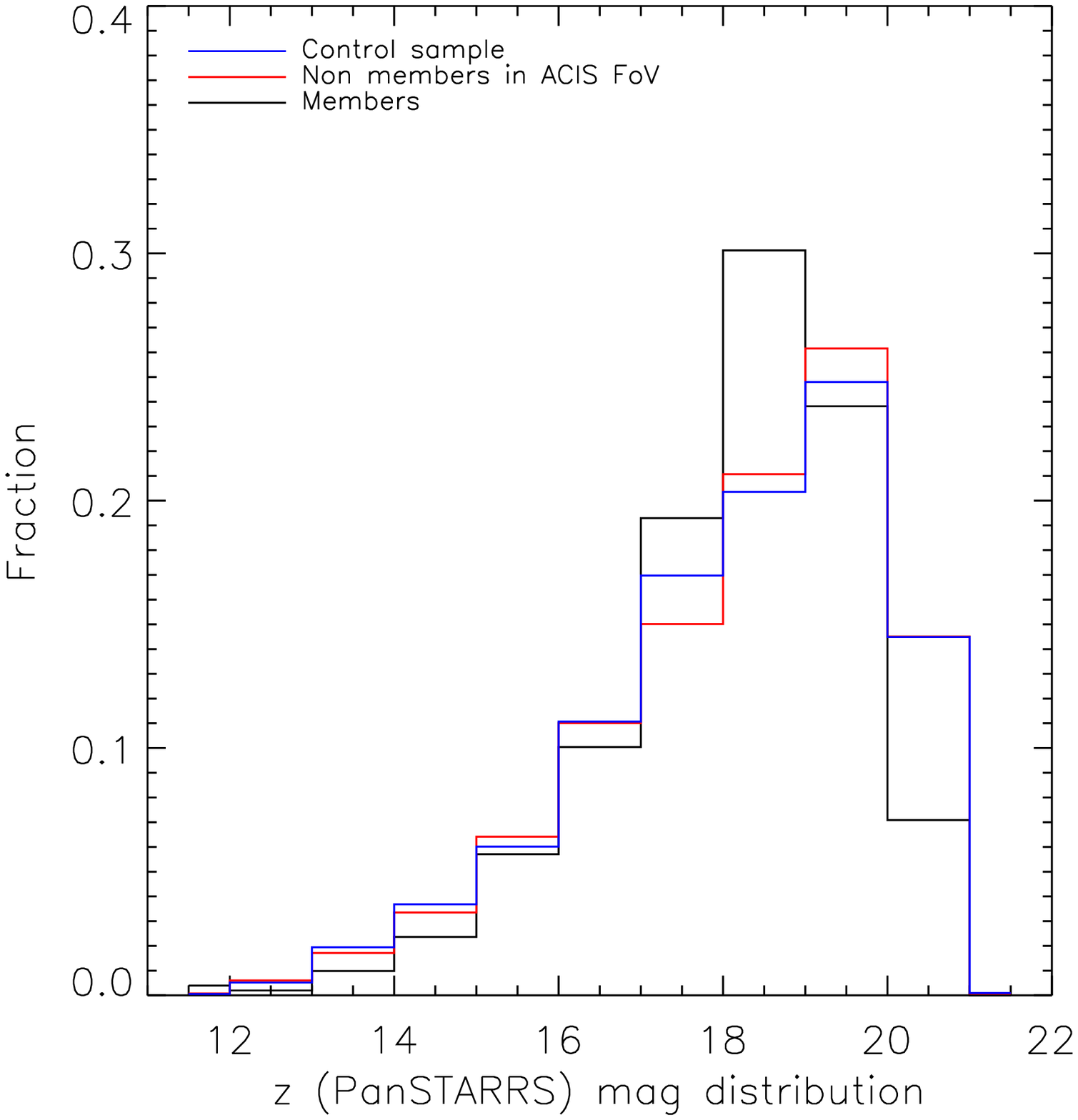}
	\caption{Magnitude distribution at [3.6] and $z_{PanSTARRS}$ bands for the ``control'', ``ACIS FoV'', and ``members'' samples}
	\label{magdis_img}
    \end{figure*}
         
         Fig. \ref{magdis_img} shows the normalized magnitude distributions of the three samples at [3.6] and in the $z$ Pan-STARRS bands. These two bands were selected since they are those in which most of the members have good photometry (in particular, the 82\% and 77\% of the members have errors smaller than 0.1$^m$ in [3.6] and $z_{PanSTARRS}$, respectively). The [3.6] band is an useful test since the effects of extinction are negligible. The fraction of missed members that can be estimated from the comparison between the ``control'' and ``ACIS FoV'' samples is of about 3\% at $\sim$15$^m$ and 5\% at $\sim$16$^m$. For fainter magnitudes, then the intrinsic incompleteness of the IRAC catalog becomes dominant. Only at the faintest magnitude bin, the ``control'' sample becomes more numerous of the ``ACIS FoV'' sample because of the highly extinguished background population. Because of the larger impact of extinction, the distributions of $z_{PanSTARRS}$ magnitudes are slightly more messy, but they suggest that the selection of members may be significantly incomplete for magnitudes fainter than 18$^m$. We can thus consider the interval 13$^m \leq z_{PanSTARRS} \leq$18$^m$ to be fairly complete. The Appendix \ref{AppH} shows the magnitude distributions of the three samples in all the photometric bands studied in this work. \par

    \subsection{Disk fraction}
    \label{df_sect}
    
    We can now calculate the disk fraction in Dolidze~25 taking into consideration the completeness of our sample. In the whole ACIS field we have selected 222 stars with disks and 424 members without disks. To account for multiple matches, we count once all the candidate disk-less members with multiple OIR counterpart, and the disk-bearing sources with multiple infrared counterpart. In this way, the numbers become 218 and 384 for disk-bearing and disk-less stars, respectively, resulting in an average disk fraction of 34\%$\pm$2\%. By considering separately the populations of the central cluster and the northern rim, the resulting disk fractions are 30\%$\pm$3\% and 43\%$\pm$3\%, respectively. This difference is expected from both the presence of a slightly younger population along the rim and for the decline of sensitivity in the ACIS-I detector at large off-axis angles. \par
    
        
    In the previous section we have found that our list is fairly complete in the $z_{PanSTARRS}$ magnitude range between 13$^m$ and 18$^m$. In this magnitude range, the disk fraction is equal to 34\%$\pm$4\% (93 unique disk-less, 48 unique disk-bearing stars). Following the low-metallicity PARSEC isochrone at log(age)=6.2$\,$[Myrs], this magnitude range corresponds to stars more massive than 1.5$\,M_{\odot}$. However, Table \ref{massDF_table} shows that the resulting value of disk fraction does not strongly changes by adopting different cuts in stellar masses. By considering only those mass intervals resulting in an error in disk fraction smaller than 0.05, the values of disk fraction ranges from 0.304$\pm$0.041 to 0.364$\pm$0.049.

    \begin{table*}[!h]
\caption{Disk fraction resulting after applying different cuts in stellar masses.}             
\label{massDF_table}      
\centering                          
\begin{tabular}{c | c c c c c c c}        
\hline\hline                 
\diagbox{to}{from}            & 0.6$\,$M$_{\odot}$& 0.8$\,$M$_{\odot}$& 0.9$\,$M$_{\odot}$& 1.0$\,$M$_{\odot}$& 1.2$\,$M$_{\odot}$& 1.5$\,$M$_{\odot}$& 1.8$\,$M$_{\odot}$\\    
\hline
0.8$\,$M$_{\odot}$          &18, 44       &             &             &             &             &             &             \\
                            &0.29$\pm$0.06&             &             &             &             &             &             \\
\hline
0.9$\,$M$_{\odot}$         &27, 61       &9, 17        &             &             &             &             &             \\
                           &0.31$\pm$0.05&0.35$\pm$0.9 &             &             &             &             &             \\
\hline
1.0$\,$M$_{\odot}$         &38, 87       &20, 43       &11, 26       &             &             &             &             \\
                           &0.30$\pm$0.04&0.32$\pm$0.06&0.30$\pm$0.08&             &             &             &             \\
\hline
1.2$\,$M$_{\odot}$         &54, 121      &36, 77       &27, 60       &16, 34       &             &             &             \\
                           &0.31$\pm$0.04&0.32$\pm$0.04&0.31$\pm$0.05&0.32$\pm$0.07&             &             &             \\
\hline
1.5$\,$M$_{\odot}$         &73, 148      &55, 104      &46, 87       &35, 61       &19, 27       &             &             \\
                           &0.33$\pm$0.03&0.35$\pm$0.04&0.35$\pm$0.04&0.37$\pm$0.05&0.41$\pm$0.07&             &             \\
\hline
1.8$\,$M$_{\odot}$         &82, 168      &64, 124      &55, 107      &44, 81       &28, 47       &9, 20        &             \\
                           &0.33$\pm$0.03&0.34$\pm$0.04&0.34$\pm$0.04&0.35$\pm$0.04&0.37$\pm$0.06&0.31$\pm$0.09&             \\
\hline
2.0$\,$M$_{\odot}$         &85, 175      &67, 131      &58, 114      &47, 88       &31, 54       &12, 27       &3, 7         \\
                           &0.33$\pm$0.03&0.34$\pm$0.03&0.34$\pm$0.04&0.35$\pm$0.04&0.37$\pm$0.05&0.31$\pm$0.07&0.30$\pm$0.15\\
\hline
\hline
\multicolumn{8}{l}{Each cell shows the number of stars with and without disks and the disk fraction.} \\ 
\end{tabular}
\end{table*}

\begin{figure}[]
	\centering	
	\includegraphics[width=9cm]{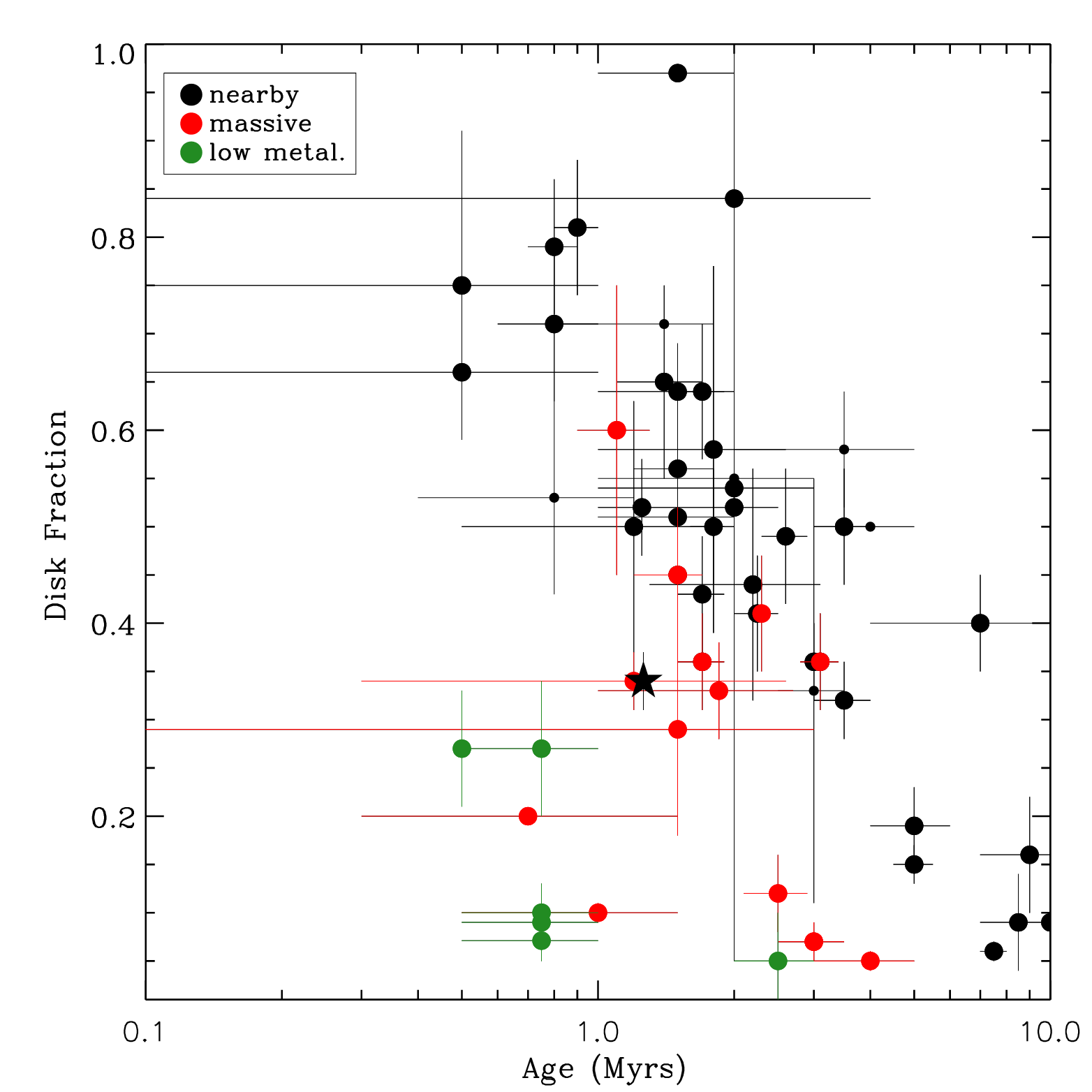}
	\caption{Disk fraction vs. age of 58 clusters with age between 0 and 10 Myrs. Nearby clusters, e.g. closer than 1$\,$kpc to the Sun, are marked with large black dots, massive clusters are marked with red dots, while the low-metallicity star forming regions studied by \citet{YasuiKTS2010} in green. The average estimate of the disk fraction in Dolidze~25 we obtained accounting for completeness is marked with star symbols.}
	\label{cluster_df}
    \end{figure}

    We can now compare the disk fraction of Dolidze~25, by adopting the value of 34\%$\pm$4\%, with that of other 58 clusters younger than 10$\,$Myrs providing a wide range of star forming environments. These clusters are listed in the table in Appendix \ref{AppD}, providing their ages, distances, disk fractions and references. In those cases where different estimates were available from different authors, we favored estimates based on selections of disk-bearing members from infrared photometry and/or disk-less members from X-ray observations or from proper motions/radial velocities, marking the relevant publication in bold in Table \ref{clusters_table}. Results are shown in Fig. \ref{cluster_df}, where we marked separately: clusters closer than 1$\,$kpc to the Sun, whose disk fraction estimate is expected to be less affected by incompleteness and that are expected to have a metallicity more similar to the solar values with respect to more distant clusters; massive clusters where evidence are found supporting a rapid dispersal of disks due to massive stars or close stellar encounters; low metallicity clusters in the outer Galaxy. For the massive clusters we show an average value of the disk fraction, which typically is about 15\%-20\% higher than the values measured in the cluster core. The ``nearby'' clusters follow the well-known narrow decline of the disk fraction with the age. The disk fraction in Dolidze~25 is more than 15\%-20\% lower than the ``nearby'' clusters with similar age. We incidentally note that Dolidze~25 has a disk fraction similar to the average value observed in coeval massive clusters with age between 1 and 2 Myrs, where externally induced photoevaporation and close encounters have induced a fast dispersal of protoplanetary disks, such as NGC6611 \citep{GuarcelloMPP2010}, CygnusOB2 \citep{GuarcelloDWA2016arXiv}, NGC2244 \citep{BalogMRS2007}, and Pismis24 \citep{FangBKH2012}. This comparison thus suggests that disk dispersal in Dolidze~25 occur faster, on average, than in coeval clusters within 1$\,$kpc from the Sun, with timescales similar to those occurring in massive clusters. In Fig. \ref{cluster_df} it is also evident that the disk fraction observed in Dolidze~25 is larger than that found in the low-metallicity clusters in the outer Galaxy by \citet{YasuiKTS2010}. Two hypothesis may explain this discrepancy: The difficulty of obtaining a complete census of members of the distant young clusters still in the embedded phase in the outer Galaxy, or a more important impact of metallicity over the disk dispersal timescale in the outer Galaxy. Future observations may help us discriminating between these two possibilities. \par 

    \subsection{Can O stars photoevaporate disks in Dolidze~25?}
    \label{photo_sect}
    
In this section we verify whether the dispersal timescale in Dolidze~25 may have been affected by externally induced photoevaporation. As explained in Sect. \ref{intro_env}, photoevaporation can be induced externally by the UV radiation emitted by nearby massive stars. Dolidze~25 hosts 10 OB stars \citep{MoffatVogt1975AAS.20.85M}, among which only five are O stars: The O6V star S33, the O7V star S17, the O7.5V star S15, and the two O9.7V stars S1 and S12. In order to estimate the intracluster UV field, we adopted as FUV and EUV fluxes emitted by these stars the values corresponding to their spectral classes provided by \citet{MartinSH2005}. We have thus calculated the total FUV and EUV local field at the position of each selected cluster member by projecting and summing the contributions from these O stars. In this calculation, we used the projected distances from the cluster members to the O stars, which results in an overestimate of the real incident flux.

%

    \begin{figure}[]
	\centering	
	\includegraphics[width=9cm]{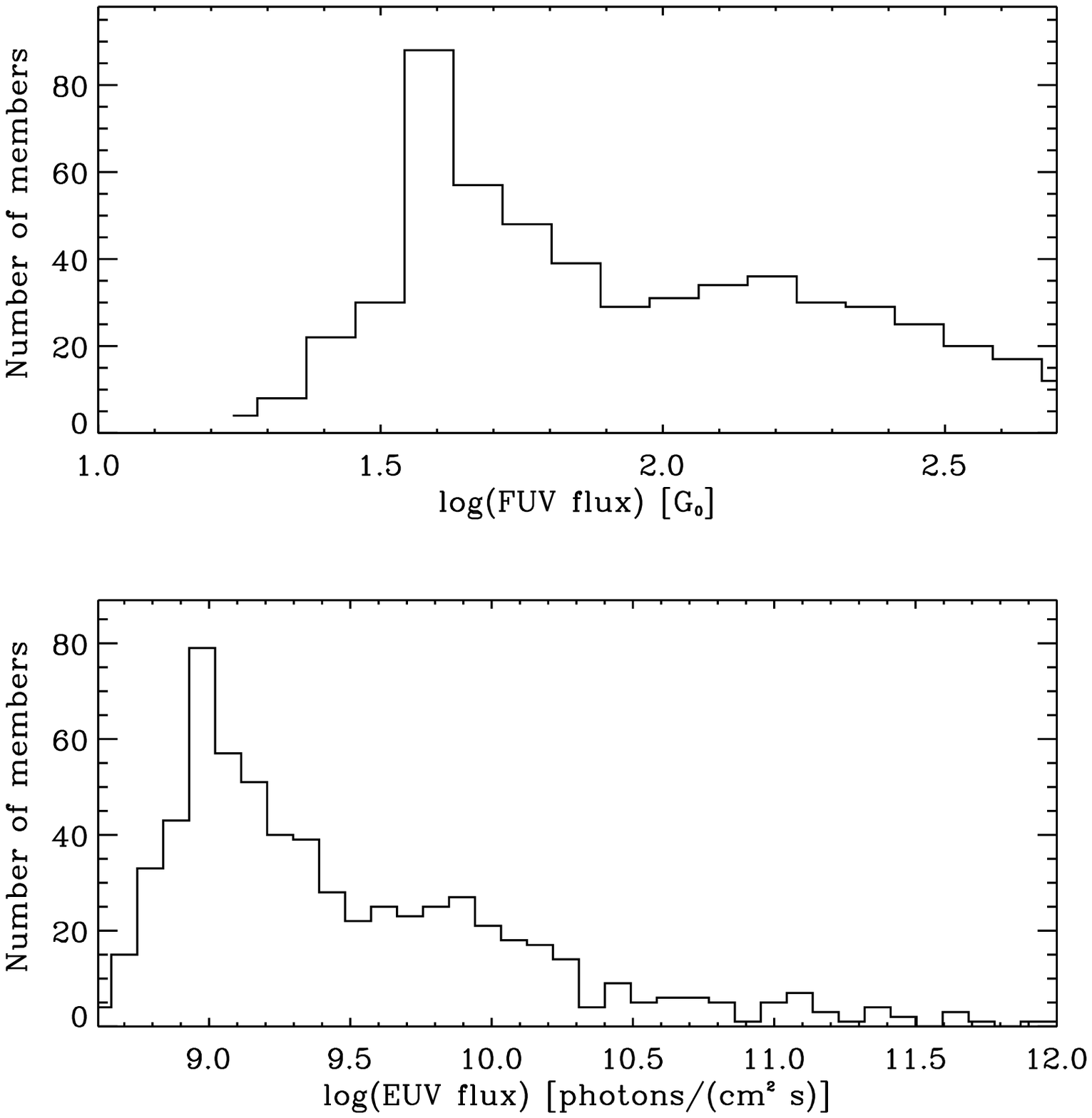}
	\caption{Histogram of the FUV (top panel) and EUV (bottom panel) fields experienced by each candidate member. The FUV fluxes are in units of the Habing flux G$_0$, with 1$\,$G$_0$=1.6$\times$10$^{-3}\,$erg/cm$^2$/s (1.7$\,$G$_0$ is equal to the average interstellar UV field in the solar neighborhood in the 912-2000$\,$\AA$\,$  band, \citealt{Habing1968}). The vast majority of candidate cluster members experiences very low values of local UV field.}
	\label{uvhisto_fig}
    \end{figure}

    The resulting distributions of local UV fields at the positions of the candidate cluster members are shown in Fig. \ref{uvhisto_fig}. By comparing these values with those experienced by the members of the young cluster NGC~6611 \citep[Fig. 15 in][]{GuarcelloDWD2013ApJ}, where the disk fraction decreases only in the cluster core where there are about 50 OB stars \citep{HillenbrandMSM1993}, it is possible to verify that the intracluster UV field in Dolidze~25 is similar to that in the outskirts of NGC~6611, where no environmental effects on disk dispersal timescales are observed. The low disk fraction in Dolidze~25 is thus more likely a consequence of a faster dispersal of disks due to their low metallicity. \par



\section{Conclusions}

In this paper we have analyzed two specific $Chandra$/ACIS-I observations (76.67 and 68.44$\,$ksec long) and archival data of the open cluster Dolidze~25, in order to calculate the disk fraction of the cluster and compare it with that of other clusters and associations younger than 10$\,$Myrs. Our work is motivated by the fact that Dolidze~25 is part of a star-forming region among those with the lowest metallicity in the Galaxy \citep{LennonDFG1990,Fitzsimmons1992MNRAS.259.489F,Negueruela2015AA.584A.77N}. Protoplanetary disks in low metallicity environments are expected to evolve in a different timescale than those formed in high-metallicity environments. However, the results obtained to date on the effects of metallicity on disk dispersal are quite controversial notwithstanding the importance of this topic. For instance, any effect of metallicity on disk dispersal timescale would mean that disks evolution and planet formation in remote epochs of our Galaxy, when metallicity was lower than present-day, was different than today. \par

    On one hand, observational studies of star forming regions in the outer Galaxy, characterized by very low metallicity, have demonstrated that disks can be dispersed very quickly, with disks fraction $\leq$20\% even at 1$\,$Myr and almost zero after 2$\,$Myrs \citep{YasuiKTS2009,YasuiKTS2010,Yasui2016AJ.151.115Y,Yasui2016AJ.151.50Y}. Their conclusions have found a strong theoretical confirmation in the models presented by \citet{ErcolanoClarke2010MNRAS.402.2735E}. On the other hand, the selection of accreting disks older than 10$\,$Myrs in the Magellanic Clouds performed by several authors \citep{SpezziDPS2012,deMarchi2010ApJ.715.1D,deMarchi2017ApJ.846.110D,Biazzo2019ApJ.875.51B} suggests that accretion timescales can be longer, despite accretion rates being more intense, in low-metallicity environments. This could be due to several effects, such as a less effective impact of radiation pressure on the inner disks in low metallicity environments \citep{deMarchi2017ApJ.846.110D}, or a lower disk temperature due to the smaller dust opacity that results in longer accretion timescales \citep{Durisen2007prpl.conf.607D}. It must be noted that the authors of these studies have adopted different diagnostics to select young stars. In fact, in the observational works by Yasui and collaborators, disks are selected from NIR photometry, thus being sensitive to the micron-size dust emission from disks, while the studies on the Magellanic Clouds are based on photometric evidences of accretion in disks, which however are not selected in infrared.  \par

In this work, we have compiled a multi-band catalog of the sources within 0.5$\deg$ from Dolidze~25 by combining the X-ray catalog with an extensive list of available optical and infrared catalogs of the region: VPHAS+, IPHAS, Pan-STARRS, the optical catalog published by \citet{DelgadoDA2010}, Gaia/DR2 and EDR3, CoRoT, LAMOST, 2MASS/PSC, UKIDSS/DR10, the Spitzer/IRAC catalog obtained by \citet{Puga2009AA.503.107P}, and the AllWISE Source Catalog. Catalogs were combined by adopting a close-neighborhood approach or a Maximum-Likelihood approach, in the latter case considering available photometry together with the angular separation between sources of the various catalogs. Multiple coincidences, false positives, and false negatives resulting from the matching procedure were properly treated in order to reduce their impact on the final multi-band catalog, counting 101722 entries. \par
    
We have selected 667 disk-bearing stars populating different recent star-forming sites of Sh2-284, already discovered by \citet{Puga2009AA.503.107P}: Together with Dolidze~25 at the center of the cavity, we have studied the clusters $Cl2$ and $Cl3$, and the bright rimmed clouds $RN$, $RS$, and $RE$. These stars were selected by adopting criteria based on the Spitzer/IRAC, WISE, and $JHK$ colors, together with specific criteria defined in order to select and discard foreground and background contaminants. Inside the ACIS FoV, centered on Dolidze~25, we found 222 stars with disks. The disk-less population of Dolidze~25 (424 sources) was instead selected among the 618 sources detected in X-rays, discarding those without optical or infrared counterpart, expected to be strongly contaminated by extragalactic sources, and X-ray+OIR sources with optical and infrared colors typical of foreground and background sources. \par

    The spatial distribution of the candidate young stars associated with Dolidze~25 and Sh2-284 confirms the existence of multiple regions in the area, and does not support the presence of mass segregation. The main concentrations in the central cavity are Dolidze~25, and the young stars associated with the northern rim of the cavity cleared by the cluster itself. We have derived an extinction map of the whole area and verified that an average difference of less than 1$\,$mag of extinction exists between the central and the northern regions of the cavity. We estimated masses and ages of the candidate members. We do not find convincing evidence supporting a deviation of the IMF in Dolidze~25 from the universal slope due to the low-metallicity of the cluster. We also estimated the median age of the cluster equal to log(age)$\rm _{median}$=6.2 [Myrs], with a standard deviation of 0.3 [Myrs]. \par
    
Our estimate of the disk fraction of the cluster slightly changes when adopting different selections aimed at minimizing the effects of incompleteness. We adopted an average value of $\sim$34\%. By collecting the estimate of the disk fractions of 58 clusters and associations younger than 10$\,$Myrs, we found evidence supporting a lower disk fraction of Dolidze~25 with respect to star forming environments with solar metallicity and similar age. In particular, the disk fraction in Dolidze~25 is similar to those found in massive clusters, where disk dispersal in the proximity of massive stars is accelerated by the externally induced disk photoevaporation. Since the massive population of Dolidze~25 (counting only five O-late stars) does not produce an intracluster UV field intense enough to induce disk photoevaporation, such a difference is more likely due to the low metallicity of the cluster rather than other environment feedback. \par

    Our conclusions depend on the reliability of the estimate of the age of the cluster. We estimated stellar ages by fitting suitable isochrones to the distribution of cluster members in color-magnitude diagrams. This procedure can be affected by important uncertainties, and it also depends on the reliability of the adopted models to describe the pre-main sequence phase. This can have an important impact to our conclusions: For instance, with a mean cluster age of 3$\,$Myrs rather than 1.2$\,$Myrs, our estimate of disk fraction would be similar to those of the other clusters. Bearing in mind this caveat, we claim that disk fraction in Dolidze~25 are likely to be affected by the low metallicity of this star-forming region.

%
%
    
\begin{acknowledgements}
The authors acknowledge the referee for his/hers comments and careful reading of our manuscript, that helped us improving our paper. KB has been supported by the project PRIN-INAF-MAIN-STREAM 2018 ``Protoplanetary disks seen through the eyes of new-generation instruments''. This work is based on: VPHAS+ data, which are based on observations made with ESO Telescopes at the La Silla or Paranal Observatories under programme ID(s) 177.D-3023(B), 177.D-3023(C), 177.D-3023(D), 177.D-3023(E); IPHAS data, obtained from observations (www.iphas.org) carried out at the Isaac Newton Telescope (INT), operating on the island of La Palma by the Isaac Newton Group in the Spanish Observatorio del Roque de los Muchachos of the Instituto de Astrofisica de Canarias; data from the Pan-STARRS1 Surveys (PS1), which have been made possible through contributions by the Institute for Astronomy, the University of Hawaii, the Pan-STARRS Project Office, the Max-Planck Society and its participating institutes, the Max Planck Institute for Astronomy, Heidelberg and the Max Planck Institute for Extraterrestrial Physics, Garching, The Johns Hopkins University, Durham University, the University of Edinburgh, the Queen's University Belfast, the Harvard-Smithsonian Center for Astrophysics, the Las Cumbres Observatory Global Telescope Network Incorporated, the National Central University of Taiwan, the Space Telescope Science Institute, and the National Aeronautics and Space Administration under Grant No. NNX08AR22G issued through the Planetary Science Division of the NASA Science Mission Directorate, the National Science Foundation Grant No. AST-1238877, the University of Maryland, Eotvos Lorand University (ELTE), and the Los Alamos National Laboratory; data from the European Space Agency (ESA) mission {\it Gaia} (\url{https://www.cosmos.esa.int/gaia}), processed by the {\it Gaia} Data Processing and Analysis Consortium (DPAC, \url{https://www.cosmos.esa.int/web/gaia/dpac/consortium}), with funds provided by national institutions, in particular the institutions participating in the {\it Gaia} Multilateral Agreement; data products from the Two Micron All Sky Survey, which is a joint project of the University of Massachusetts and the Infrared Processing and Analysis Center/California Institute of Technology, funded by the National Aeronautics and Space Administration and the National Science Foundation; data obtained as part of the UKIRT Infrared Deep Sky Survey; data from observations made with the Spitzer Space Telescope, which was operated by the Jet Propulsion Laboratory, California Institute of Technology under a contract with NASA; data products from the Wide-field Infrared Survey Explorer, which is a joint project of the University of California, Los Angeles, and the Jet Propulsion Laboratory/California Institute of Technology, funded by the National Aeronautics and Space Administration. The authors also acknowledge the use of the TOPCAT software, described in \citet{Taylor2005ASPC.347.29T}, and the VizieR catalogue access tool, CDS, Strasbourg, France (DOI : 10.26093/cds/vizier), described in \citet{Ochsenbein2000AAS.143.23O}.

\end{acknowledgements}
\newpage

\addcontentsline{toc}{section}{\bf Bibliografia}
\bibliographystyle{aa}
\bibliography{biblio}

\begin{thebibliography}{168}
\expandafter\ifx\csname natexlab\endcsname\relax\def\natexlab#1{#1}\fi

\bibitem[{{Affer} {et~al.}(2012){Affer}, {Micela}, {Favata}, \&
  {Flaccomio}}]{Affer2012MNRAS.424.11A}
{Affer}, L., {Micela}, G., {Favata}, F., \& {Flaccomio}, E. 2012, \mnras, 424,
  11

\bibitem[{{Alcal{\'a}} {et~al.}(2008){Alcal{\'a}}, {Spezzi}, {Chapman},
  {Evans}, {Huard}, {J{\o}rgensen}, {Mer{\'\i}n}, {Stapelfeldt}, {Covino},
  {Frasca}, {Gandolfi}, \& {Oliveira}}]{Alcala2008ApJ.676.427A}
{Alcal{\'a}}, J.~M., {Spezzi}, L., {Chapman}, N., {et~al.} 2008, \apj, 676, 427

\bibitem[{{Allen} {et~al.}(2012){Allen}, {Gutermuth}, {Kryukova}, {Megeath},
  {Pipher}, {Naylor}, {Jeffries}, {Wolk}, {Spitzbart}, \&
  {Muzerolle}}]{Allen2012ApJ.750.125A}
{Allen}, T.~S., {Gutermuth}, R.~A., {Kryukova}, E., {et~al.} 2012, \apj, 750,
  125

\bibitem[{{Allison} {et~al.}(2009){Allison}, {Goodwin}, {Parker}, {Portegies
  Zwart}, {de Grijs}, \& {Kouwenhoven}}]{Allison2009MNRAS.395.1449A}
{Allison}, R.~J., {Goodwin}, S.~P., {Parker}, R.~J., {et~al.} 2009, \mnras,
  395, 1449

\bibitem[{{Armitage} {et~al.}(2003){Armitage}, {Clarke}, \&
  {Palla}}]{Armitage2003MNRAS.342.1139A}
{Armitage}, P.~J., {Clarke}, C.~J., \& {Palla}, F. 2003, \mnras, 342, 1139

\bibitem[{{Baglin} {et~al.}(2006){Baglin}, {Auvergne}, {Barge}, {Deleuil},
  {Catala}, {Michel}, {Weiss}, \& {COROT Team}}]{BaglinABD2006ESASP}
{Baglin}, A., {Auvergne}, M., {Barge}, P., {et~al.} 2006, in ESA Special
  Publication, Vol. 1306, The CoRoT Mission Pre-Launch Status - Stellar
  Seismology and Planet Finding, ed. M.~{Fridlund}, A.~{Baglin}, J.~{Lochard},
  \& L.~{Conroy}, 33

\bibitem[{{Bai} \& {Stone}(2013)}]{BaiStones2013ApJ.769.76B}
{Bai}, X.-N. \& {Stone}, J.~M. 2013, \apj, 769, 76

\bibitem[{{Bai} {et~al.}(2016){Bai}, {Ye}, {Goodman}, \&
  {Yuan}}]{Bai2016ApJ.818.152B}
{Bai}, X.-N., {Ye}, J., {Goodman}, J., \& {Yuan}, F. 2016, \apj, 818, 152

\bibitem[{{Bailer-Jones} {et~al.}(2018){Bailer-Jones}, {Rybizki}, {Fouesneau},
  {Mantelet}, \& {Andrae}}]{Bailer-Jones2018AJ.156.58B}
{Bailer-Jones}, C.~A.~L., {Rybizki}, J., {Fouesneau}, M., {Mantelet}, G., \&
  {Andrae}, R. 2018, \aj, 156, 58

\bibitem[{{Balbus} \& {Hawley}(1991)}]{BalbusHawley1991}
{Balbus}, S.~A. \& {Hawley}, J.~F. 1991, \apj, 376, 214

\bibitem[{{Bally} {et~al.}(2000){Bally}, {O'Dell}, \&
  {McCaughrean}}]{BallyOM2000}
{Bally}, J., {O'Dell}, C.~R., \& {McCaughrean}, M.~J. 2000, \aj, 119, 2919

\bibitem[{{Balog} {et~al.}(2007){Balog}, {Muzerolle}, {Rieke}, {Su}, {Young},
  \& {Megeath}}]{BalogMRS2007}
{Balog}, Z., {Muzerolle}, J., {Rieke}, G.~H., {et~al.} 2007, \apj, 660, 1532

\bibitem[{{Balog} {et~al.}(2006){Balog}, {Rieke}, {Su}, {Muzerolle}, \&
  {Young}}]{BalogRSM2006}
{Balog}, Z., {Rieke}, G.~H., {Su}, K.~Y.~L., {Muzerolle}, J., \& {Young}, E.~T.
  2006, \apjl, 650, L83

\bibitem[{{Barentsen} {et~al.}(2014){Barentsen}, {Farnhill}, {Drew}, \&
  et~al.}]{BarentsenFDG2014}
{Barentsen}, G., {Farnhill}, H.~J., {Drew}, J.~E., \& et~al. 2014, \mnras, 444,
  3230

\bibitem[{{Barentsen} {et~al.}(2011){Barentsen}, {Vink}, {Drew}, {Greimel},
  {Wright}, {Drake}, {Martin}, {Valdivielso}, \& {Corradi}}]{BarentsenVDG2011}
{Barentsen}, G., {Vink}, J.~S., {Drew}, J.~E., {et~al.} 2011, \mnras, 415, 103

\bibitem[{{Barrow} {et~al.}(1985){Barrow}, {Bhavsar}, \&
  {Sonoda}}]{BarrowBS1985}
{Barrow}, J.~D., {Bhavsar}, S.~P., \& {Sonoda}, D.~H. 1985, \mnras, 216, 17

\bibitem[{{Baume} {et~al.}(1999){Baume}, {V{\'a}zquez}, \&
  {Feinstein}}]{Baume1999AAS.137.233B}
{Baume}, G., {V{\'a}zquez}, R.~A., \& {Feinstein}, A. 1999, \aaps, 137, 233

\bibitem[{{Beccari} {et~al.}(2015){Beccari}, {De Marchi}, {Panagia}, {Valenti},
  {Carraro}, {Romaniello}, {Zoccali}, \& {Weidner}}]{Beccari2015AA.574A.44B}
{Beccari}, G., {De Marchi}, G., {Panagia}, N., {et~al.} 2015, \aap, 574, A44

\bibitem[{{Biazzo} {et~al.}(2019){Biazzo}, {Beccari}, {De Marchi}, \&
  {Panagia}}]{Biazzo2019ApJ.875.51B}
{Biazzo}, K., {Beccari}, G., {De Marchi}, G., \& {Panagia}, N. 2019, \apj, 875,
  51

\bibitem[{{Bik} {et~al.}(2014){Bik}, {Stolte}, {Gennaro}, {Brandner},
  {Gouliermis}, {Hu{\ss}mann}, {Tognelli}, {Rochau}, {Henning}, {Adamo},
  {Beuther}, {Pasquali}, \& {Wang}}]{Bik2014AA.561A.12B}
{Bik}, A., {Stolte}, A., {Gennaro}, M., {et~al.} 2014, \aap, 561, A12

\bibitem[{{Brescia} {et~al.}(2015){Brescia}, {Cavuoti}, \&
  {Longo}}]{Brescia2015MNRAS.450.3893B}
{Brescia}, M., {Cavuoti}, S., \& {Longo}, G. 2015, \mnras, 450, 3893

\bibitem[{{Bressan} {et~al.}(2012){Bressan}, {Marigo}, {Girardi}, {Salasnich},
  {Dal Cero}, {Rubele}, \& {Nanni}}]{BressanMGS2012MNRAS}
{Bressan}, A., {Marigo}, P., {Girardi}, L., {et~al.} 2012, \mnras, 427, 127

\bibitem[{{Brice{\~n}o}(2009)}]{Briceno2009RMxAC.35.27B}
{Brice{\~n}o}, C. 2009, in Revista Mexicana de Astronomia y Astrofisica
  Conference Series, Vol.~35, Revista Mexicana de Astronomia y Astrofisica
  Conference Series, 27--32

\bibitem[{{Brice{\~n}o} {et~al.}(2007){Brice{\~n}o}, {Hartmann},
  {Hern{\'a}ndez}, {Calvet}, {Vivas}, {Furesz}, \&
  {Szentgyorgyi}}]{Briceno2007ApJ.661.1119B}
{Brice{\~n}o}, C., {Hartmann}, L., {Hern{\'a}ndez}, J., {et~al.} 2007, \apj,
  661, 1119

\bibitem[{{Broos} {et~al.}(2010){Broos}, {Townsley}, {Feigelson}, {Getman},
  {Bauer}, \& {Garmire}}]{BroosTFG2010}
{Broos}, P.~S., {Townsley}, L.~K., {Feigelson}, E.~D., {et~al.} 2010, \apj,
  714, 1582

\bibitem[{{Calvet} \& {Gullbring}(1998)}]{CalvetGullbring1998}
{Calvet}, N. \& {Gullbring}, E. 1998, \apj, 509, 802

\bibitem[{{Cardelli} {et~al.}(1989){Cardelli}, {Clayton}, \&
  {Mathis}}]{CardelliCM1989}
{Cardelli}, J.~A., {Clayton}, G.~C., \& {Mathis}, J.~S. 1989, \apj, 345, 245

\bibitem[{{Carone} {et~al.}(2012){Carone}, {Gandolfi}, {Cabrera}, {Hatzes},
  {Deeg}, {Csizmadia}, {P{\"a}tzold}, {Weingrill}, {Aigrain}, {Alonso},
  {Alapini}, {Almenara}, {Auvergne}, {Baglin}, {Barge}, {Bonomo}, {Bord{\'e}},
  {Bouchy}, {Bruntt}, {Carpano}, {Cochran}, {Deleuil}, {D{\'\i}az}, {Dreizler},
  {Dvorak}, {Eisl{\"o}ffel}, {Eigm{\"u}ller}, {Endl}, {Erikson},
  {Ferraz-Mello}, {Fridlund}, {Gazzano}, {Gibson}, {Gillon}, {Gondoin},
  {Grziwa}, {G{\"u}nther}, {Guillot}, {Hartmann}, {Havel}, {H{\'e}brard},
  {Jorda}, {Kabath}, {L{\'e}ger}, {Llebaria}, {Lammer}, {Lovis}, {MacQueen},
  {Mayor}, {Mazeh}, {Moutou}, {Nortmann}, {Ofir}, {Ollivier}, {Parviainen},
  {Pepe}, {Pont}, {Queloz}, {Rabus}, {Rauer}, {R{\'e}gulo}, {Renner}, {de La
  Reza}, {Rouan}, {Santerne}, {Samuel}, {Schneider}, {Shporer}, {Stecklum},
  {Tal-Or}, {Tingley}, {Udry}, \& {Wuchterl}}]{Carone2012AA.538A.112C}
{Carone}, L., {Gandolfi}, D., {Cabrera}, J., {et~al.} 2012, \aap, 538, A112

\bibitem[{{Carpenter} {et~al.}(2006){Carpenter}, {Mamajek}, {Hillenbrand}, \&
  {Meyer}}]{CarpenterMHM2006ApJ}
{Carpenter}, J.~M., {Mamajek}, E.~E., {Hillenbrand}, L.~A., \& {Meyer}, M.~R.
  2006, \apjl, 651, L49

\bibitem[{{Chambers} {et~al.}(2016){Chambers}, {Magnier}, {Metcalfe},
  {Flewelling}, {Huber}, {Waters}, {Denneau}, {Draper}, {Farrow}, {Finkbeiner},
  {Holmberg}, {Koppenhoefer}, {Price}, {Rest}, {Saglia}, {Schlafly}, {Smartt},
  {Sweeney}, {Wainscoat}, {Burgett}, {Chastel}, {Grav}, {Heasley}, {Hodapp},
  {Jedicke}, {Kaiser}, {Kudritzki}, {Luppino}, {Lupton}, {Monet}, {Morgan},
  {Onaka}, {Shiao}, {Stubbs}, {Tonry}, {White}, {Ba{\~n}ados}, {Bell},
  {Bender}, {Bernard}, {Boegner}, {Boffi}, {Botticella}, {Calamida},
  {Casertano}, {Chen}, {Chen}, {Cole}, {Deacon}, {Frenk}, {Fitzsimmons},
  {Gezari}, {Gibbs}, {Goessl}, {Goggia}, {Gourgue}, {Goldman}, {Grant},
  {Grebel}, {Hambly}, {Hasinger}, {Heavens}, {Heckman}, {Henderson}, {Henning},
  {Holman}, {Hopp}, {Ip}, {Isani}, {Jackson}, {Keyes}, {Koekemoer}, {Kotak},
  {Le}, {Liska}, {Long}, {Lucey}, {Liu}, {Martin}, {Masci}, {McLean}, {Mindel},
  {Misra}, {Morganson}, {Murphy}, {Obaika}, {Narayan}, {Nieto-Santisteban},
  {Norberg}, {Peacock}, {Pier}, {Postman}, {Primak}, {Rae}, {Rai}, {Riess},
  {Riffeser}, {Rix}, {R{\"o}ser}, {Russel}, {Rutz}, {Schilbach}, {Schultz},
  {Scolnic}, {Strolger}, {Szalay}, {Seitz}, {Small}, {Smith}, {Soderblom},
  {Taylor}, {Thomson}, {Taylor}, {Thakar}, {Thiel}, {Thilker}, {Unger},
  {Urata}, {Valenti}, {Wagner}, {Walder}, {Walter}, {Watters}, {Werner},
  {Wood-Vasey}, \& {Wyse}}]{Chambers2016arXiv161205560C}
{Chambers}, K.~C., {Magnier}, E.~A., {Metcalfe}, N., {et~al.} 2016, arXiv
  e-prints [\eprint[arXiv]{1612.05560}]

\bibitem[{{Clarke} \& {Pringle}(1993)}]{ClarkePringle1993}
{Clarke}, C.~J. \& {Pringle}, J.~E. 1993, \mnras, 261, 190

\bibitem[{{COROT Team}(2016)}]{COROT2014yCat.102028C}
{COROT Team}. 2016, VizieR Online Data Catalog, B/corot

\bibitem[{{Cusano} {et~al.}(2011){Cusano}, {Ripepi}, {Alcal{\'a}}, {Gandolfi},
  {Marconi}, {Degl'Innocenti}, {Palla}, {Guenther}, {Bernabei}, {Covino},
  {Neiner}, {Puga}, \& {Hony}}]{CusanoRAG2011}
{Cusano}, F., {Ripepi}, V., {Alcal{\'a}}, J.~M., {et~al.} 2011, \mnras, 410,
  227

\bibitem[{{Cutri} \& {et al.}(2012)}]{Cutri2012yCat.2311.0C}
{Cutri}, R.~M. \& {et al.} 2012, VizieR Online Data Catalog, 2311

\bibitem[{{Cutri} {et~al.}(2003){Cutri}, {Skrutskie}, {van Dyk}, \&
  et~al.}]{CutriSDB2003}
{Cutri}, R.~M., {Skrutskie}, M.~F., {van Dyk}, S., \& et~al. 2003, {2MASS All
  Sky Catalog of point sources.}, ed. {Cutri, R.~M., Skrutskie, M.~F., van Dyk,
  S., Beichman, C.~A., Carpenter, J.~M., Chester, T., Cambresy, L., Evans, T.,
  Fowler, J., Gizis, J., Howard, E., Huchra, J., Jarrett, T., Kopan, E.~L.,
  Kirkpatrick, J.~D., Light, R.~M., Marsh, K.~A., McCallon, H., Schneider, S.,
  Stiening, R., Sykes, M., Weinberg, M., Wheaton, W.~A., Wheelock, S., \&
  Zacarias, N.}

\bibitem[{{Dahm} \& {Hillenbrand}(2007)}]{DahmHillenbrand2007AJ.133.2072D}
{Dahm}, S.~E. \& {Hillenbrand}, L.~A. 2007, \aj, 133, 2072

\bibitem[{{Damiani} {et~al.}(1997){Damiani}, {Maggio}, {Micela}, \&
  {Sciortino}}]{Damiani1997ApJ.483.350D}
{Damiani}, F., {Maggio}, A., {Micela}, G., \& {Sciortino}, S. 1997, \apj, 483,
  350

\bibitem[{{Damiani} {et~al.}(2016){Damiani}, {Micela}, \&
  {Sciortino}}]{Damiani2016AA.596A.82D}
{Damiani}, F., {Micela}, G., \& {Sciortino}, S. 2016, \aap, 596, A82

\bibitem[{{Damiani} {et~al.}(2006){Damiani}, {Prisinzano}, {Micela}, \&
  {Sciortino}}]{Damiani2006}
{Damiani}, F., {Prisinzano}, L., {Micela}, G., \& {Sciortino}, S. 2006, \aap,
  459, 477

\bibitem[{{De Marchi} {et~al.}(2017){De Marchi}, {Panagia}, \&
  {Beccari}}]{deMarchi2017ApJ.846.110D}
{De Marchi}, G., {Panagia}, N., \& {Beccari}, G. 2017, \apj, 846, 110

\bibitem[{{De Marchi} {et~al.}(2010){De Marchi}, {Panagia}, \&
  {Romaniello}}]{deMarchi2010ApJ.715.1D}
{De Marchi}, G., {Panagia}, N., \& {Romaniello}, M. 2010, \apj, 715, 1

\bibitem[{{De Marchi} {et~al.}(2011){De Marchi}, {Panagia}, {Romaniello},
  {Sabbi}, {Sirianni}, {Prada Moroni}, \&
  {Degl'Innocenti}}]{deMarchi2011ApJ.740.11D}
{De Marchi}, G., {Panagia}, N., {Romaniello}, M., {et~al.} 2011, \apj, 740, 11

\bibitem[{{Debosscher} {et~al.}(2009){Debosscher}, {Sarro}, {L{\'o}pez},
  {Deleuil}, {Aerts}, {Auvergne}, {Baglin}, {Baudin}, {Chadid}, {Charpinet},
  {Cuypers}, {De Ridder}, {Garrido}, {Hubert}, {Janot-Pacheco}, {Jorda},
  {Kaiser}, {Kallinger}, {Kollath}, {Maceroni}, {Mathias}, {Michel}, {Moutou},
  {Neiner}, {Ollivier}, {Samadi}, {Solano}, {Surace}, {Vandenbussche}, \&
  {Weiss}}]{Debosscher2009AA.506.519D}
{Debosscher}, J., {Sarro}, L.~M., {L{\'o}pez}, M., {et~al.} 2009, \aap, 506,
  519

\bibitem[{{Delgado} {et~al.}(2010){Delgado}, {Djupvik}, \&
  {Alfaro}}]{DelgadoDA2010}
{Delgado}, A.~J., {Djupvik}, A.~A., \& {Alfaro}, E.~J. 2010, \aap, 509, A104

\bibitem[{{Drew} {et~al.}(2014){Drew}, {Gonzalez-Solares}, {Greimel}, {Irwin},
  {K{\"u}pc{\"u} Yoldas}, {Lewis}, {Barentsen}, {Eisl{\"o}ffel}, {Farnhill},
  {Martin}, {Walsh}, {Walton}, {Mohr-Smith}, {Raddi}, {Sale}, {Wright},
  {Groot}, {Barlow}, {Corradi}, {Drake}, {Fabregat}, {Frew}, {G{\"a}nsicke},
  {Knigge}, {Mampaso}, {Morris}, {Naylor}, {Parker}, {Phillipps}, {Ruhland},
  {Steeghs}, {Unruh}, {Vink}, {Wesson}, \&
  {Zijlstra}}]{Drew2014MNRAS.440.2036D}
{Drew}, J.~E., {Gonzalez-Solares}, E., {Greimel}, R., {et~al.} 2014, \mnras,
  440, 2036

\bibitem[{{Drew} {et~al.}(2016){Drew}, {Gonzalez-Solares}, {Greimel}, {Irwin},
  {Kupcu Yoldas}, {Lewis}, {Barentsen}, {Eisloffel}, {Farnhill}, {Martin},
  {Walsh}, {Walton}, {Mohr-Smith}, {Raddi}, {Sale}, {Wright}, {Groot},
  {Barlow}, {Corradi}, {Drake}, {Fabregat}, {Frew}, {Gansicke}, {Knigge},
  {Mampaso}, {Morris}, {Naylor}, {Parker}, {Phillipps}, {Ruhland}, {Steeghs},
  {Unruh}, {Vink}, {Wesson}, \& {Zijlstra}}]{Drew2016yCat.2341.0D}
{Drew}, J.~E., {Gonzalez-Solares}, E., {Greimel}, R., {et~al.} 2016, VizieR
  Online Data Catalog, 2341

\bibitem[{{Drew} {et~al.}(2005){Drew}, {Greimel}, {Irwin}, \&
  et~al.}]{DrewGIA2005}
{Drew}, J.~E., {Greimel}, R., {Irwin}, M.~J., \& et~al. 2005, \mnras, 362, 753

\bibitem[{{Dullemond} \& {Dominik}(2005)}]{DullemondDomink2005AA.434.971D}
{Dullemond}, C.~P. \& {Dominik}, C. 2005, \aap, 434, 971

\bibitem[{{Durisen} {et~al.}(2007){Durisen}, {Boss}, {Mayer}, {Nelson},
  {Quinn}, \& {Rice}}]{Durisen2007prpl.conf.607D}
{Durisen}, R.~H., {Boss}, A.~P., {Mayer}, L., {et~al.} 2007, in Protostars and
  Planets V, ed. B.~{Reipurth}, D.~{Jewitt}, \& K.~{Keil}, 607

\bibitem[{{Dutta} {et~al.}(2015){Dutta}, {Mondal}, {Jose}, {Das}, {Samal}, \&
  {Ghosh}}]{Dutta2015MNRAS.454.3597D}
{Dutta}, S., {Mondal}, S., {Jose}, J., {et~al.} 2015, \mnras, 454, 3597

\bibitem[{{Ercolano} \& {Clarke}(2010)}]{ErcolanoClarke2010MNRAS.402.2735E}
{Ercolano}, B. \& {Clarke}, C.~J. 2010, \mnras, 402, 2735

\bibitem[{{Ercolano} {et~al.}(2009){Ercolano}, {Clarke}, \&
  {Drake}}]{ErcolanoCLarkeDrake2009ApJ.699.1639E}
{Ercolano}, B., {Clarke}, C.~J., \& {Drake}, J.~J. 2009, \apj, 699, 1639

\bibitem[{{Ercolano} {et~al.}(2008){Ercolano}, {Drake}, {Raymond}, \&
  {Clarke}}]{ErcolanoDRC2008ApJ}
{Ercolano}, B., {Drake}, J.~J., {Raymond}, J.~C., \& {Clarke}, C.~C. 2008,
  \apj, 688, 398

\bibitem[{{Esteban} {et~al.}(2013){Esteban}, {Carigi}, {Copetti},
  {Garc{\'\i}a-Rojas}, {Mesa-Delgado}, {Casta{\~n}eda}, \&
  {P{\'e}quignot}}]{Esteban2013MNRAS.433.382E}
{Esteban}, C., {Carigi}, L., {Copetti}, M.~V.~F., {et~al.} 2013, \mnras, 433,
  382

\bibitem[{{Fang} {et~al.}(2016){Fang}, {Kim}, {Pascucci}, {Apai}, \&
  {Manara}}]{Fang2016ApJ.833L.16F}
{Fang}, M., {Kim}, J.~S., {Pascucci}, I., {Apai}, D., \& {Manara}, C.~F. 2016,
  \apjl, 833, L16

\bibitem[{{Fang} {et~al.}(2013){Fang}, {Kim}, {van Boekel}, {Sicilia-Aguilar},
  {Henning}, \& {Flaherty}}]{Fang2013ApJS.207.5F}
{Fang}, M., {Kim}, J.~S., {van Boekel}, R., {et~al.} 2013, \apjs, 207, 5

\bibitem[{{Fang} {et~al.}(2012){Fang}, {van Boekel}, {King}, {Henning},
  {Bouwman}, {Doi}, {Okamoto}, {Roccatagliata}, \&
  {Sicilia-Aguilar}}]{FangBKH2012}
{Fang}, M., {van Boekel}, R., {King}, R.~R., {et~al.} 2012, \aap, 539, A119

\bibitem[{{Feigelson} {et~al.}(2013){Feigelson}, {Townsley}, {Broos}, {Busk},
  {Getman}, {King}, {Kuhn}, {Naylor}, {Povich}, {Baddeley}, {Bate},
  {Indebetouw}, {Luhman}, {McCaughrean}, {Pittard}, {Pudritz}, {Sills}, {Song},
  \& {Wadsley}}]{Feigelson2013ApJS.209.26F}
{Feigelson}, E.~D., {Townsley}, L.~K., {Broos}, P.~S., {et~al.} 2013, \apjs,
  209, 26

\bibitem[{{Fischer} \& {Valenti}(2005)}]{FischerValenti2005ApJ.622.1102F}
{Fischer}, D.~A. \& {Valenti}, J. 2005, \apj, 622, 1102

\bibitem[{{Fitzsimmons} {et~al.}(1992){Fitzsimmons}, {Dufton}, \&
  {Rolleston}}]{Fitzsimmons1992MNRAS.259.489F}
{Fitzsimmons}, A., {Dufton}, P.~L., \& {Rolleston}, W.~R.~J. 1992, \mnras, 259,
  489

\bibitem[{{Flaherty} \& {Muzerolle}(2008)}]{FlahertyMuzerolle2008AJ.135.966F}
{Flaherty}, K.~M. \& {Muzerolle}, J. 2008, \aj, 135, 966

\bibitem[{{Freeman} {et~al.}(2002){Freeman}, {Kashyap}, {Rosner}, \&
  {Lamb}}]{Freeman2002ApJS.138.185F}
{Freeman}, P.~E., {Kashyap}, V., {Rosner}, R., \& {Lamb}, D.~Q. 2002, \apjs,
  138, 185

\bibitem[{{Fruscione} {et~al.}(2006){Fruscione}, {McDowell}, {Allen},
  {Brickhouse}, {Burke}, {Davis}, {Durham}, {Elvis}, {Galle}, {Harris},
  {Huenemoerder}, {Houck}, {Ishibashi}, {Karovska}, {Nicastro}, {Noble},
  {Nowak}, {Primini}, {Siemiginowska}, {Smith}, \&
  {Wise}}]{Fruscione2006SPIE.6270E.1VF}
{Fruscione}, A., {McDowell}, J.~C., {Allen}, G.~E., {et~al.} 2006, in
  \procspie, Vol. 6270, Society of Photo-Optical Instrumentation Engineers
  (SPIE) Conference Series, 62701V

\bibitem[{{Gaia Collaboration} {et~al.}(2016){Gaia Collaboration}, {Prusti},
  {de Bruijne}, {Brown}, {Vallenari}, {Babusiaux}, {Bailer-Jones}, {Bastian},
  {Biermann}, {Evans}, {Eyer}, {Jansen}, {Jordi}, {Klioner}, {Lammers},
  {Lindegren}, {Luri}, {Mignard}, {Milligan}, {Panem}, {Poinsignon},
  {Pourbaix}, {Randich}, {Sarri}, {Sartoretti}, {Siddiqui}, {Soubiran},
  {Valette}, {van Leeuwen}, {Walton}, {Aerts}, {Arenou}, {Cropper}, {Drimmel},
  {H{\o}g}, {Katz}, {Lattanzi}, {O'Mullane}, {Grebel}, {Holland}, {Huc},
  {Passot}, {Bramante}, {Cacciari}, {Casta{\~n}eda}, {Chaoul}, {Cheek}, {De
  Angeli}, {Fabricius}, {Guerra}, {Hern{\'a}ndez}, {Jean-Antoine-Piccolo},
  {Masana}, {Messineo}, {Mowlavi}, {Nienartowicz}, {Ord{\'o}{\~n}ez-Blanco},
  {Panuzzo}, {Portell}, {Richards}, {Riello}, {Seabroke}, {Tanga},
  {Th{\'e}venin}, {Torra}, {Els}, {Gracia-Abril}, {Comoretto},
  {Garcia-Reinaldos}, {Lock}, {Mercier}, {Altmann}, {Andrae}, {Astraatmadja},
  {Bellas-Velidis}, {Benson}, {Berthier}, {Blomme}, {Busso}, {Carry},
  {Cellino}, {Clementini}, {Cowell}, {Creevey}, {Cuypers}, {Davidson}, {De
  Ridder}, {de Torres}, {Delchambre}, {Dell'Oro}, {Ducourant}, {Fr{\'e}mat},
  {Garc{\'\i}a-Torres}, {Gosset}, {Halbwachs}, {Hambly}, {Harrison}, {Hauser},
  {Hestroffer}, {Hodgkin}, {Huckle}, {Hutton}, {Jasniewicz}, {Jordan},
  {Kontizas}, {Korn}, {Lanzafame}, {Manteiga}, {Moitinho}, {Muinonen},
  {Osinde}, {Pancino}, {Pauwels}, {Petit}, {Recio-Blanco}, {Robin}, {Sarro},
  {Siopis}, {Smith}, {Smith}, {Sozzetti}, {Thuillot}, {van Reeven}, {Viala},
  {Abbas}, {Abreu Aramburu}, {Accart}, {Aguado}, {Allan}, {Allasia},
  {Altavilla}, {{\'A}lvarez}, {Alves}, {Anderson}, {Andrei}, {Anglada Varela},
  {Antiche}, {Antoja}, {Ant{\'o}n}, {Arcay}, {Atzei}, {Ayache}, {Bach},
  {Baker}, {Balaguer-N{\'u}{\~n}ez}, {Barache}, {Barata}, {Barbier}, {Barblan},
  {Baroni}, {Barrado y Navascu{\'e}s}, {Barros}, {Barstow}, {Becciani},
  {Bellazzini}, {Bellei}, {Bello Garc{\'\i}a}, {Belokurov}, {Bendjoya},
  {Berihuete}, {Bianchi}, {Bienaym{\'e}}, {Billebaud}, {Blagorodnova},
  {Blanco-Cuaresma}, {Boch}, {Bombrun}, {Borrachero}, {Bouquillon}, {Bourda},
  {Bouy}, {Bragaglia}, {Breddels}, {Brouillet}, {Br{\"u}semeister},
  {Bucciarelli}, {Budnik}, {Burgess}, {Burgon}, {Burlacu}, {Busonero}, {Buzzi},
  {Caffau}, {Cambras}, {Campbell}, {Cancelliere}, {Cantat-Gaudin}, {Carlucci},
  {Carrasco}, {Castellani}, {Charlot}, {Charnas}, {Charvet}, {Chassat},
  {Chiavassa}, {Clotet}, {Cocozza}, {Collins}, {Collins}, {Costigan}, {Crifo},
  {Cross}, {Crosta}, {Crowley}, {Dafonte}, {Damerdji}, {Dapergolas}, {David},
  {David}, {De Cat}, {de Felice}, {de Laverny}, {De Luise}, {De March}, {de
  Martino}, {de Souza}, {Debosscher}, {del Pozo}, {Delbo}, {Delgado},
  {Delgado}, {di Marco}, {Di Matteo}, {Diakite}, {Distefano}, {Dolding}, {Dos
  Anjos}, {Drazinos}, {Dur{\'a}n}, {Dzigan}, {Ecale}, {Edvardsson}, {Enke},
  {Erdmann}, {Escolar}, {Espina}, {Evans}, {Eynard Bontemps}, {Fabre},
  {Fabrizio}, {Faigler}, {Falc{\~a}o}, {Farr{\`a}s Casas}, {Faye}, {Federici},
  {Fedorets}, {Fern{\'a}ndez-Hern{\'a}ndez}, {Fernique}, {Fienga}, {Figueras},
  {Filippi}, {Findeisen}, {Fonti}, {Fouesneau}, {Fraile}, {Fraser}, {Fuchs},
  {Furnell}, {Gai}, {Galleti}, {Galluccio}, {Garabato}, {Garc{\'\i}a-Sedano},
  {Gar{\'e}}, {Garofalo}, {Garralda}, {Gavras}, {Gerssen}, {Geyer}, {Gilmore},
  {Girona}, {Giuffrida}, {Gomes}, {Gonz{\'a}lez-Marcos},
  {Gonz{\'a}lez-N{\'u}{\~n}ez}, {Gonz{\'a}lez-Vidal}, {Granvik}, {Guerrier},
  {Guillout}, {Guiraud}, {G{\'u}rpide}, {Guti{\'e}rrez-S{\'a}nchez}, {Guy},
  {Haigron}, {Hatzidimitriou}, {Haywood}, {Heiter}, {Helmi}, {Hobbs},
  {Hofmann}, {Holl}, {Holland}, {Hunt}, {Hypki}, {Icardi}, {Irwin}, {Jevardat
  de Fombelle}, {Jofr{\'e}}, {Jonker}, {Jorissen}, {Julbe}, {Karampelas},
  {Kochoska}, {Kohley}, {Kolenberg}, {Kontizas}, {Koposov}, {Kordopatis},
  {Koubsky}, {Kowalczyk}, {Krone-Martins}, {Kudryashova}, {Kull}, {Bachchan},
  {Lacoste-Seris}, {Lanza}, {Lavigne}, {Le Poncin-Lafitte}, {Lebreton},
  {Lebzelter}, {Leccia}, {Leclerc}, {Lecoeur-Taibi}, {Lemaitre}, {Lenhardt},
  {Leroux}, {Liao}, {Licata}, {Lindstr{\o}m}, {Lister}, {Livanou}, {Lobel},
  {L{\"o}ffler}, {L{\'o}pez}, {Lopez-Lozano}, {Lorenz}, {Loureiro},
  {MacDonald}, {Magalh{\~a}es Fernandes}, {Managau}, {Mann}, {Mantelet},
  {Marchal}, {Marchant}, {Marconi}, {Marie}, {Marinoni}, {Marrese},
  {Marschalk{\'o}}, {Marshall}, {Mart{\'\i}n-Fleitas}, {Martino}, {Mary},
  {Matijevi{\v{c}}}, {Mazeh}, {McMillan}, {Messina}, {Mestre}, {Michalik},
  {Millar}, {Miranda}, {Molina}, {Molinaro}, {Molinaro}, {Moln{\'a}r},
  {Moniez}, {Montegriffo}, {Monteiro}, {Mor}, {Mora}, {Morbidelli}, {Morel},
  {Morgenthaler}, {Morley}, {Morris}, {Mulone}, {Muraveva}, {Musella},
  {Narbonne}, {Nelemans}, {Nicastro}, {Noval}, {Ord{\'e}novic},
  {Ordieres-Mer{\'e}}, {Osborne}, {Pagani}, {Pagano}, {Pailler}, {Palacin},
  {Palaversa}, {Parsons}, {Paulsen}, {Pecoraro}, {Pedrosa}, {Pentik{\"a}inen},
  {Pereira}, {Pichon}, {Piersimoni}, {Pineau}, {Plachy}, {Plum}, {Poujoulet},
  {Pr{\v{s}}a}, {Pulone}, {Ragaini}, {Rago}, {Rambaux}, {Ramos-Lerate},
  {Ranalli}, {Rauw}, {Read}, {Regibo}, {Renk}, {Reyl{\'e}}, {Ribeiro},
  {Rimoldini}, {Ripepi}, {Riva}, {Rixon}, {Roelens}, {Romero-G{\'o}mez},
  {Rowell}, {Royer}, {Rudolph}, {Ruiz-Dern}, {Sadowski}, {Sagrist{\`a}
  Sell{\'e}s}, {Sahlmann}, {Salgado}, {Salguero}, {Sarasso}, {Savietto},
  {Schnorhk}, {Schultheis}, {Sciacca}, {Segol}, {Segovia}, {Segransan},
  {Serpell}, {Shih}, {Smareglia}, {Smart}, {Smith}, {Solano}, {Solitro},
  {Sordo}, {Soria Nieto}, {Souchay}, {Spagna}, {Spoto}, {Stampa}, {Steele},
  {Steidelm{\"u}ller}, {Stephenson}, {Stoev}, {Suess}, {S{\"u}veges}, {Surdej},
  {Szabados}, {Szegedi-Elek}, {Tapiador}, {Taris}, {Tauran}, {Taylor},
  {Teixeira}, {Terrett}, {Tingley}, {Trager}, {Turon}, {Ulla}, {Utrilla},
  {Valentini}, {van Elteren}, {Van Hemelryck}, {van Leeuwen}, {Varadi},
  {Vecchiato}, {Veljanoski}, {Via}, {Vicente}, {Vogt}, {Voss}, {Votruba},
  {Voutsinas}, {Walmsley}, {Weiler}, {Weingrill}, {Werner}, {Wevers},
  {Whitehead}, {Wyrzykowski}, {Yoldas}, {{\v{Z}}erjal}, {Zucker}, {Zurbach},
  {Zwitter}, {Alecu}, {Allen}, {Allende Prieto}, {Amorim},
  {Anglada-Escud{\'e}}, {Arsenijevic}, {Azaz}, {Balm}, {Beck}, {Bernstein},
  {Bigot}, {Bijaoui}, {Blasco}, {Bonfigli}, {Bono}, {Boudreault}, {Bressan},
  {Brown}, {Brunet}, {Bunclark}, {Buonanno}, {Butkevich}, {Carret}, {Carrion},
  {Chemin}, {Ch{\'e}reau}, {Corcione}, {Darmigny}, {de Boer}, {de Teodoro}, {de
  Zeeuw}, {Delle Luche}, {Domingues}, {Dubath}, {Fodor}, {Fr{\'e}zouls},
  {Fries}, {Fustes}, {Fyfe}, {Gallardo}, {Gallegos}, {Gardiol}, {Gebran},
  {Gomboc}, {G{\'o}mez}, {Grux}, {Gueguen}, {Heyrovsky}, {Hoar}, {Iannicola},
  {Isasi Parache}, {Janotto}, {Joliet}, {Jonckheere}, {Keil}, {Kim},
  {Klagyivik}, {Klar}, {Knude}, {Kochukhov}, {Kolka}, {Kos}, {Kutka}, {Lainey},
  {LeBouquin}, {Liu}, {Loreggia}, {Makarov}, {Marseille}, {Martayan},
  {Martinez-Rubi}, {Massart}, {Meynadier}, {Mignot}, {Munari}, {Nguyen},
  {Nordlander}, {Ocvirk}, {O'Flaherty}, {Olias Sanz}, {Ortiz}, {Osorio},
  {Oszkiewicz}, {Ouzounis}, {Palmer}, {Park}, {Pasquato}, {Peltzer}, {Peralta},
  {P{\'e}turaud}, {Pieniluoma}, {Pigozzi}, {Poels}, {Prat}, {Prod'homme},
  {Raison}, {Rebordao}, {Risquez}, {Rocca-Volmerange}, {Rosen}, {Ruiz-Fuertes},
  {Russo}, {Sembay}, {Serraller Vizcaino}, {Short}, {Siebert}, {Silva},
  {Sinachopoulos}, {Slezak}, {Soffel}, {Sosnowska}, {Strai{\v{z}}ys}, {ter
  Linden}, {Terrell}, {Theil}, {Tiede}, {Troisi}, {Tsalmantza}, {Tur},
  {Vaccari}, {Vachier}, {Valles}, {Van Hamme}, {Veltz}, {Virtanen}, {Wallut},
  {Wichmann}, {Wilkinson}, {Ziaeepour}, \& {Zschocke}}]{Gaia2016AA.595A.1G}
{Gaia Collaboration}, {Prusti}, T., {de Bruijne}, J.~H.~J., {et~al.} 2016,
  \aap, 595, A1

\bibitem[{{Galli} {et~al.}(2019){Galli}, {Loinard}, {Bouy}, {Sarro},
  {Ortiz-Le{\'o}n}, {Dzib}, {Olivares}, {Heyer}, {Hernandez},
  {Rom{\'a}n-Z{\'u}{\~n}iga}, {Kounkel}, \& {Covey}}]{Galli2019AA.630A.137G}
{Galli}, P.~A.~B., {Loinard}, L., {Bouy}, H., {et~al.} 2019, \aap, 630, A137

\bibitem[{{Gorti} \& {Hollenbach}(2009)}]{GortiHollenbach2009ApJ.690.1539G}
{Gorti}, U. \& {Hollenbach}, D. 2009, \apj, 690, 1539

\bibitem[{{Guarcello} {et~al.}(2010{\natexlab{a}}){Guarcello}, {Damiani},
  {Micela}, {Peres}, {Prisinzano}, \& {Sciortino}}]{GuarcelloDMP2010}
{Guarcello}, M.~G., {Damiani}, F., {Micela}, G., {et~al.} 2010{\natexlab{a}},
  \aap, 521, A18+

\bibitem[{{Guarcello} {et~al.}(2016){Guarcello}, {Drake}, {Wright},
  {Albacete-Colombo}, {Clarke}, {Ercolano}, {Flaccomio}, {Kashyap}, {Micela},
  {Naylor}, {Schneider}, {Sciortino}, \& {Vink}}]{GuarcelloDWA2016arXiv}
{Guarcello}, M.~G., {Drake}, J.~J., {Wright}, N.~J., {et~al.} 2016, ArXiv
  e-prints, 1605.01773

\bibitem[{{Guarcello} {et~al.}(2013){Guarcello}, {Drake}, {Wright}, {Drew},
  {Gutermuth}, {Hora}, {Naylor}, {Aldcroft}, {Fruscione},
  {Garc{\'{\i}}a-Alvarez}, {Kashyap}, \& {King}}]{GuarcelloDWD2013ApJ}
{Guarcello}, M.~G., {Drake}, J.~J., {Wright}, N.~J., {et~al.} 2013, \apj, 773,
  135

\bibitem[{{Guarcello} {et~al.}(2014){Guarcello}, {Drake}, {Wright},
  {Garc{\'{\i}}a-Alvarez}, \& {Kraemer}}]{GuarcelloDWG2014}
{Guarcello}, M.~G., {Drake}, J.~J., {Wright}, N.~J., {Garc{\'{\i}}a-Alvarez},
  D., \& {Kraemer}, K.~E. 2014, \apj, 793, 56

\bibitem[{{Guarcello} {et~al.}(2015){Guarcello}, {Drake}, {Wright}, {Naylor},
  {Flaccomio}, {Kashyap}, \& {Garcia-Alvarez}}]{GuarcelloDWN2015}
{Guarcello}, M.~G., {Drake}, J.~J., {Wright}, N.~J., {et~al.} 2015, ArXiv
  e-prints [\eprint[arXiv]{1501.03761}]

\bibitem[{{Guarcello} {et~al.}(2009){Guarcello}, {Micela}, {Damiani}, {Peres},
  {Prisinzano}, \& {Sciortino}}]{GuarcelloMDP2009}
{Guarcello}, M.~G., {Micela}, G., {Damiani}, F., {et~al.} 2009, \aap, 496, 453

\bibitem[{{Guarcello} {et~al.}(2010{\natexlab{b}}){Guarcello}, {Micela},
  {Peres}, {Prisinzano}, \& {Sciortino}}]{GuarcelloMPP2010}
{Guarcello}, M.~G., {Micela}, G., {Peres}, G., {Prisinzano}, L., \&
  {Sciortino}, S. 2010{\natexlab{b}}, \aap, 521, A61

\bibitem[{{Guarcello} {et~al.}(2007){Guarcello}, {Prisinzano}, {Micela},
  {Damiani}, {Peres}, \& {Sciortino}}]{GuarcelloPMD2007}
{Guarcello}, M.~G., {Prisinzano}, L., {Micela}, G., {et~al.} 2007, \aap, 462,
  245

\bibitem[{{Guenther} {et~al.}(2012){Guenther}, {Gandolfi}, {Sebastian},
  {Deleuil}, {Moutou}, \& {Cusano}}]{Guenther2012AA.543A.125G}
{Guenther}, E.~W., {Gandolfi}, D., {Sebastian}, D., {et~al.} 2012, \aap, 543,
  A125

\bibitem[{{Gutermuth} {et~al.}(2009){Gutermuth}, {Megeath}, {Myers}, {Allen},
  {Pipher}, \& {Fazio}}]{GutermuthMMA2009ApJ}
{Gutermuth}, R.~A., {Megeath}, S.~T., {Myers}, P.~C., {et~al.} 2009, \apjs,
  184, 18

\bibitem[{{Habing}(1968)}]{Habing1968}
{Habing}, H.~J. 1968, \bain, 19, 421

\bibitem[{{Haisch} {et~al.}(2001){Haisch}, {Lada}, \& {Lada}}]{HaischLL2001}
{Haisch}, Jr., K.~E., {Lada}, E.~A., \& {Lada}, C.~J. 2001, \apjl, 553, L153

\bibitem[{{Hartmann}(2009)}]{Hartmann2009apsf.book.H}
{Hartmann}, L. 2009, {Accretion Processes in Star Formation: Second Edition}

\bibitem[{{Hartmann} {et~al.}(2006){Hartmann}, {D'Alessio}, {Calvet}, \&
  {Muzerolle}}]{Hartmann2006ApJ.648.484H}
{Hartmann}, L., {D'Alessio}, P., {Calvet}, N., \& {Muzerolle}, J. 2006, \apj,
  648, 484

\bibitem[{{Helled} {et~al.}(2014){Helled}, {Bodenheimer}, {Podolak}, {Boley},
  {Meru}, {Nayakshin}, {Fortney}, {Mayer}, {Alibert}, \&
  {Boss}}]{Helled2014prpl.conf.643H}
{Helled}, R., {Bodenheimer}, P., {Podolak}, M., {et~al.} 2014, in Protostars
  and Planets VI, ed. H.~{Beuther}, R.~S. {Klessen}, C.~P. {Dullemond}, \&
  T.~{Henning}, 643

\bibitem[{{Hern{\'a}ndez} {et~al.}(2008){Hern{\'a}ndez}, {Hartmann}, {Calvet},
  {Jeffries}, {Gutermuth}, {Muzerolle}, \&
  {Stauffer}}]{Hernandez2008ApJ.686.1195H}
{Hern{\'a}ndez}, J., {Hartmann}, L., {Calvet}, N., {et~al.} 2008, \apj, 686,
  1195

\bibitem[{{Hern{\'a}ndez} {et~al.}(2007){Hern{\'a}ndez}, {Hartmann}, {Megeath},
  {Gutermuth}, {Muzerolle}, {Calvet}, {Vivas}, {Brice{\~n}o}, {Allen},
  {Stauffer}, {Young}, \& {Fazio}}]{HernandezHMG2007}
{Hern{\'a}ndez}, J., {Hartmann}, L., {Megeath}, T., {et~al.} 2007, \apj, 662,
  1067

\bibitem[{{Hern{\'a}ndez} {et~al.}(2010){Hern{\'a}ndez}, {Morales-Calderon},
  {Calvet}, {Hartmann}, {Muzerolle}, {Gutermuth}, {Luhman}, \&
  {Stauffer}}]{Hernandez2010ApJ.722.1226H}
{Hern{\'a}ndez}, J., {Morales-Calderon}, M., {Calvet}, N., {et~al.} 2010, \apj,
  722, 1226

\bibitem[{{Hillenbrand} {et~al.}(1993){Hillenbrand}, {Massey}, {Strom}, \&
  {Merrill}}]{HillenbrandMSM1993}
{Hillenbrand}, L.~A., {Massey}, P., {Strom}, S.~E., \& {Merrill}, K.~M. 1993,
  \aj, 106, 1906

\bibitem[{{Hubickyj} {et~al.}(2005){Hubickyj}, {Bodenheimer}, \&
  {Lissauer}}]{Hubickyj2005Icar.179.415H}
{Hubickyj}, O., {Bodenheimer}, P., \& {Lissauer}, J.~J. 2005, \icarus, 179, 415

\bibitem[{{Jeffries} {et~al.}(2017){Jeffries}, {Jackson}, {Franciosini}, {Rand
  ich}, {Barrado}, {Frasca}, {Klutsch}, {Lanzafame}, {Prisinzano}, {Sacco},
  {Gilmore}, {Vallenari}, {Alfaro}, {Koposov}, {Pancino}, {Bayo}, {Casey},
  {Costado}, {Damiani}, {Hourihane}, {Lewis}, {Jofre}, {Magrini}, {Monaco},
  {Morbidelli}, {Worley}, {Zaggia}, \& {Zwitter}}]{Jeffries2017MNRAS.464.1456J}
{Jeffries}, R.~D., {Jackson}, R.~J., {Franciosini}, E., {et~al.} 2017, \mnras,
  464, 1456

\bibitem[{{Johnstone} {et~al.}(1998){Johnstone}, {Hollenbach}, \&
  {Bally}}]{JohnstoneHB1998}
{Johnstone}, D., {Hollenbach}, D., \& {Bally}, J. 1998, \apj, 499, 758

\bibitem[{{Jose} {et~al.}(2016){Jose}, {Kim}, {Herczeg}, {Samal}, {Bieging},
  {Meyer}, \& {Sherry}}]{Jose2016ApJ.822.49J}
{Jose}, J., {Kim}, J.~S., {Herczeg}, G.~J., {et~al.} 2016, \apj, 822, 49

\bibitem[{{Kalari} \& {Vink}(2015)}]{Kalari2015ApJ.800.113K}
{Kalari}, V.~M. \& {Vink}, J.~S. 2015, \apj, 800, 113

\bibitem[{{Kim} {et~al.}(2016){Kim}, {Clarke}, {Fang}, \&
  {Facchini}}]{Kim2016ApJ.826L.15K}
{Kim}, J.~S., {Clarke}, C.~J., {Fang}, M., \& {Facchini}, S. 2016, \apjl, 826,
  L15

\bibitem[{{Koenig} \& {Leisawitz}(2014)}]{KoenigLeisawitz2014ApJ.791.131K}
{Koenig}, X.~P. \& {Leisawitz}, D.~T. 2014, \apj, 791, 131

\bibitem[{{Kounkel} {et~al.}(2018){Kounkel}, {Covey}, {Su{\'a}rez},
  {Rom{\'a}n-Z{\'u}{\~n}iga}, {Hernandez}, {Stassun}, {Jaehnig}, {Feigelson},
  {Pe{\~n}a Ram{\'\i}rez}, {Roman-Lopes}, {Da Rio}, {Stringfellow}, {Kim},
  {Borissova}, {Fern{\'a}ndez-Trincado}, {Burgasser},
  {Garc{\'\i}a-Hern{\'a}ndez}, {Zamora}, {Pan}, \&
  {Nitschelm}}]{Kounkel2018AJ.156.84K}
{Kounkel}, M., {Covey}, K., {Su{\'a}rez}, G., {et~al.} 2018, \aj, 156, 84

\bibitem[{{Kraus} {et~al.}(2017){Kraus}, {Herczeg}, {Rizzuto}, {Mann},
  {Slesnick}, {Carpenter}, {Hillenbrand}, \& {Mamajek}}]{Kraus2017ApJ.838.150K}
{Kraus}, A.~L., {Herczeg}, G.~J., {Rizzuto}, A.~C., {et~al.} 2017, \apj, 838,
  150

\bibitem[{{Kroupa} \& {Weidner}(2003)}]{KroupaWeidner2003ApJ.598.1076K}
{Kroupa}, P. \& {Weidner}, C. 2003, \apj, 598, 1076

\bibitem[{{Lada} {et~al.}(2006){Lada}, {Muench}, {Luhman}, {Allen}, {Hartmann},
  {Megeath}, {Myers}, {Fazio}, {Wood}, {Muzerolle}, {Rieke}, {Siegler}, \&
  {Young}}]{LadaMLA2006}
{Lada}, C.~J., {Muench}, A.~A., {Luhman}, K.~L., {et~al.} 2006, \aj, 131, 1574

\bibitem[{{Lawrence} {et~al.}(2007){Lawrence}, {Warren}, {Almaini}, {Edge},
  {Hambly}, {Jameson}, {Lucas}, {Casali}, {Adamson}, {Dye}, {Emerson},
  {Foucaud}, {Hewett}, {Hirst}, {Hodgkin}, {Irwin}, {Lodieu}, {McMahon},
  {Simpson}, {Smail}, {Mortlock}, \& {Folger}}]{LawrenceWAE2007}
{Lawrence}, A., {Warren}, S.~J., {Almaini}, O., {et~al.} 2007, \mnras, 379,
  1599

\bibitem[{{Lennon} {et~al.}(1990){Lennon}, {Dufton}, {Fitzsimmons}, {Gehren},
  \& {Nissen}}]{LennonDFG1990}
{Lennon}, D.~J., {Dufton}, P.~L., {Fitzsimmons}, A., {Gehren}, T., \& {Nissen},
  P.~E. 1990, \aap, 240, 349

\bibitem[{{Luhman} \& {Mamajek}(2012)}]{LuhmanMamajek2012ApJ}
{Luhman}, K.~L. \& {Mamajek}, E.~E. 2012, \apj, 758, 31

\bibitem[{{Luo} {et~al.}(2015){Luo}, {Zhao}, {Zhao}, {Deng}, {Liu}, {Jing},
  {Wang}, {Zhang}, {Shi}, {Cui}, {Chu}, {Li}, {Bai}, {Cai}, {Cao}, {Cao},
  {Carlin}, {Chen}, {Chen}, {Chen}, {Chen}, {Chen}, {Chen}, {Chen},
  {Christlieb}, {Chu}, {Cui}, {Dong}, {Du}, {Fan}, {Feng}, {Fu}, {Gao}, {Gong},
  {Gu}, {Guo}, {Han}, {He}, {Hou}, {Hou}, {Hou}, {Hu}, {Hu}, {Hu}, {Huo},
  {Jia}, {Jiang}, {Jiang}, {Jiang}, {Jin}, {Kong}, {Kong}, {Lei}, {Li}, {Li},
  {Li}, {Li}, {Li}, {Li}, {Li}, {Li}, {Li}, {Li}, {Li}, {Li}, {Liang}, {Lin},
  {Liu}, {Liu}, {Liu}, {Liu}, {Lu}, {Luo}, {Mao}, {Newberg}, {Ni}, {Qi}, {Qi},
  {Shen}, {Shi}, {Song}, {Song}, {Su}, {Su}, {Tang}, {Tao}, {Tian}, {Wang},
  {Wang}, {Wang}, {Wang}, {Wang}, {Wang}, {Wang}, {Wang}, {Wang}, {Wang},
  {Wang}, {Wang}, {Wang}, {Wang}, {Wang}, {Wang}, {Wang}, {Wang}, {Wang},
  {Wang}, {Wei}, {Wei}, {Wu}, {Wu}, {Wu}, {Wu}, {Wu}, {Xing}, {Xu}, {Xu}, {Xu},
  {Yan}, {Yang}, {Yang}, {Yang}, {Yang}, {Yao}, {Yu}, {Yuan}, {Yuan}, {Yuan},
  {Yuan}, {Zhai}, {Zhang}, {Zhang}, {Zhang}, {Zhang}, {Zhang}, {Zhang},
  {Zhang}, {Zhang}, {Zhao}, {Zhou}, {Zhou}, {Zhu}, {Zhu}, {Zou}, \&
  {Zuo}}]{Luo2015arXiv150501570L}
{Luo}, A.-L., {Zhao}, Y.-H., {Zhao}, G., {et~al.} 2015, arXiv e-prints
  [\eprint[arXiv]{1505.01570}]

\bibitem[{{Luo} {et~al.}(2016){Luo}, {Zhao}, {Zhao}, {Deng}, {Liu}, {Jing},
  {Wang}, {Zhang}, {Shi}, {Cui}, {Chu}, {Li}, {Bai}, {Wu}, {Cai}, {Cao}, {Cao},
  {Carlin}, {Chen}, {Chen}, {Chen}, {Chen}, {Chen}, {Chen}, {Chen},
  {Christlieb}, {Chu}, {Cui}, {Dong}, {Du}, {Fan}, {Feng}, {Fu}, {Gao}, {Gong},
  {Gu}, {Guo}, {Han}, {He}, {Hou}, {Hou}, {Hou}, {Hu}, {Hu}, {Hu}, {Huo},
  {Jia}, {Jiang}, {Jiang}, {Jiang}, {Jin}, {Kong}, {Kong}, {Lei}, {Li}, {Li},
  {Li}, {Li}, {Li}, {Li}, {Li}, {Li}, {Li}, {Li}, {Li}, {Li}, {Liang}, {Lin},
  {Liu}, {Liu}, {Liu}, {Liu}, {Lu}, {Luo}, {Mao}, {Newberg}, {Ni}, {Qi}, {Qi},
  {Shen}, {Shi}, {Song}, {Song}, {Su}, {Su}, {Tang}, {Tao}, {Tian}, {Wang},
  {Wang}, {Wang}, {Wang}, {Wang}, {Wang}, {Wang}, {Wang}, {Wang}, {Wang},
  {Wang}, {Wang}, {Wang}, {Wang}, {Wang}, {Wang}, {Wang}, {Wang}, {Wang},
  {Wang}, {Wei}, {Wei}, {Wu}, {Wu}, {Wu}, {Wu}, {Xing}, {Xu}, {Xu}, {Xu},
  {Yan}, {Yang}, {Yang}, {Yang}, {Yang}, {Yao}, {Yu}, {Yuan}, {Yuan}, {Yuan},
  {Yuan}, {Zhai}, {Zhang}, {Zhang}, {Zhang}, {Zhang}, {Zhang}, {Zhang},
  {Zhang}, {Zhang}, {Zhao}, {Zhou}, {Zhou}, {Zhu}, {Zhu}, {Zou}, \&
  {Zuo}}]{Luo2016yCat.5149.0L}
{Luo}, A.-L., {Zhao}, Y.-H., {Zhao}, G., {et~al.} 2016, VizieR Online Data
  Catalog, 5149

\bibitem[{{Maeder} {et~al.}(1999){Maeder}, {Grebel}, \&
  {Mermilliod}}]{Maeder1999AA.346.459M}
{Maeder}, A., {Grebel}, E.~K., \& {Mermilliod}, J.-C. 1999, \aap, 346, 459

\bibitem[{{Manzo-Mart{\'\i}nez} {et~al.}(2020){Manzo-Mart{\'\i}nez}, {Calvet},
  {Hern{\'a}ndez}, {Lizano}, {Hern{\'a}ndez}, {Miller}, {Mauc{\'o}},
  {Brice{\~n}o}, \& {D'Alessio}}]{ManzoMartinez2020ApJ.893.56M}
{Manzo-Mart{\'\i}nez}, E., {Calvet}, N., {Hern{\'a}ndez}, J., {et~al.} 2020,
  \apj, 893, 56

\bibitem[{{Mari{\~n}as} {et~al.}(2013){Mari{\~n}as}, {Lada}, {Teixeira}, \&
  {Lada}}]{Marinas2013ApJ.772.81M}
{Mari{\~n}as}, N., {Lada}, E.~A., {Teixeira}, P.~S., \& {Lada}, C.~J. 2013,
  \apj, 772, 81

\bibitem[{{Martins} {et~al.}(2005){Martins}, {Schaerer}, \&
  {Hillier}}]{MartinSH2005}
{Martins}, F., {Schaerer}, D., \& {Hillier}, D.~J. 2005, \aap, 436, 1049

\bibitem[{{Megeath} {et~al.}(2005){Megeath}, {Hartmann}, {Luhman}, \&
  {Fazio}}]{Megeath2005ApJ.634L.113M}
{Megeath}, S.~T., {Hartmann}, L., {Luhman}, K.~L., \& {Fazio}, G.~G. 2005,
  \apjl, 634, L113

\bibitem[{{Mer{\'\i}n} {et~al.}(2008){Mer{\'\i}n}, {J{\o}rgensen}, {Spezzi},
  {Alcal{\'a}}, {Evans}, {Harvey}, {Prusti}, {Chapman}, {Huard}, {van
  Dishoeck}, \& {Comer{\'o}n}}]{Merin2008ApJS.177.551M}
{Mer{\'\i}n}, B., {J{\o}rgensen}, J., {Spezzi}, L., {et~al.} 2008, \apjs, 177,
  551

\bibitem[{{Mesa-Delgado} {et~al.}(2016){Mesa-Delgado}, {Zapata}, {Henney},
  {Puzia}, \& {Tsamis}}]{Mesa-Delgado2016ApJ.825L.16M}
{Mesa-Delgado}, A., {Zapata}, L., {Henney}, W.~J., {Puzia}, T.~H., \& {Tsamis},
  Y.~G. 2016, \apjl, 825, L16

\bibitem[{{Moffat} \& {Vogt}(1975)}]{MoffatVogt1975AAS.20.85M}
{Moffat}, A.~F.~J. \& {Vogt}, N. 1975, \aaps, 20, 85

\bibitem[{{Montmerle}(1996)}]{Montmerle1996}
{Montmerle}, T. 1996, in Astronomical Society of the Pacific Conference Series,
  Vol. 109, Cool Stars, Stellar Systems, and the Sun, ed. {R.~Pallavicini \&
  A.~K.~Dupree}, 405--+

\bibitem[{{Morales-Calder{\'o}n} {et~al.}(2011){Morales-Calder{\'o}n},
  {Stauffer}, {Hillenbrand}, {Gutermuth}, {Song}, {Rebull}, {Plavchan},
  {Carpenter}, {Whitney}, {Covey}, {Alves de Oliveira}, {Winston},
  {McCaughrean}, {Bouvier}, {Guieu}, {Vrba}, {Holtzman}, {Marchis}, {Hora},
  {Wasserman}, {Terebey}, {Megeath}, {Guinan}, {Forbrich}, {Hu{\'e}lamo},
  {Riviere-Marichalar}, {Barrado}, {Stapelfeldt}, {Hern{\'a}ndez}, {Allen},
  {Ardila}, {Bayo}, {Favata}, {James}, {Werner}, \&
  {Wood}}]{Morales-CalderonSHG2011}
{Morales-Calder{\'o}n}, M., {Stauffer}, J.~R., {Hillenbrand}, L.~A., {et~al.}
  2011, \apj, 733, 50

\bibitem[{{Muzerolle} {et~al.}(1998){Muzerolle}, {Hartmann}, \&
  {Calvet}}]{MuzerolleHC1998}
{Muzerolle}, J., {Hartmann}, L., \& {Calvet}, N. 1998, \aj, 116, 455

\bibitem[{{Nakatani} {et~al.}(2018{\natexlab{a}}){Nakatani}, {Hosokawa},
  {Yoshida}, {Nomura}, \& {Kuiper}}]{Nakatani2018ApJ.857.57N}
{Nakatani}, R., {Hosokawa}, T., {Yoshida}, N., {Nomura}, H., \& {Kuiper}, R.
  2018{\natexlab{a}}, \apj, 857, 57

\bibitem[{{Nakatani} {et~al.}(2018{\natexlab{b}}){Nakatani}, {Hosokawa},
  {Yoshida}, {Nomura}, \& {Kuiper}}]{Nakatani2018ApJ.865.75N}
{Nakatani}, R., {Hosokawa}, T., {Yoshida}, N., {Nomura}, H., \& {Kuiper}, R.
  2018{\natexlab{b}}, \apj, 865, 75

\bibitem[{{Negueruela} {et~al.}(2015){Negueruela}, {Sim{\'o}n-D{\'\i}az},
  {Lorenzo}, {Castro}, \& {Herrero}}]{Negueruela2015AA.584A.77N}
{Negueruela}, I., {Sim{\'o}n-D{\'\i}az}, S., {Lorenzo}, J., {Castro}, N., \&
  {Herrero}, A. 2015, \aap, 584, A77

\bibitem[{{Ochsenbein} {et~al.}(2000){Ochsenbein}, {Bauer}, \&
  {Marcout}}]{Ochsenbein2000AAS.143.23O}
{Ochsenbein}, F., {Bauer}, P., \& {Marcout}, J. 2000, \aaps, 143, 23

\bibitem[{{O'dell} \& {Wen}(1994)}]{ODellW1994}
{O'dell}, C.~R. \& {Wen}, Z. 1994, \apj, 436, 194

\bibitem[{{O'Donnell}(1994)}]{Donnell1994ApJ}
{O'Donnell}, J.~E. 1994, \apj, 422, 158

\bibitem[{{Olczak} {et~al.}(2012){Olczak}, {Kaczmarek}, {Harfst}, {Pfalzner},
  \& {Portegies Zwart}}]{OlczakKHP2012}
{Olczak}, C., {Kaczmarek}, T., {Harfst}, S., {Pfalzner}, S., \& {Portegies
  Zwart}, S. 2012, \apj, 756, 123

\bibitem[{Paradis \& Schliep(2018)}]{Paradis10.1093/bioinformatics/bty633}
Paradis, E. \& Schliep, K. 2018, Bioinformatics, 35, 526

\bibitem[{{Park} \& {Sung}(2002)}]{ParkSung2002}
{Park}, B.-G. \& {Sung}, H. 2002, \aj, 123, 892

\bibitem[{{Pecaut} \& {Mamajek}(2016)}]{PecautMamajek2016MNRAS.461.794P}
{Pecaut}, M.~J. \& {Mamajek}, E.~E. 2016, \mnras, 461, 794

\bibitem[{{Pfalzner} {et~al.}(2005){Pfalzner}, {Umbreit}, \&
  {Henning}}]{PfalznerUH2005}
{Pfalzner}, S., {Umbreit}, S., \& {Henning}, T. 2005, \apj, 629, 526

\bibitem[{{Pizzolato} {et~al.}(2001){Pizzolato}, {Ventura}, {D'Antona},
  {Maggio}, {Micela}, \& {Sciortino}}]{Pizzolato2001AA.373.597P}
{Pizzolato}, N., {Ventura}, P., {D'Antona}, F., {et~al.} 2001, \aap, 373, 597

\bibitem[{{Pollack} {et~al.}(1996){Pollack}, {Hubickyj}, {Bodenheimer},
  {Lissauer}, {Podolak}, \& {Greenzweig}}]{PollackHBL1996Icar}
{Pollack}, J.~B., {Hubickyj}, O., {Bodenheimer}, P., {et~al.} 1996, icarus,
  124, 62

\bibitem[{{Prisinzano} {et~al.}(2011){Prisinzano}, {Sanz-Forcada}, {Micela},
  {Caramazza}, {Guarcello}, {Sciortino}, \&
  {Testi}}]{Prisinzano2011AA.527A.77P}
{Prisinzano}, L., {Sanz-Forcada}, J., {Micela}, G., {et~al.} 2011, \aap, 527,
  A77

\bibitem[{{Puga} {et~al.}(2009){Puga}, {Hony}, {Neiner}, {Lenorzer}, {Hubert},
  {Waters}, {Cusano}, \& {Ripepi}}]{Puga2009AA.503.107P}
{Puga}, E., {Hony}, S., {Neiner}, C., {et~al.} 2009, \aap, 503, 107

\bibitem[{{Rafferty} {et~al.}(2011){Rafferty}, {Brandt}, {Alexander}, {Xue},
  {Bauer}, {Lehmer}, {Luo}, \& {Papovich}}]{Rafferty2011ApJ.742.3R}
{Rafferty}, D.~A., {Brandt}, W.~N., {Alexander}, D.~M., {et~al.} 2011, \apj,
  742, 3

\bibitem[{{Rapson} {et~al.}(2014){Rapson}, {Pipher}, {Gutermuth}, {Megeath},
  {Allen}, {Myers}, \& {Allen}}]{Rapson2014ApJ.794.124R}
{Rapson}, V.~A., {Pipher}, J.~L., {Gutermuth}, R.~A., {et~al.} 2014, \apj, 794,
  124

\bibitem[{{Reiter} \& {Parker}(2019)}]{ReiterParker2019MNRAS.486.4354R}
{Reiter}, M. \& {Parker}, R.~J. 2019, \mnras, 486, 4354

\bibitem[{{Ribas} {et~al.}(2014){Ribas}, {Mer{\'\i}n}, {Bouy}, \&
  {Maud}}]{Ribas2014AA.561A.54R}
{Ribas}, {\'A}., {Mer{\'\i}n}, B., {Bouy}, H., \& {Maud}, L.~T. 2014, \aap,
  561, A54

\bibitem[{{Richert} {et~al.}(2015){Richert}, {Feigelson}, {Getman}, \&
  {Kuhn}}]{RichertFGK2015}
{Richert}, A.~J.~W., {Feigelson}, E.~D., {Getman}, K.~V., \& {Kuhn}, M.~A.
  2015, \apj, 811, 10

\bibitem[{{Richert} {et~al.}(2018){Richert}, {Getman}, {Feigelson}, {Kuhn},
  {Broos}, {Povich}, {Bate}, \& {Garmire}}]{Richert2018MNRAS.477.5191R}
{Richert}, A.~J.~W., {Getman}, K.~V., {Feigelson}, E.~D., {et~al.} 2018,
  \mnras, 477, 5191

\bibitem[{{Robitaille}(2017)}]{Robitaille2017AA.600A.11R}
{Robitaille}, T.~P. 2017, \aap, 600, A11

\bibitem[{{Roccatagliata} {et~al.}(2011){Roccatagliata}, {Bouwman}, {Henning},
  {Gennaro}, {Feigelson}, {Kim}, {Sicilia-Aguilar}, \&
  {Lawson}}]{RoccatagliataBHG2011}
{Roccatagliata}, V., {Bouwman}, J., {Henning}, T., {et~al.} 2011, \apj, 733,
  113

\bibitem[{{Rolleston} {et~al.}(2000){Rolleston}, {Smartt}, {Dufton}, \&
  {Ryans}}]{Rolleston2000AA.363.537R}
{Rolleston}, W.~R.~J., {Smartt}, S.~J., {Dufton}, P.~L., \& {Ryans}, R.~S.~I.
  2000, \aap, 363, 537

\bibitem[{{Rugel} {et~al.}(2018){Rugel}, {Fedele}, \&
  {Herczeg}}]{Rugel2018AA.609A.70R}
{Rugel}, M., {Fedele}, D., \& {Herczeg}, G. 2018, \aap, 609, A70

\bibitem[{{Sebastian} {et~al.}(2012){Sebastian}, {Guenther}, {Schaffenroth},
  {Gand olfi}, {Geier}, {Heber}, {Deleuil}, \&
  {Moutou}}]{Sebastian2012AA.541A.34S}
{Sebastian}, D., {Guenther}, E.~W., {Schaffenroth}, V., {et~al.} 2012, \aap,
  541, A34

\bibitem[{{Sharpless}(1959)}]{Sharpless1959ApJS.4.257S}
{Sharpless}, S. 1959, \apjs, 4, 257

\bibitem[{{Sicilia-Aguilar} {et~al.}(2006{\natexlab{a}}){Sicilia-Aguilar},
  {Hartmann}, {Calvet}, {Megeath}, {Muzerolle}, {Allen}, {D'Alessio},
  {Mer{\'\i}n}, {Stauffer}, {Young}, \&
  {Lada}}]{Sicilia-Aguilar2006ApJ.638.897S}
{Sicilia-Aguilar}, A., {Hartmann}, L., {Calvet}, N., {et~al.}
  2006{\natexlab{a}}, \apj, 638, 897

\bibitem[{{Sicilia-Aguilar} {et~al.}(2006{\natexlab{b}}){Sicilia-Aguilar},
  {Hartmann}, {F{\"u}r{\'e}sz}, {Henning}, {Dullemond}, \&
  {Brandner}}]{Sicilia-Aguilar2006AJ.132.2135S}
{Sicilia-Aguilar}, A., {Hartmann}, L.~W., {F{\"u}r{\'e}sz}, G., {et~al.}
  2006{\natexlab{b}}, \aj, 132, 2135

\bibitem[{{Simon} {et~al.}(2013){Simon}, {Bai}, {Armitage}, {Stone}, \&
  {Beckwith}}]{Simon2013ApJ.775.73S}
{Simon}, J.~B., {Bai}, X.-N., {Armitage}, P.~J., {Stone}, J.~M., \& {Beckwith},
  K. 2013, \apj, 775, 73

\bibitem[{{Smith} {et~al.}(2011){Smith}, {Dunne}, {Maddox}, {Eales},
  {Bonfield}, {Jarvis}, {Sutherland}, {Fleuren}, {Rigby}, {Thompson}, {Baldry},
  {Bamford}, {Buttiglione}, {Cava}, {Clements}, {Cooray}, {Croom}, {Dariush},
  {de Zotti}, {Driver}, {Dunlop}, {Fritz}, {Hill}, {Hopkins}, {Hopwood},
  {Ibar}, {Ivison}, {Jones}, {Kelvin}, {Leeuw}, {Liske}, {Loveday}, {Madore},
  {Norberg}, {Panuzzo}, {Pascale}, {Pohlen}, {Popescu}, {Prescott}, {Robotham},
  {Rodighiero}, {Scott}, {Seibert}, {Sharp}, {Temi}, {Tuffs}, {van der Werf},
  \& {van Kampen}}]{SmithDJB2011}
{Smith}, D.~J.~B., {Dunne}, L., {Maddox}, S.~J., {et~al.} 2011, \mnras, 416,
  857

\bibitem[{{Spezzi} {et~al.}(2012){Spezzi}, {de Marchi}, {Panagia},
  {Sicilia-Aguilar}, \& {Ercolano}}]{SpezziDPS2012}
{Spezzi}, L., {de Marchi}, G., {Panagia}, N., {Sicilia-Aguilar}, A., \&
  {Ercolano}, B. 2012, \mnras, 421, 78

\bibitem[{{Spezzi} {et~al.}(2015){Spezzi}, {Petr-Gotzens}, {Alcal{\'a}},
  {J{\o}rgensen}, {Stanke}, {Lombardi}, \& {Alves}}]{Spezzi2015AA.581A.140S}
{Spezzi}, L., {Petr-Gotzens}, M.~G., {Alcal{\'a}}, J.~M., {et~al.} 2015, \aap,
  581, A140

\bibitem[{{Stassun} {et~al.}(2007){Stassun}, {van den Berg}, \&
  {Feigelson}}]{StassunBF2007ApJ}
{Stassun}, K.~G., {van den Berg}, M., \& {Feigelson}, E. 2007, \apj, 660, 704

\bibitem[{{Steinhausen} \& {Pfalzner}(2014)}]{SteinhausenPfalzner2014AA}
{Steinhausen}, M. \& {Pfalzner}, S. 2014, \aap, 565, A32

\bibitem[{{Stelzer} \& {Scholz}(2009)}]{StelzerScholz2009AA.507.227S}
{Stelzer}, B. \& {Scholz}, A. 2009, \aap, 507, 227

\bibitem[{{Stern} {et~al.}(2005){Stern}, {Eisenhardt}, {Gorjian}, {Kochanek},
  {Caldwell}, {Eisenstein}, {Brodwin}, {Brown}, {Cool}, {Dey}, {Green},
  {Jannuzi}, {Murray}, {Pahre}, \& {Willner}}]{Stern2005}
{Stern}, D., {Eisenhardt}, P., {Gorjian}, V., {et~al.} 2005, \apj, 631, 163

\bibitem[{{Sung} {et~al.}(2013){Sung}, {Sana}, \&
  {Bessell}}]{Sung2013AJ.145.37S}
{Sung}, H., {Sana}, H., \& {Bessell}, M.~S. 2013, \aj, 145, 37

\bibitem[{{Sung} {et~al.}(2009){Sung}, {Stauffer}, \& {Bessell}}]{SungSB2009}
{Sung}, H., {Stauffer}, J.~R., \& {Bessell}, M.~S. 2009, \aj, 138, 1116

\bibitem[{{Suzuki} {et~al.}(2010){Suzuki}, {Muto}, \&
  {Inutsuka}}]{Suzuki2010ApJ.718.1289S}
{Suzuki}, T.~K., {Muto}, T., \& {Inutsuka}, S.-i. 2010, \apj, 718, 1289

\bibitem[{{Taylor}(2005)}]{Taylor2005ASPC.347.29T}
{Taylor}, M.~B. 2005, in Astronomical Society of the Pacific Conference Series,
  Vol. 347, Astronomical Data Analysis Software and Systems XIV, ed.
  P.~{Shopbell}, M.~{Britton}, \& R.~{Ebert}, 29

\bibitem[{{Thies} {et~al.}(2010){Thies}, {Kroupa}, {Goodwin}, {Stamatellos}, \&
  {Whitworth}}]{ThiesKGS2010}
{Thies}, I., {Kroupa}, P., {Goodwin}, S.~P., {Stamatellos}, D., \& {Whitworth},
  A.~P. 2010, \apj, 717, 577

\bibitem[{{Treister} {et~al.}(2006){Treister}, {Urry}, {Van Duyne},
  {Dickinson}, {Chary}, {Alexander}, {Bauer}, {Natarajan}, {Lira}, \&
  {Grogin}}]{Treister2006ApJ.640.603T}
{Treister}, E., {Urry}, C.~M., {Van Duyne}, J., {et~al.} 2006, \apj, 640, 603

\bibitem[{{Turbide} \& {Moffat}(1993)}]{TurbideMoffat1993AJ.105.1831T}
{Turbide}, L. \& {Moffat}, A. F.~J. 1993, \aj, 105, 1831

\bibitem[{{Usatov}(2018)}]{Usatov2018JAD.24.3U}
{Usatov}, M. 2018, Journal of Astronomical Data, 24, 3

\bibitem[{{Vincke} {et~al.}(2015){Vincke}, {Breslau}, \&
  {Pfalzner}}]{VinckeBP2015}
{Vincke}, K., {Breslau}, A., \& {Pfalzner}, S. 2015, \aap, 577, A115

\bibitem[{{Wang} {et~al.}(2008){Wang}, {Townsley}, {Feigelson}, {Broos},
  {Getman}, {Rom{\'a}n-Z{\'u}{\~n}iga}, \& {Lada}}]{Wang2008ApJ.675.464W}
{Wang}, J., {Townsley}, L.~K., {Feigelson}, E.~D., {et~al.} 2008, \apj, 675,
  464

\bibitem[{{Weidner} {et~al.}(2010){Weidner}, {Kroupa}, \&
  {Bonnell}}]{Weidner2010MNRAS.401.275W}
{Weidner}, C., {Kroupa}, P., \& {Bonnell}, I.~A.~D. 2010, \mnras, 401, 275

\bibitem[{{Winston} {et~al.}(2007){Winston}, {Megeath}, {Wolk}, {Muzerolle},
  {Gutermuth}, {Hora}, {Allen}, {Spitzbart}, {Myers}, \&
  {Fazio}}]{Winston2007ApJ.669.493W}
{Winston}, E., {Megeath}, S.~T., {Wolk}, S.~J., {et~al.} 2007, \apj, 669, 493

\bibitem[{{Wright} {et~al.}(2010){Wright}, {Eisenhardt}, {Mainzer}, {Ressler},
  {Cutri}, {Jarrett}, {Kirkpatrick}, {Padgett}, {McMillan}, {Skrutskie},
  {Stanford}, {Cohen}, {Walker}, {Mather}, {Leisawitz}, {Gautier}, {McLean},
  {Benford}, {Lonsdale}, {Blain}, {Mendez}, {Irace}, {Duval}, {Liu}, {Royer},
  {Heinrichsen}, {Howard}, {Shannon}, {Kendall}, {Walsh}, {Larsen}, {Cardon},
  {Schick}, {Schwalm}, {Abid}, {Fabinsky}, {Naes}, \&
  {Tsai}}]{Wright2010AJ.140.1868W}
{Wright}, E.~L., {Eisenhardt}, P.~R.~M., {Mainzer}, A.~K., {et~al.} 2010, \aj,
  140, 1868

\bibitem[{{Wright} {et~al.}(2012){Wright}, {Drake}, {Drew}, {Guarcello},
  {Gutermuth}, {Hora}, \& {Kraemer}}]{WrightDDG2012}
{Wright}, N.~J., {Drake}, J.~J., {Drew}, J.~E., {et~al.} 2012, \apjl, 746, L21

\bibitem[{{Wright} {et~al.}(2014){Wright}, {Drake}, {Guarcello}, {Aldcroft},
  {Kashyap}, {Damiani}, {DePasquale}, \& {Fruscione}}]{WrightDGA2014}
{Wright}, N.~J., {Drake}, J.~J., {Guarcello}, M.~G., {et~al.} 2014, ArXiv
  e-prints, 14086579 [\eprint[arXiv]{1408.6579}]

\bibitem[{{Yasui} {et~al.}(2016{\natexlab{a}}){Yasui}, {Kobayashi}, {Saito}, \&
  {Izumi}}]{Yasui2016AJ.151.115Y}
{Yasui}, C., {Kobayashi}, N., {Saito}, M., \& {Izumi}, N. 2016{\natexlab{a}},
  \aj, 151, 115

\bibitem[{{Yasui} {et~al.}(2016{\natexlab{b}}){Yasui}, {Kobayashi}, {Tokunaga},
  {Saito}, \& {Izumi}}]{Yasui2016AJ.151.50Y}
{Yasui}, C., {Kobayashi}, N., {Tokunaga}, A.~T., {Saito}, M., \& {Izumi}, N.
  2016{\natexlab{b}}, \aj, 151, 50

\bibitem[{{Yasui} {et~al.}(2009){Yasui}, {Kobayashi}, {Tokunaga}, {Saito}, \&
  {Tokoku}}]{YasuiKTS2009}
{Yasui}, C., {Kobayashi}, N., {Tokunaga}, A.~T., {Saito}, M., \& {Tokoku}, C.
  2009, \apj, 705, 54

\bibitem[{{Yasui} {et~al.}(2010){Yasui}, {Kobayashi}, {Tokunaga}, {Saito}, \&
  {Tokoku}}]{YasuiKTS2010}
{Yasui}, C., {Kobayashi}, N., {Tokunaga}, A.~T., {Saito}, M., \& {Tokoku}, C.
  2010, \apjl, 723, L113

\end{thebibliography}

%
%
\newpage
\begin{appendix} 
\begin{onecolumn}

\section{The catalog of X-ray sources in Dolidze~25}
\label{app_xraycat}

In this appendix we describe the catalog of the X-ray sources in Dolidze~25.

\begin{table}[!h]
\caption{X-ray catalog of Dolidze~25}             
\label{match_table}      
\centering                          
\begin{tabular}{c c c c}        
\hline\hline                 
Index & Field & Units & Description \\    
\hline                        
1   &   Member\_ID                & ...               &   Stellar ID in the members catalog       \\
2   &   ACIS\_DES                 & ...               &   X-ray identifier                 \\
3   &   RA                        & deg               &   Right Ascension                  \\
4   &   Dec                       & deg               &   Declination                      \\
5   &   Counts\_B                 & counts            &   Counts in the broad band         \\
6   &   Counts\_S                 & counts            &   Counts in the soft band                 \\
7   &   Counts\_H                 & counts            &   Counts in the hard band                 \\
8   &   OFFAXIS\_ANGLE            & arcmin            &   Off-axis angle                   \\
9   &   SRC\_AREA                 & pixel$^2$         &   Average aperture area for merged observations \\
10  &   PSF\_FRAC                 & ...               &   Average PSF fraction at 1.5$\,$keV for merged observations \\
11  &   PROB\_KS                  & ...               &   Smallest p-value for the non-variable null hypothesis   \\
12  &   MEAN\_ARF\_B              & cm$^2$/count      &   Mean effective area in the broad band    \\
13  &   MEAN\_ARF\_S              & cm$^2$/count      &   Mean effective area in the soft band    \\
14  &   MEAN\_ARF\_H              & cm$^2$/count      &   Mean effective area in the hard band    \\
15  &   BKG\_CNTS\_B              & counts            &   Counts in the background area in the broad band  \\
16  &   BKG\_CNTS\_S              & counts            &   Counts in the background area in the soft band  \\
17  &   BKG\_CNTS\_H              & counts            &   Counts in the background area in the hard band  \\
18  &   BACKSCAL\_B               & ...               &   Background scaling factor in the broad band \\
19  &   BACKSCAL\_S               & ...               &   Background scaling factor in the soft band \\
20  &   BACKSCAL\_H               & ...               &   Background scaling factor in the hard band \\
21  &   NET\_CNTS\_B              & counts            &   Net counts in the broad band    \\
22  &   NET\_CNTS\_S              & counts            &   Net counts in the soft band    \\
23  &   NET\_CNTS\_H              & counts            &   Net counts in the hard band    \\
24  &   NET\_CNTS\_SIGMAUP\_B     & counts            &   1-$\sigma$ upper bound of the net counts in the broad band    \\
25  &   NET\_CNTS\_SIGMAUP\_S     & counts            &   1-$\sigma$ upper bound of the net counts in the soft band    \\
26  &   NET\_CNTS\_SIGMAUP\_H     & counts            &   1-$\sigma$ upper bound of the net counts in the hard band    \\
27  &   NET\_CNTS\_SIGMALO\_B     & counts            &   1-$\sigma$ lower bound of the net counts in the broad band    \\
28  &   NET\_CNTS\_SIGMALO\_S     & counts            &   1-$\sigma$ lower bound of the net counts in the soft band    \\
29  &   NET\_CNTS\_SIGMALO\_H     & counts            &   1-$\sigma$ lower bound of the net counts in the hard band    \\
30  &   PBS\_B                    & ...               &   Log(10) p-value for the null-hypothesis of no source in the broad band   \\
31  &   PBS\_S                    & ...               &   Log(10) p-value for the null-hypothesis of no source in the soft band   \\
32  &   PBS\_H                    & ...               &   Log(10) p-value for the null-hypothesis of no source in the hard band   \\
33  &   PHOTFLUX\_B               & photons/cm$^2$/s  &   Photons flux in the broad band   \\
34  &   PHOTFLUX\_S               & photons/cm$^2$/s  &   Photons flux in the soft band   \\
35  &   PHOTFLUX\_H               & photons/cm$^2$/s  &   Photons flux in the hard band   \\
36  &   ENERMED                   & keV               &   Median photon energy             \\
37  &   ENERQ25                   & keV               &   25th percentile of the photon energy distribution            \\
38  &   ENERQ75                   & keV               &   75th percentile of the photon energy distribution            \\
\hline
\hline
\end{tabular}
\end{table}

\newpage

\section{The multi-wavelength catalog and the merging process}
\label{AppA}

In this appendix, we describe the merging process adopted to compile the multi-wavelength catalog. We merged the catalogs two by two, defining in each step a ``master'' catalog which was matched to a ``slave'' catalog. Before merging, the astrometry of each catalog was anchored to the Gaia astrometry by matching them with a ``close-neighbors'' approach, adopting a radius of 5$^{\prime\prime}$, and then correcting for the median of the differences of the celestial coordinates of the matched sources. Table \ref{match_table} shows each step of the merging sequence together with other useful information: The matched catalogs and the name of the output catalog, the number of matched sources, the expected number of spurious coincidences, the initial number of multiple matches and the resolved multiple coincidences after a visual inspection (see later). We adopted two merging methods: a ``close-neighbors'' approach, in which we searched for and merged the pairs of sources in the two catalogs separated by a spatial offset smaller than a given tolerance (the $matching$ $radius$ $r_{match}$); and a Maximum-Likelihood (ML) method, adapting to our case the algorithm described in \citet{SmithDJB2011}. \par

\begin{table*}[!b]
\caption{Intermediate steps of the catalogs merging process}             
\label{match_table}      
\centering                          
\begin{tabular}{c c c c c c c c c}        
\hline\hline                 
Step & Master & Slave & Output & Method$^1$ & Matched & Spurious & Multiple & Resolved$^2$ \\    
\hline                        
1 & VPHAS+  & IPHAS & Inter\_1 & close-neigh. (0.6$^{\prime\prime}$) & 33770 & 189 & 22 & 2 \\                
2 & Inter\_1  & Pan-STARRS & Inter\_2    & close-neigh. (0.6$^{\prime\prime}$) & 45193 & 395 & 262 & 18 \\     
3 & Inter\_2  & Gaia & Inter\_3    & close-neigh. (0.5$^{\prime\prime}$) & 36843 & 234 & 208  & \\           
4 & Inter\_3  & Delgado2010 & Inter\_4    & close-neigh. (0.7$^{\prime\prime}$) & 1107 & 18 & 4 & \\         
5 & Inter\_4  & CoRoT & Inter\_5    & close-neigh. (0.7$^{\prime\prime}$) & 2051 & 28 & 16 & \\              
6 & Inter\_5  & LAMOST & Optical    & close-neigh. (1.0$^{\prime\prime}$) & 138 & 4 & 4 & \\                 
7 & UKIDSS    & 2MASS  & Inter\_6   & close-neigh. (1.0$^{\prime\prime}$) & 14753 & 93 & 13 & 5 \\           
8 & Inter\_6  & IRAC   & Inter\_7   & close-neigh. (0.9$^{\prime\prime}$) & 43630 & 305 & 108 & 14 \\        
9 & Inter\_7  & WISE   & NIR        & ML (TH=0.975) & 13714 & $\sim$9\%   & 835 & \\                         
10 & NIR       & Optical& OIR        & close-neigh. (0.6$^{\prime\prime}$) & 65125 & 772 & 674 & \\           
11 & OIR       & $Chandra$ & XOIR    & ML (TH=0.8, three iter.) & 593 & $\sim$13\% & 150 & 71\\               
\hline
\hline
\multicolumn{8}{l}{$^1$: close-neighbors with the matching radius or Maximum-Likelihood with the adopted threshold (see the text)} \\ 
\multicolumn{8}{l}{$^2$: resolved multiple coincidences (see the text)} \\ 
\end{tabular}
\end{table*}

In the close-neighbors method the matching radius is fixed by analyzing how the expected number of false positives grows as a function of the matching radius. If the stars in the catalogs were uniformly distributed over the area, the number of spurious coincidences could be calculated easily as $N_{sp}=N_{master}N_{slave}\times A_{match}/A_{total}$, where $N_{master}$ and $N_{slave}$ are the number of master and slave sources, respectively, while $A_{match}/A_{total}$ is the ratio between the matching area $\pi r_{match}^2$ over the area covered by the catalogs. This is not our case (see Fig. \ref{spadis_img}). We thus estimated the number of expected false positives by shifting rigidly the slave catalog in four directions by 5$^\prime$. In this way we ``randomize'' the slave catalog, keeping the information of the spatial distribution of the slave sources. We then merged the master and the four randomized slave catalogs adopting a set of increasing $r_{match}$. For each value of the matching radius, we took the mean number of matches obtained with the four randomized slave catalogs. As an example, in Fig. \ref{rmatch_fig} we show how in step \#10, the distribution of total ($\rm N_{total}$), spurious ($\rm N_{spurious}$), and real ($\rm N_{real}$) coincidences vary as a function of the matching radius. The $y$ axis shows the differential increment of the coincidences (e.g. the difference between the number of matches obtained with the given $\rm r_{match}$ with those obtained with the previous matching radius). In this case, for $\rm r_{match}$ between 0.6$^{\prime\prime}$ and 0.7$^{\prime\prime}$, the increment of the estimated values of $\rm N_{spurious}$ is comparable with that of the expected real matches $\rm N_{real}$=$\rm N_{total}$-$\rm N_{spurious}$, which motivated our choice of fixing $\rm r_{match}$=0.6$^{\prime\prime}$. \par

    \begin{figure}[!b]
	\centering	
	\includegraphics[width=8.5cm]{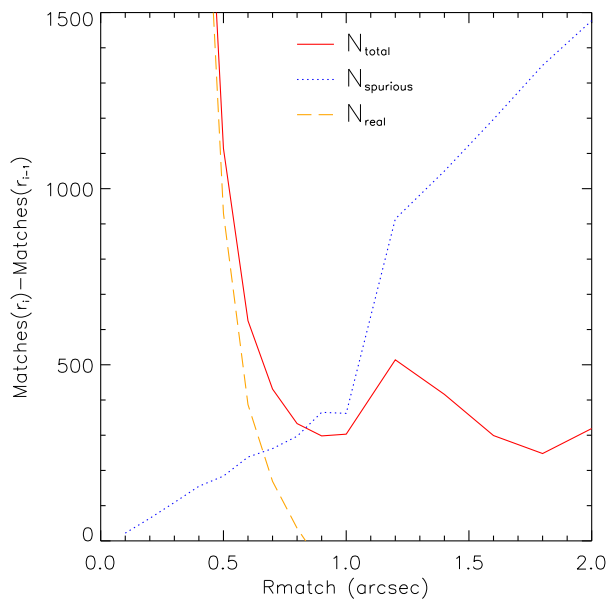}
	\caption{Distribution of total ($\rm N_{total}$), spurious ($\rm N_{spurious}$), and real ($\rm N_{real}$) coincidences in the match between the optical and the infrared catalogs. The distribution of $\rm N_{spurious}$ becomes comparable to that of $\rm N_{real}$ for $\rm r_{match}$ between 0.6$^{\prime\prime}$ and 0.7$^{\prime\prime}$}
	\label{rmatch_fig}
	\end{figure}

The close-neighbors method can fail when one of the two catalogs, typically the master, is deeper than the slave. In this case, the chances of false positives can be too high, resulting also in an excess of multiple and spurious matches. After each match, we have verified the relative depth of the matched catalog with that of the slave and master catalogs in color-magnitude diagrams, and we counted the number of coincidences with sources which are fainter than the relative magnitude limit. If this number was too high, we improved the results of the match by using a Maximum-Likelihood method which also compares the magnitude of candidate counterparts with the expected magnitude distribution of the real counterparts. We have adopted and modified the algorithm introduced by \citet{SmithDJB2011} as explained in details by \citet{GuarcelloDWN2015}. Briefly, the method is based on the calculation, for each pair of sources, of the Likelihood Ratio:

\begin{equation}
LR=\frac{q \left( m \right) f \left( r \right)}{n \left( m \right)}
\label{like_eq}
\end{equation}

where $f \left( r \right)$ is the radial distribution function of the separation between the ``master'' and ``slave'' sources, calculated as:

\begin{equation}
f \left( r \right) = \frac{1}{2 \pi \sigma_{pos}^2} exp\left( \frac{-r^2}{2\sigma_{pos}^2} \right)
\label{fr_eq}
\end{equation}

with $r$ being the positional offset between two sources, and $\sigma_{pos}$ the positional uncertainties. The quantities $q(m)$ and $n(m)$ are the magnitude probability distributions of the correlated sources (i.e.: the real counterparts) and the observed magnitude probability distribution of all the ``master'' sources in some representative bands, respectively. The distribution $n(m)$ is calculated directly from the whole master catalog, while $q(m)$ from a reliable set of expected real counterparts in the master catalog (computed as described below). After LR is calculated, it must be used to estimate the $reliability$ that the given pair of sources are true counterparts of an astronomical source. To this aim, we first calculated the distribution of $LR$ values from 200000 test ``slave'' sources uniformly distributed across the field and matched with the master sources of the pair. The {\it reliability} associated with each pair is, by definition, the probability that the given ``slave'' source is the real counterpart of the ``master'' source, and it can be calculated as: 

\begin{equation}
R_{ij}=1-\frac{N_{gt}}{N_{sim}}
\label{rij_eq}
\end{equation}

where $R_{ij}$ is the reliability that the given $ij$ pair is a real coincidence; $N_{sim}$ is the number of simulated $LR$ values; and $N_{gt}$ is the number of simulated $LR$ values larger than the one observed between the $ij$ pair $N_{gt}=N \left( LR_{simul} > LR_{ij} \right)$. This $reliability$ is then compared with a given threshold. To estimate the threshold, we merged the master catalog with a slave catalog whose coordinates were rigidly shifted in four directions, taking the mean of the resulting matches with varying the threshold. We typically fixed the threshold as the value resulting in a number of spurious coincidences which is about 10\% of the real matches. \par

    As shown in Table \ref{match_table}, we used the ML method in two matches: Steps 9 and 11. In the former (match between the ``UKIDSS+2MASS+IRAC'' catalog as master with the WISE catalog as slave), we adopted as representative bands the $JHK$ bands from UKIDSS or 2MASS (choosing for each source the one with the smallest error when they are both available) and the [3.6] band from IRAC. The expected correlated population necessary to calculate $q(m)$ is obtained from a close-neighbors match with $\rm r_{match}$=0.7$^{\prime\prime}$. In step 11 (the match between the optical-infrared catalog as master and the $Chandra$ catalog as slave), we adopted as representative bands the $r$ band from VPHAS+, IPHAS, and Pan-STARRS, the $J$ magnitude from 2MASS or UKIDSS (the one with the smallest error, when they are both available), and the [3.6] band from Spitzer/IRAC. We performed the match in three iterations, updating in each run the expected correlated population and thus the $q(m)$ distributions. In the first iteration it has been defined from a close-neighbors match with $\rm r_{match}$=1$^{\prime\prime}$, while in the second and third runs it was obtained from the ML match performed in the previous iteration. As shown in Table \ref{oirx_table}, this strategy did not improve the total number of matches, while it reduced in the second iteration the number of multiple coincidences (the $single$, $double$, and $multiple$ columns). 
    
\begin{table}
\caption{Iterations in the OIR-X match}             
\label{oirx_table}      
\centering                          
\begin{tabular}{c c c c c c}        
\hline\hline                 
Iter. & $\rm N_{total}$ & single & double & multiple & unique X-ray  \\    
\hline                        
1 & 589 & 235 & 89  & 146 & 459 \\
2 & 591 & 360 & 136 & 95  & 463 \\
3 & 593 & 360 & 136 & 97  & 463 \\
\hline
\hline
\end{tabular}
\end{table}

Each step shown in Table \ref{match_table} produced a number of multiple matches. These multiple coincidences are retained in the merged catalog and then visually inspected when possible. In cases in which the merged catalogs provide photometry in similar bands (such as VPHAS+ vs. IPHAS or 2MASS vs. UKIDSS), or in cases in which hypotheses can be made a priory on the nature of the merged sources (for instance, the X+OIR sources are expected to be mainly pre-main sequence stars), it is possible to inspect the multiple coincidences in order to ``resolve'' some of them, i.e., by separating some of the multiple coincidences that are likely false positives. This iteration was particularly important for the X+OIR match (step 11), as shown in Table \ref{match_table}. For this match, we also searched for false negatives, i.e., nearby OIR sources that likely are real counterparts of unmatched X-ray sources. We performed this search within 5$^{\prime\prime}$ from each X-ray source without OIR counterpart, and merged 9 OIR sources that from their colors and magnitudes and proximity to an unmatched X-ray source are likely pre-Main Sequence stars.

\newpage
\section{Extinction coefficients used in this work}
\label{AppB}

In this appendix we summarize the extinction coefficients $\rm C_\lambda$, where $\rm A_\lambda=C_\lambda A_V$, used in this work.

\begin{table*}[!h]
\caption{Extinction coefficients $\rm C_\lambda$}             
\label{extcoef_table}      
\centering                          
\begin{tabular}{c c c c c c c c c}        
\hline\hline                 
UBVRI       & IPHAS             & VPHAS             & Pan-STARRS & Gaia            & UKIDSS      & 2MASS       & IRAC        & WISE \\    
\hline                        
$A_U$=1.55  & $A_r$=0.86        & $A_u$=1.54          & $A_g$=1.17  & $A_G$=0.86    & $A_J$=0.29  & $A_J$=0.29  & $A_1$=0.07  & $A_{w1}$=0.07    \\
$A_B$=1.30  & $A_i$=0.65        & $A_g$=1.19          & $A_r$=0.87  & $A_{BP}$=1.07 & $A_H$=0.18  & $A_H$=0.18  & $A_2$=0.06  & $A_{w2}$=0.06    \\
$A_V$=1.01  & $A_{H\alpha}$=0.81& $A_r$=0.85          & $A_i$=0.68  & $A_{RP}$=0.65 & $A_K$=0.12  & $A_K$=0.12  & $A_3$=0.055 & $A_{w3}$=0.002   \\
$A_R$=0.82  &                   & $A_i$=0.68          & $A_z$=0.52  &               &             &             & $A_4$=0.056 & $A_{w4}$=0.00001 \\
$A_I$=0.61  &                   & $A_z$=0.50          & $A_y$=0.43  &               &             &             &             &                  \\
            &                   & $A_{H\alpha}$=0.81  &             &               &             &             &             &                  \\
\hline
\hline
\multicolumn{9}{l}{} \\ 
\end{tabular}
\end{table*}

\newpage
\section{Color-Color and Color-Magnitude diagrams used to select and classify members}
\label{AppE}

In this appendix we show the diagrams that, together with those shown in Fig. \ref{diagrams3_img}, we used to select stars with disks and discard contaminants. 

   \begin{figure*}[!b]
	\centering	
	\includegraphics[width=9cm]{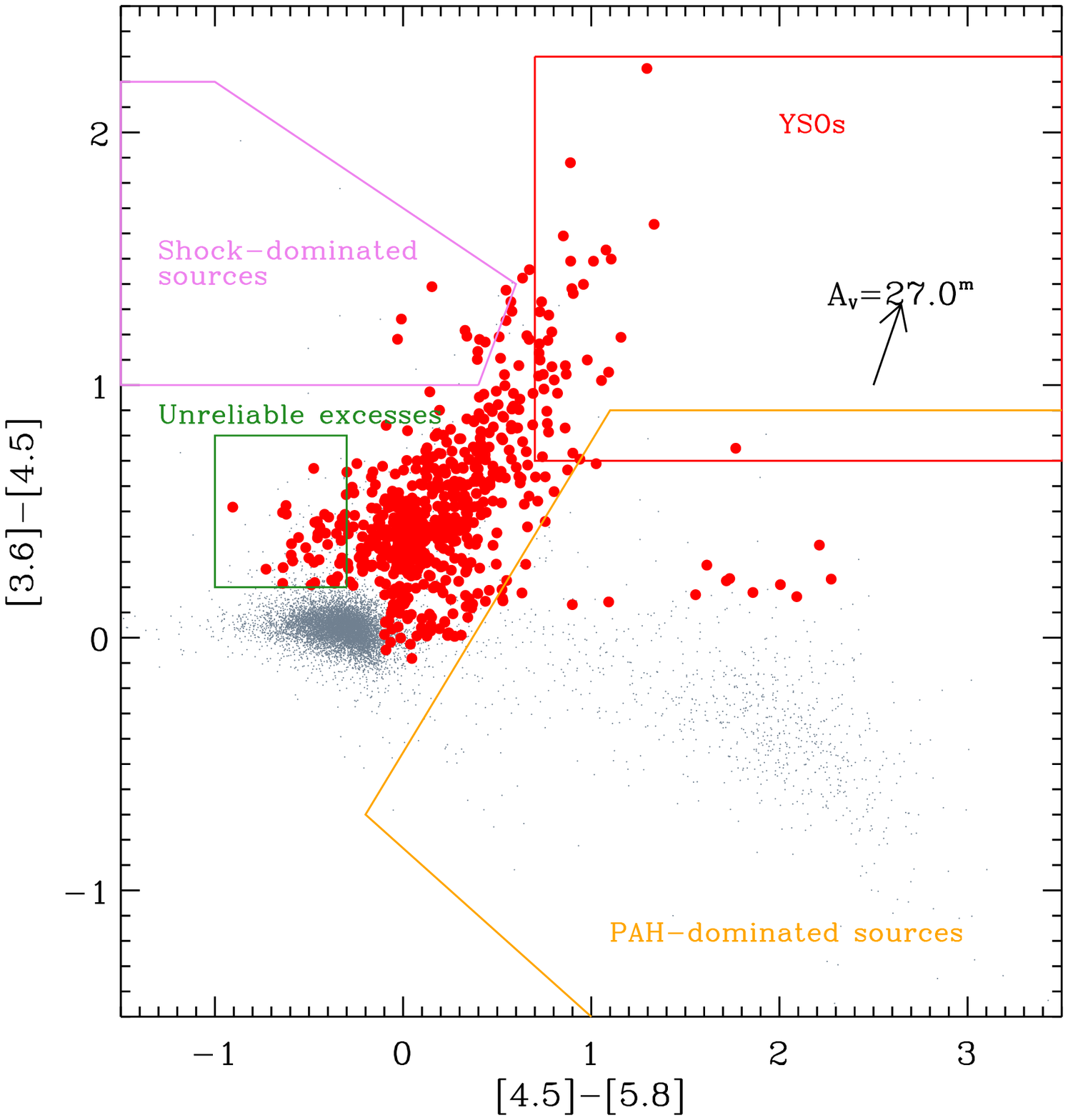}
	\includegraphics[width=9cm]{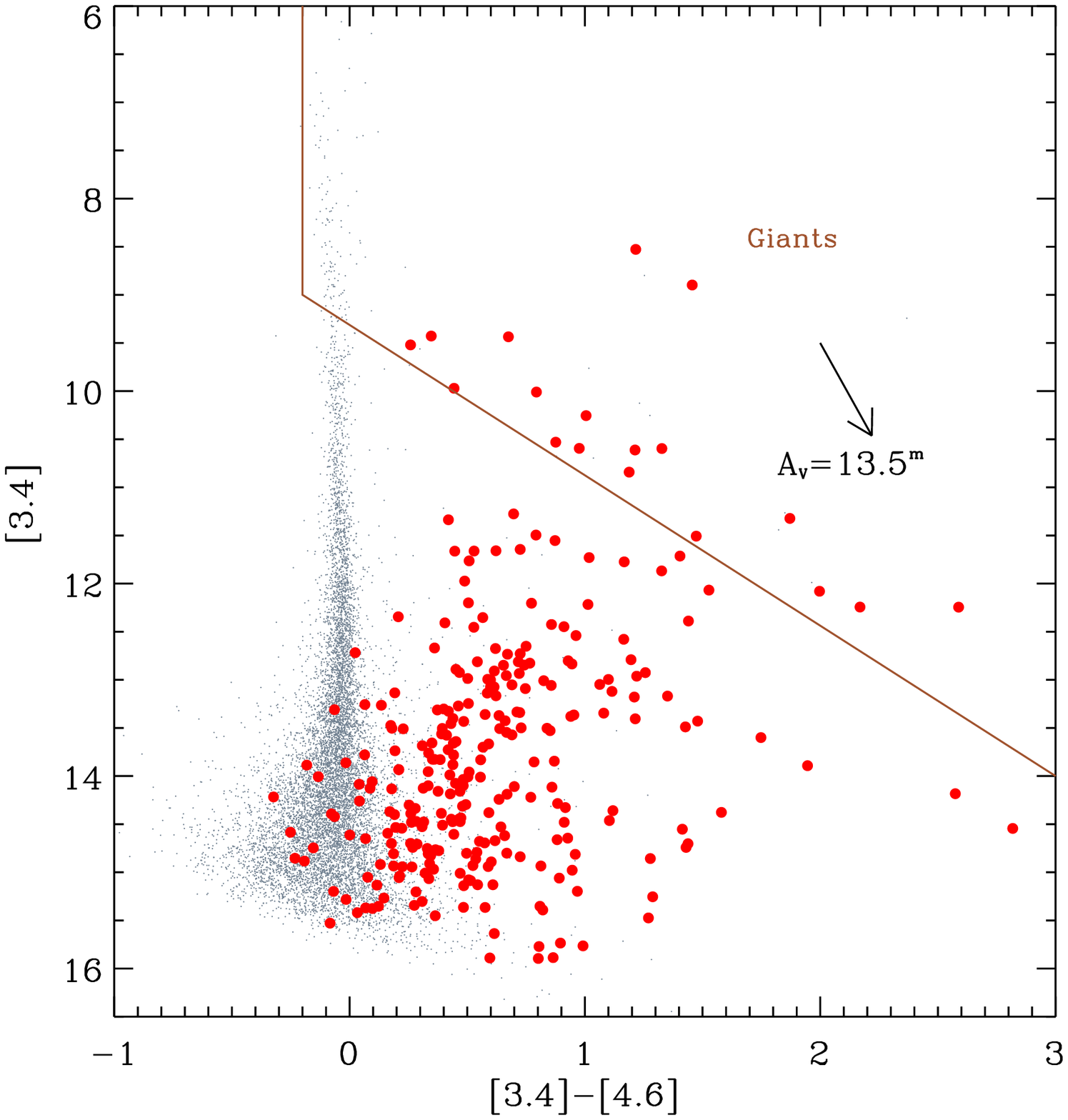}
	\includegraphics[width=9cm]{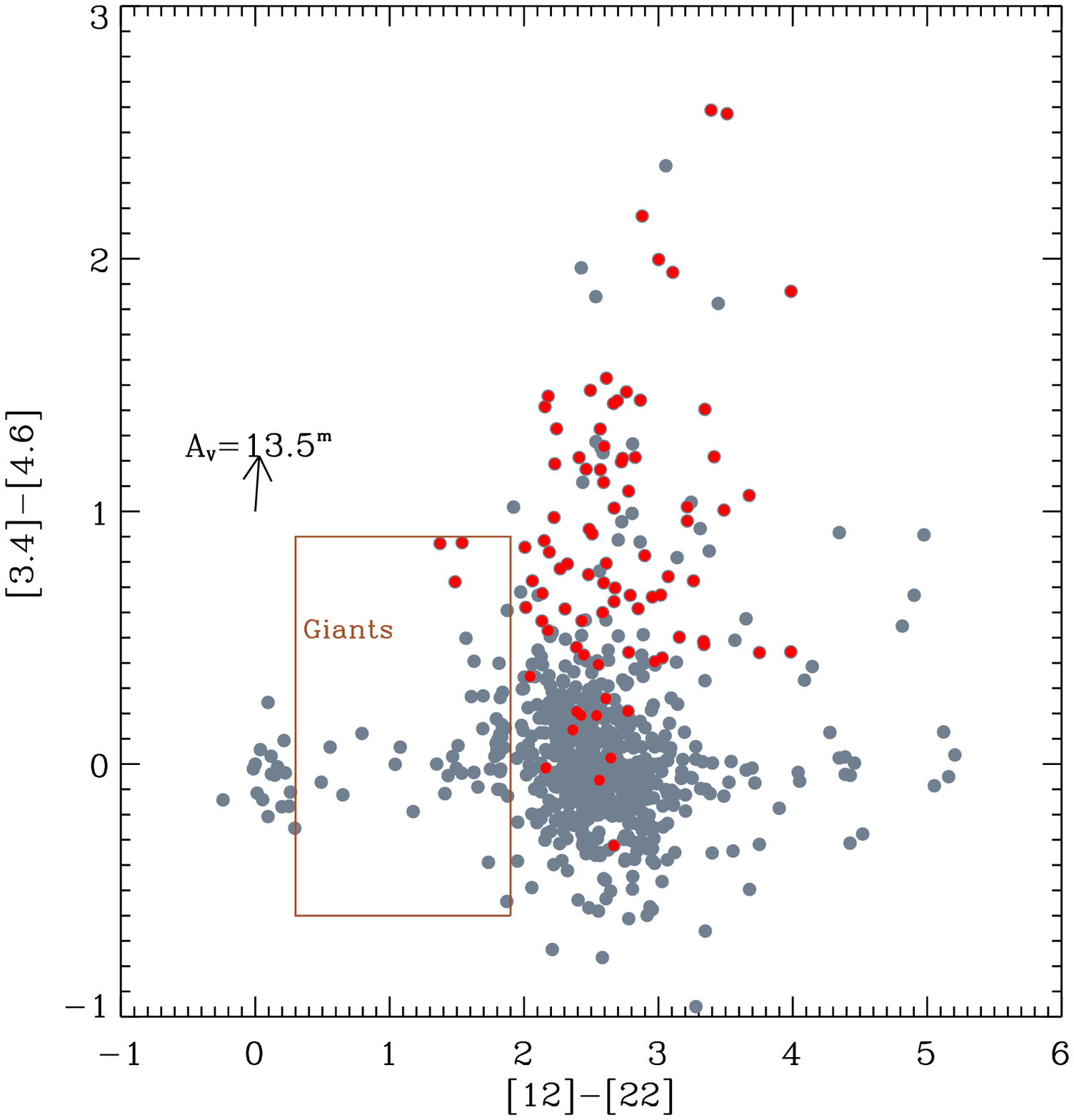}
	\includegraphics[width=9cm]{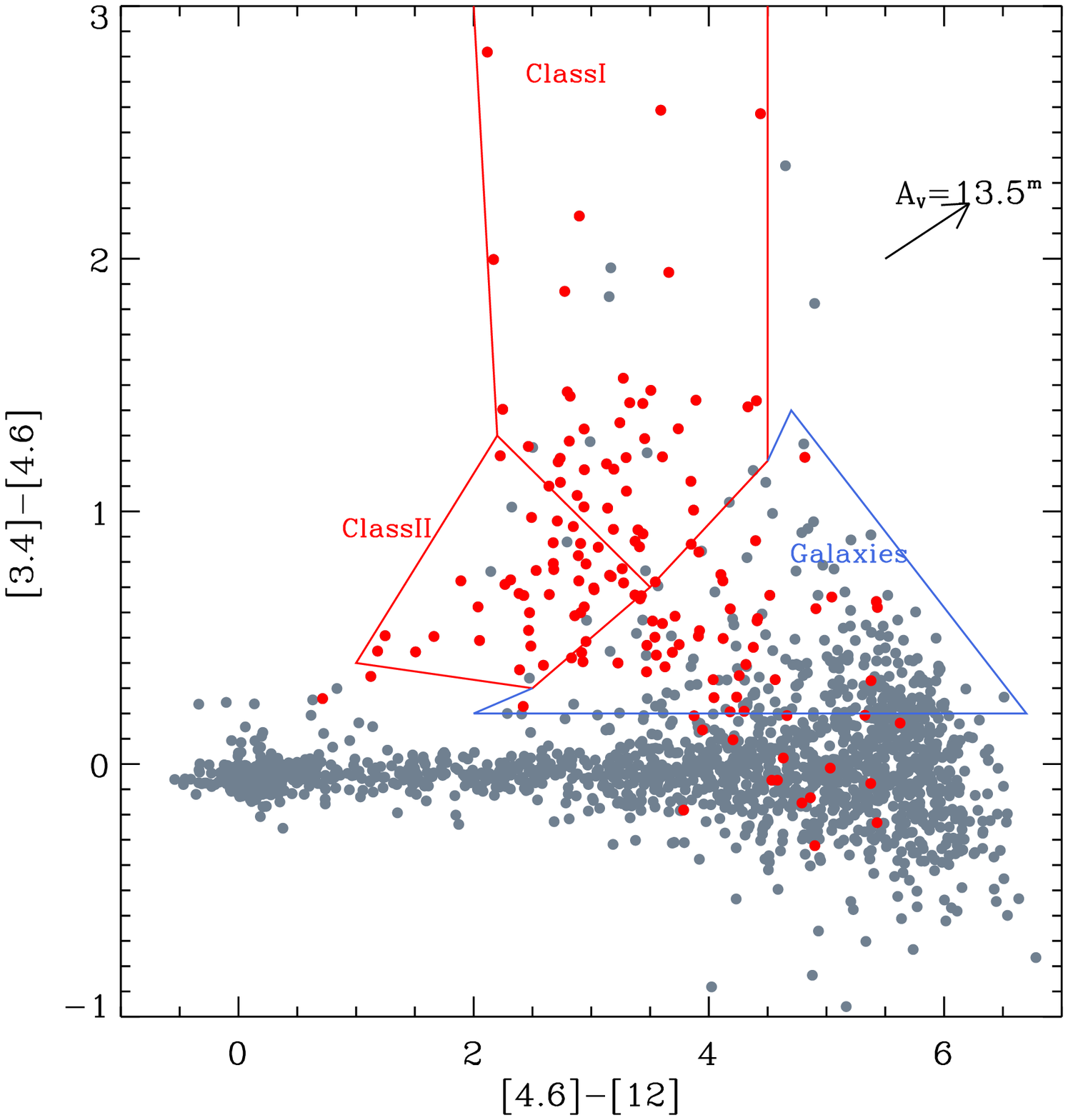}
	\caption{Infrared diagrams of all sources falling in the studied field meeting the criteria of good photometry. Figure layout and symbols as in Fig. \ref{diagrams3_img}. In these diagrams we show the loci expected to be populated by stars with disks (class~II and class~I YSOs separated), extragalactic sources, giants with circumstellar dust, unreliable stars with excesses, PAH contaminated sources, and unresolved shock knots.}
	\label{diagrams5_img}
	\end{figure*}
	
\newpage
    \begin{figure*}[]
	\centering	
	\includegraphics[width=9cm]{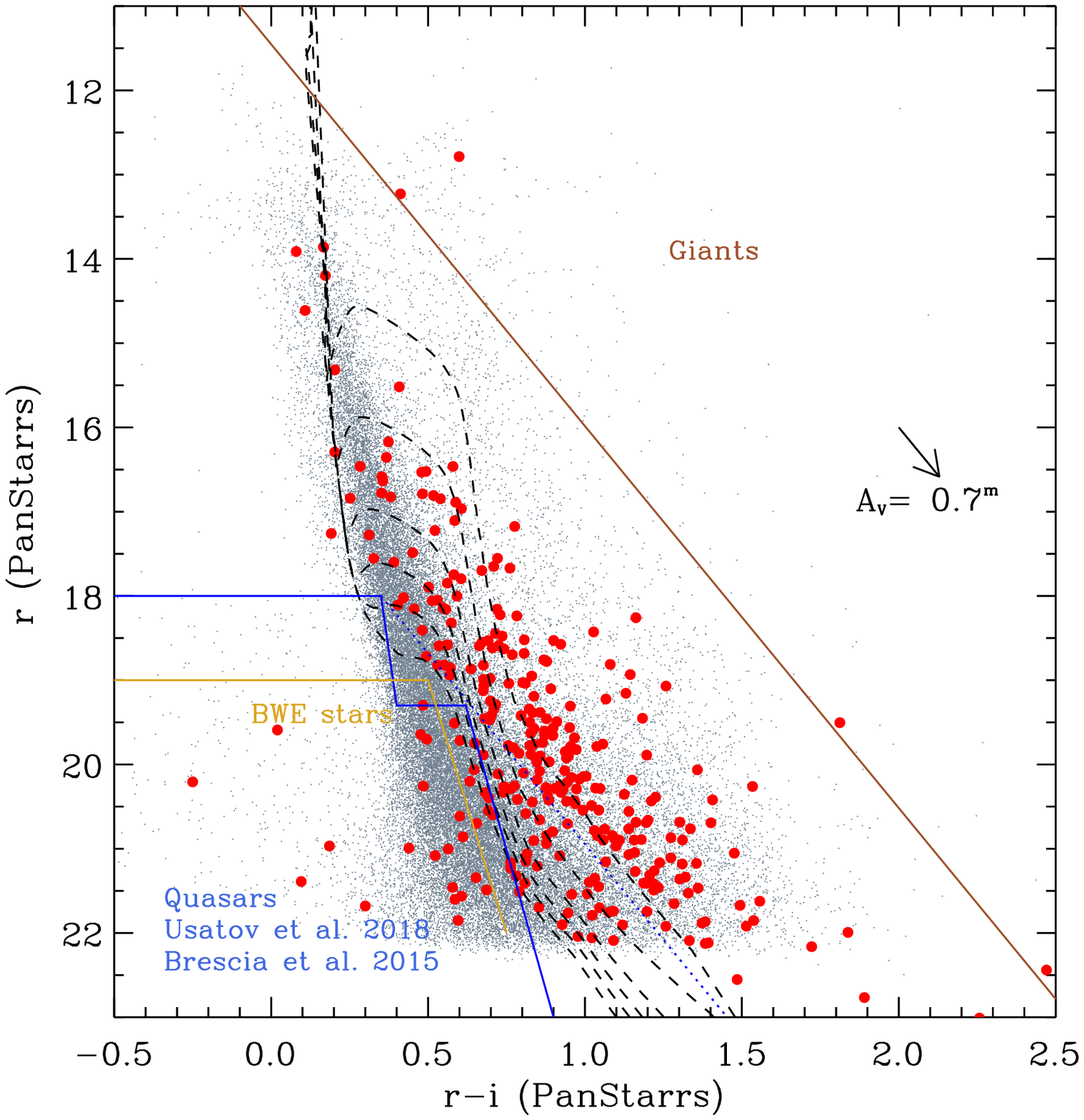}
	\includegraphics[width=9cm]{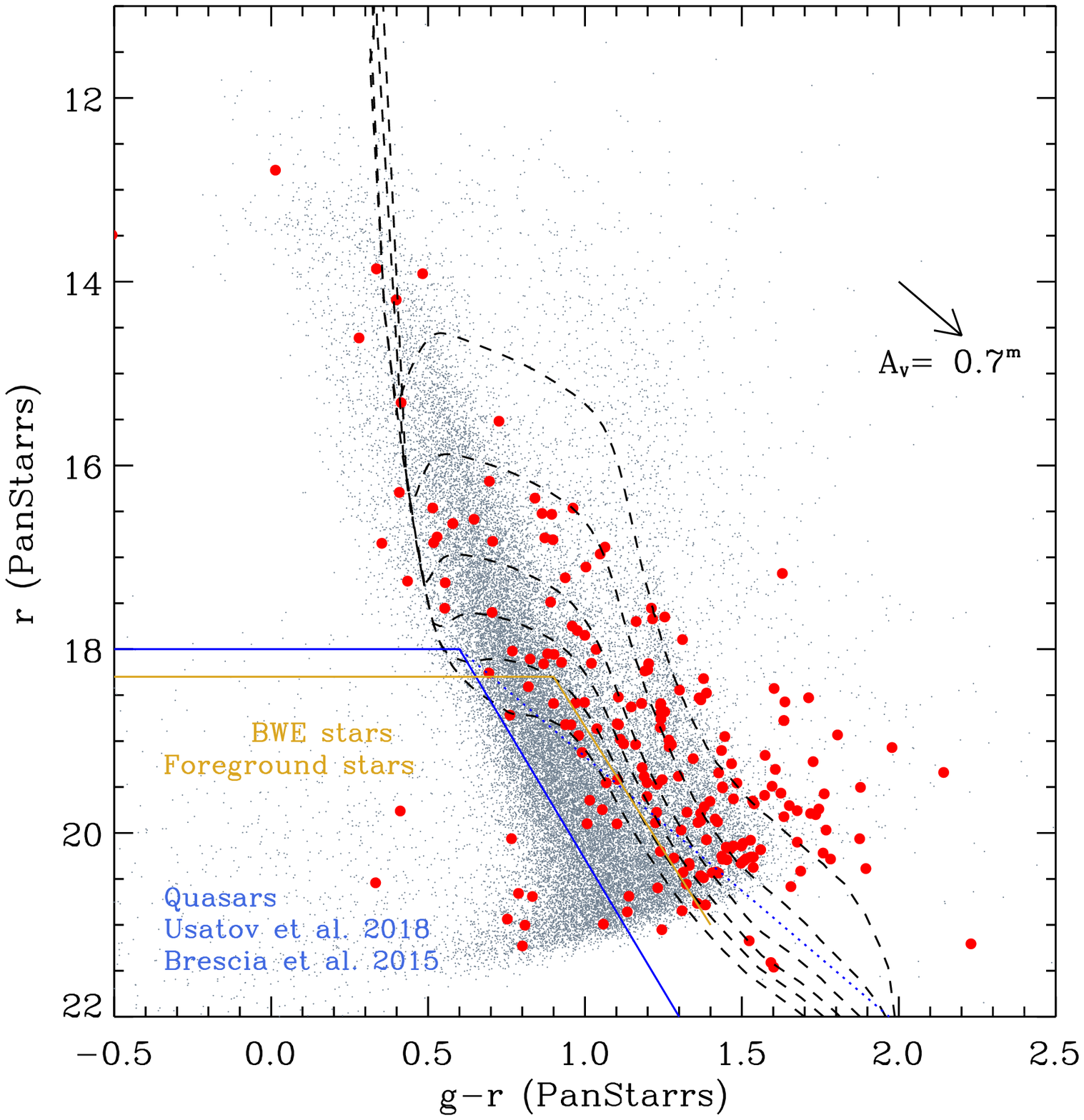}
	\includegraphics[width=9cm]{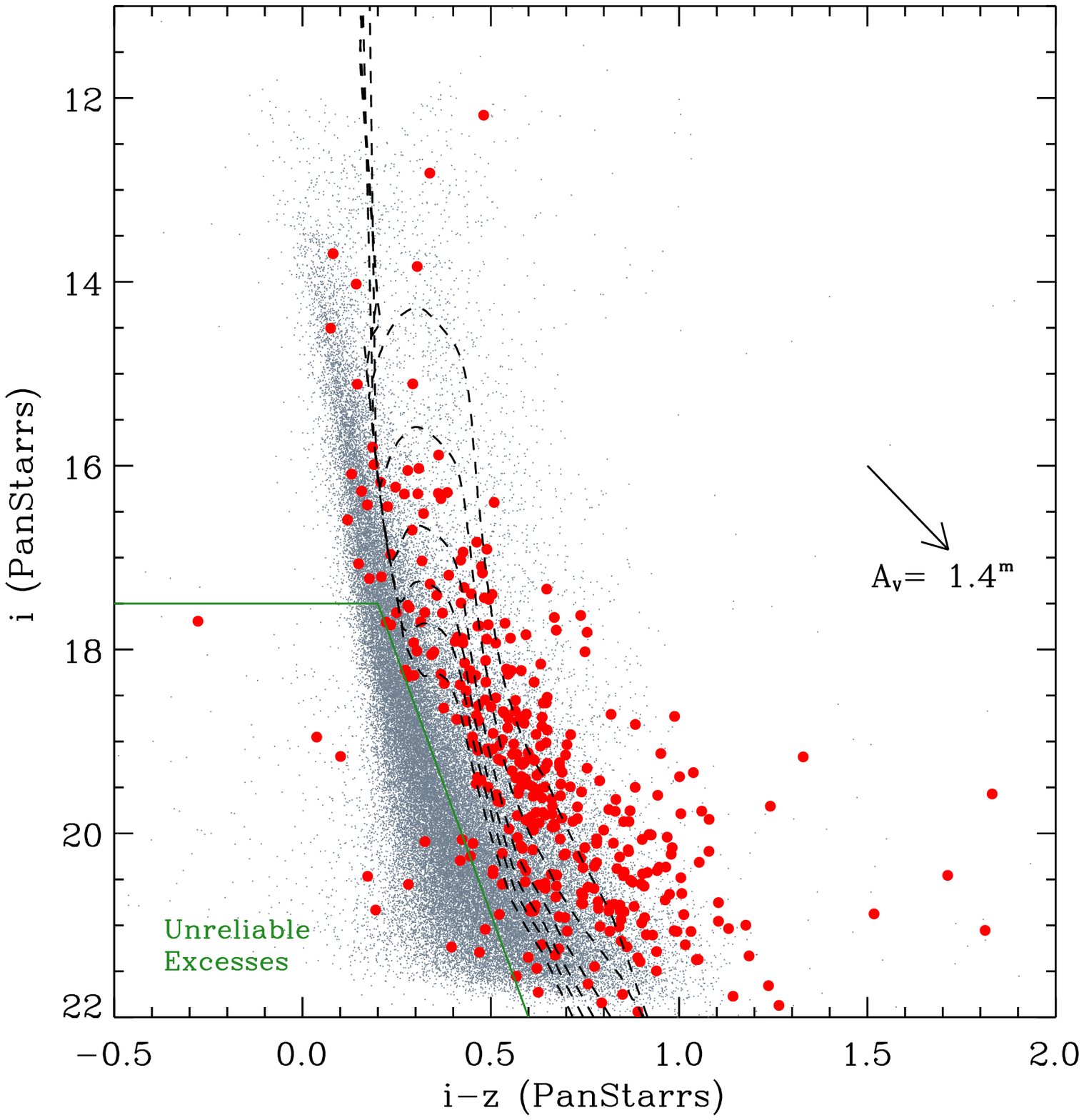}
	\includegraphics[width=9cm]{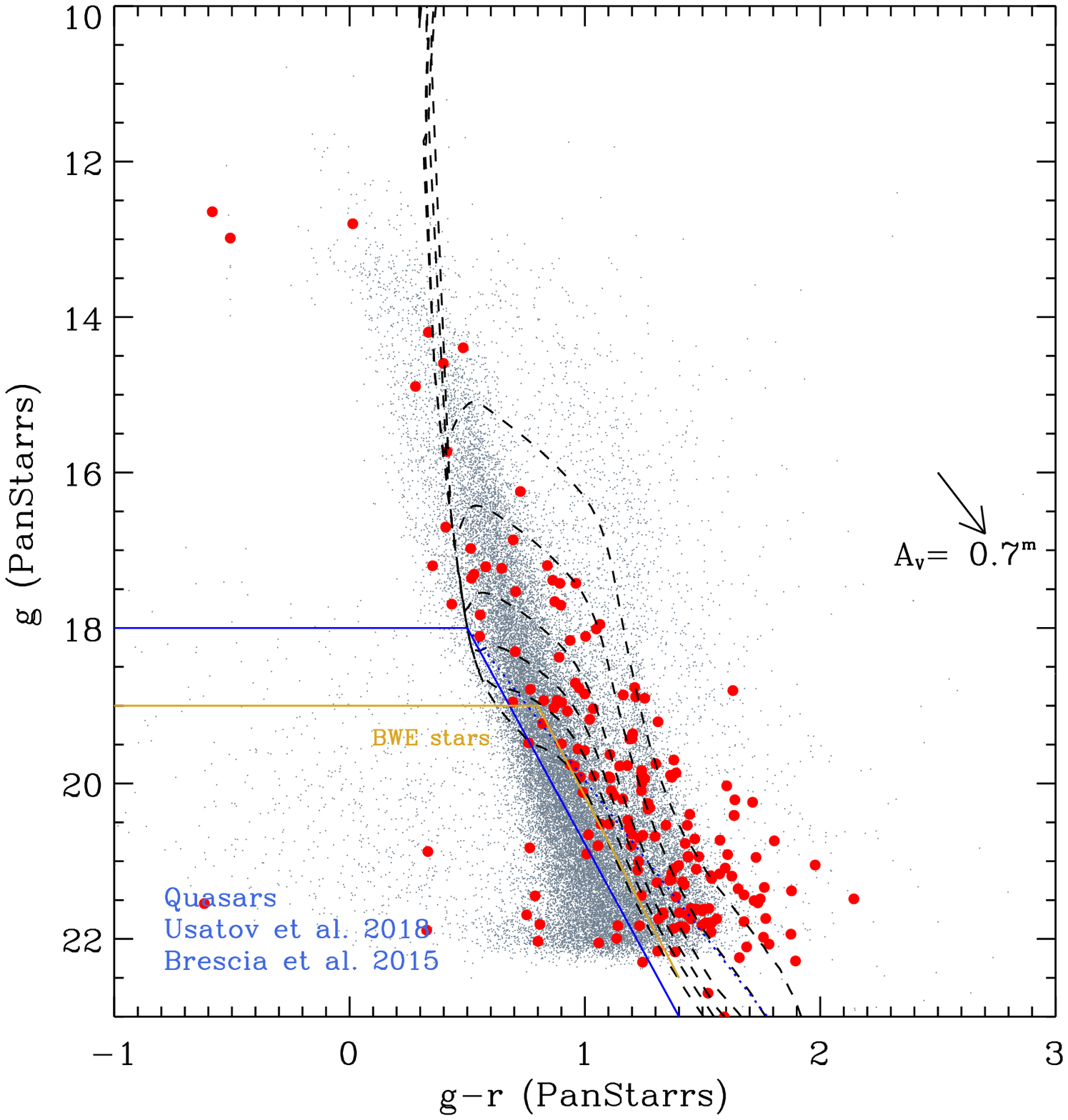}
	\caption{Pan-STARRS color-magnitude diagrams of all sources falling in the studied field meeting the criteria of good photometry.  Figure layout and symbols as in Fig. \ref{diagrams3_img}. In these diagrams we show the loci expected to be populated by blue stars with excesses, giants, and galaxies (see the text).}
	\label{diagrams1_img}
	\end{figure*}

$  $
\vspace{10cm}

\newpage
    \begin{figure*}[]
	\centering	
	\includegraphics[width=9cm]{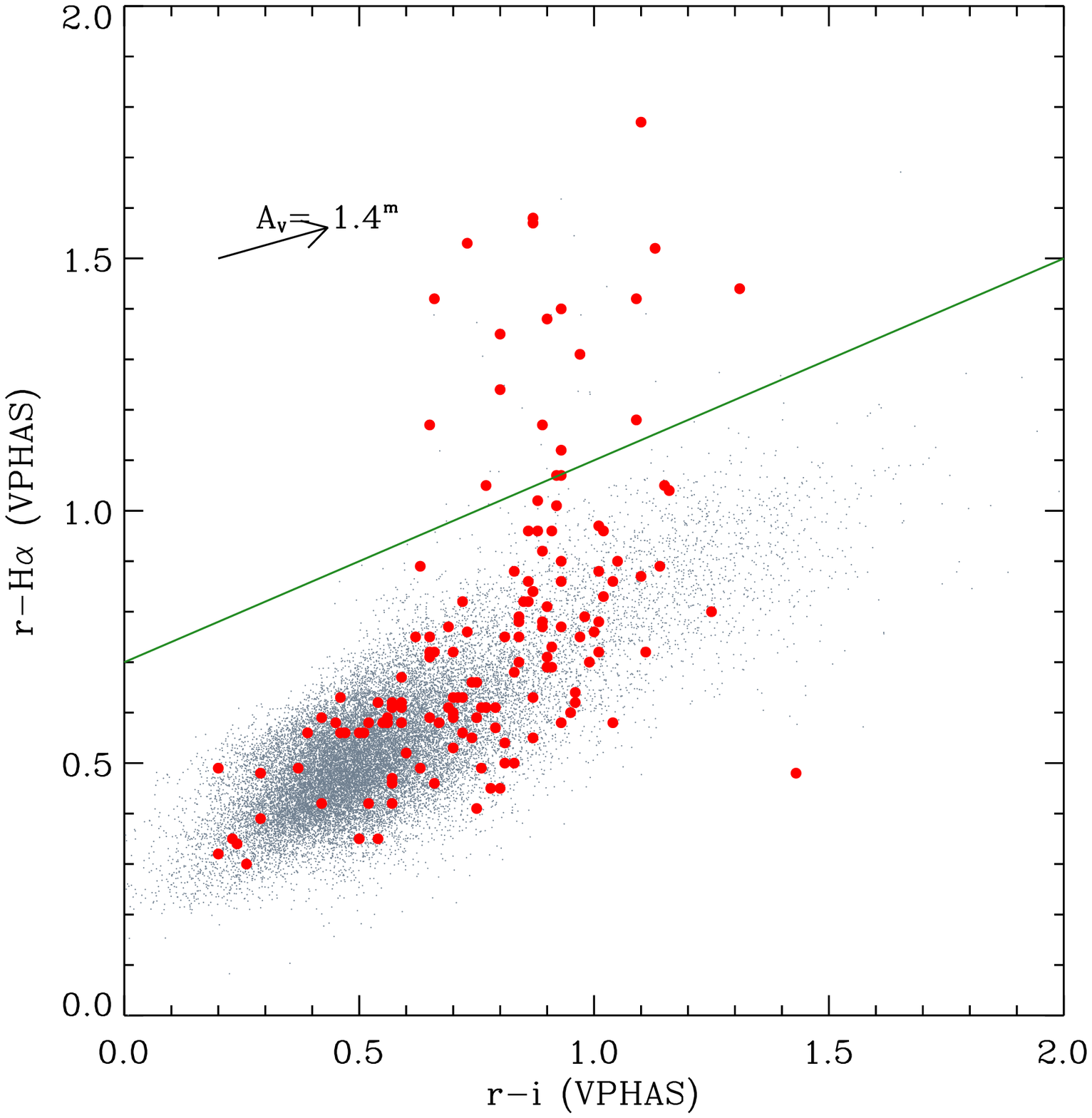}
	\includegraphics[width=9cm]{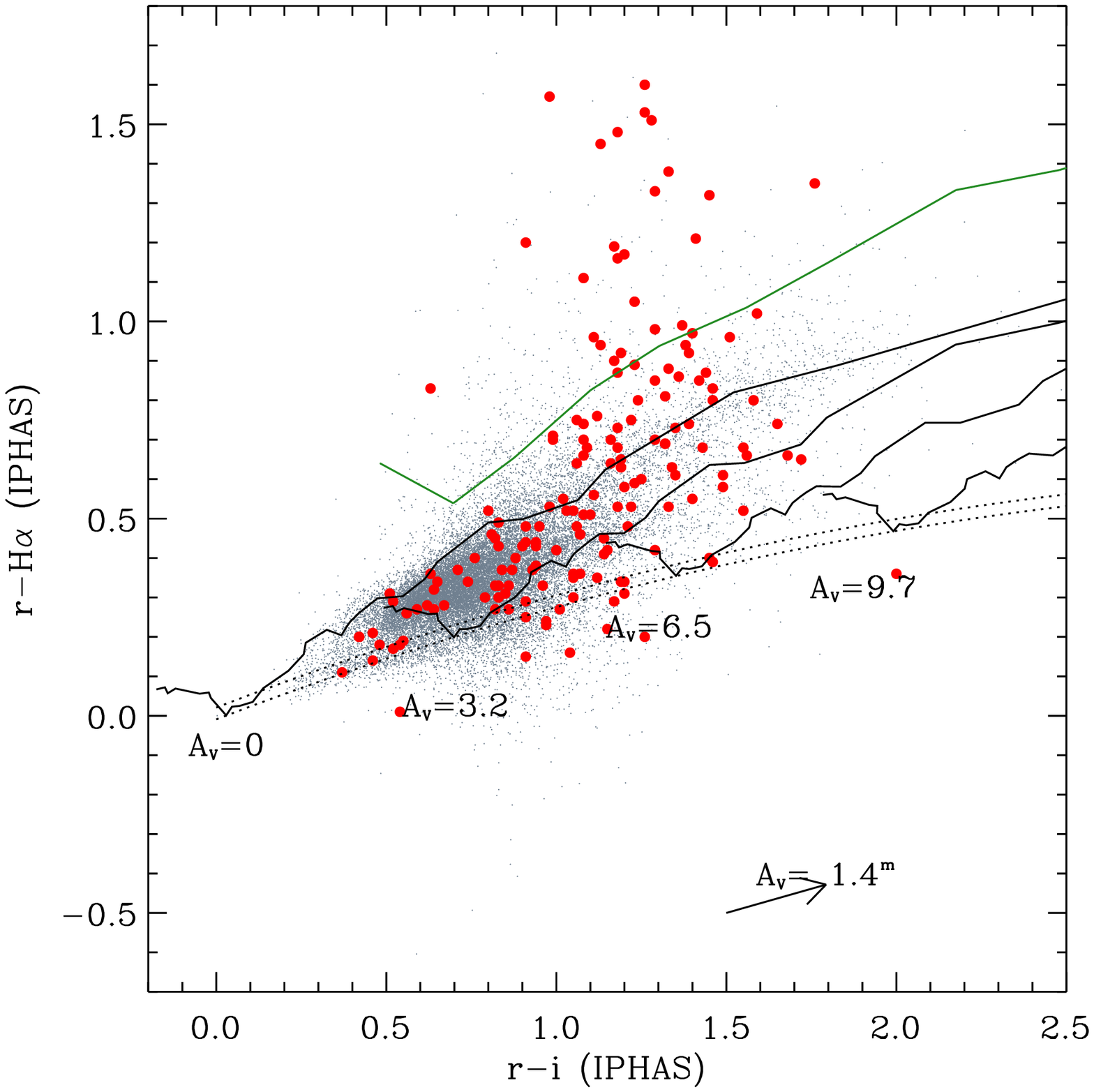}
	\includegraphics[width=9cm]{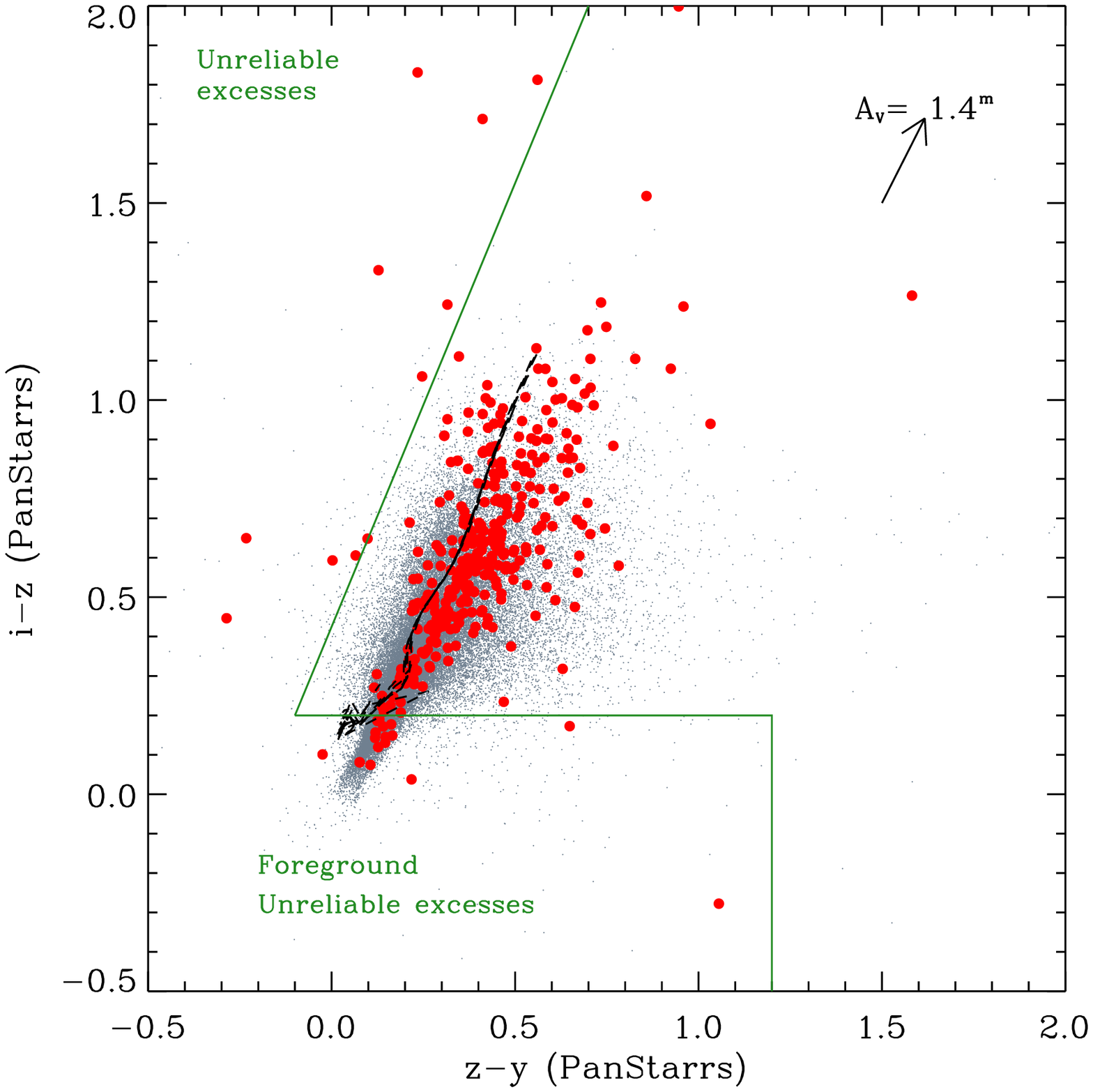}
	\includegraphics[width=9cm]{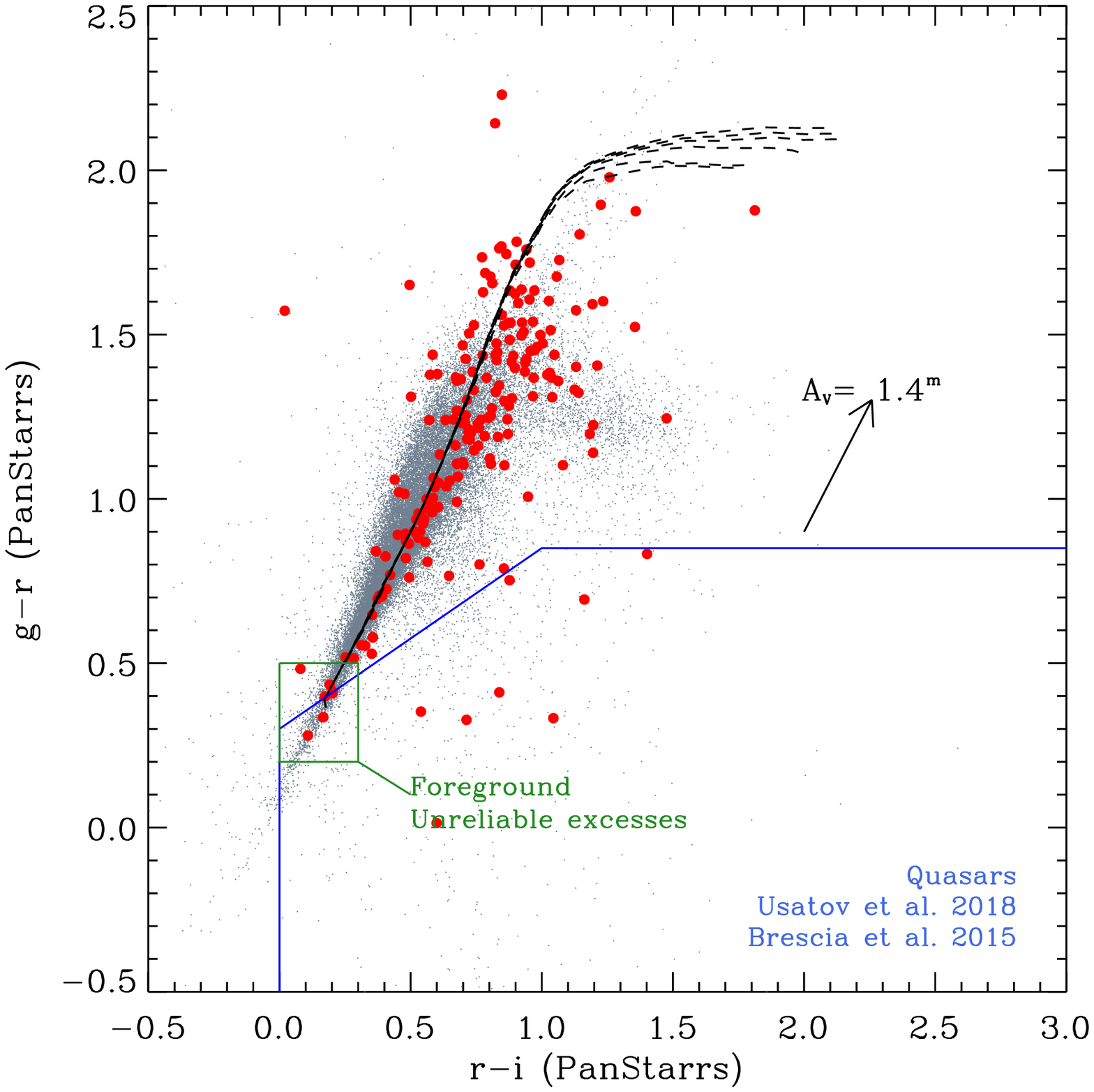}
	\caption{Optical color-color diagrams of all sources falling in the studied field meeting the criteria of good photometry. Figure layout and symbols as in Fig. \ref{diagrams3_img}. In these diagrams we show the loci expected to be populated by accreting stars with disks, stars with unreliable infrared excesses, and possible foreground stars. In the IPHAS $r-H\alpha$ vs. $r-i$ diagram, the solid black lines are ZAMS at increasing extinction, while the dashed lines mark the locus of A stars, as defined by \citet{DrewGIA2005}, while the solid green line is a ZAMS with EW$\rm_{H\alpha}$=-40$\AA$ and E$\rm_{B-V}$=1 defined by \citet{BarentsenVDG2011}.}
	\label{diagrams2_img}
	\end{figure*}

$  $
\vspace{10cm}
\newpage
    \begin{figure*}[]
	\centering	
	\includegraphics[width=9cm]{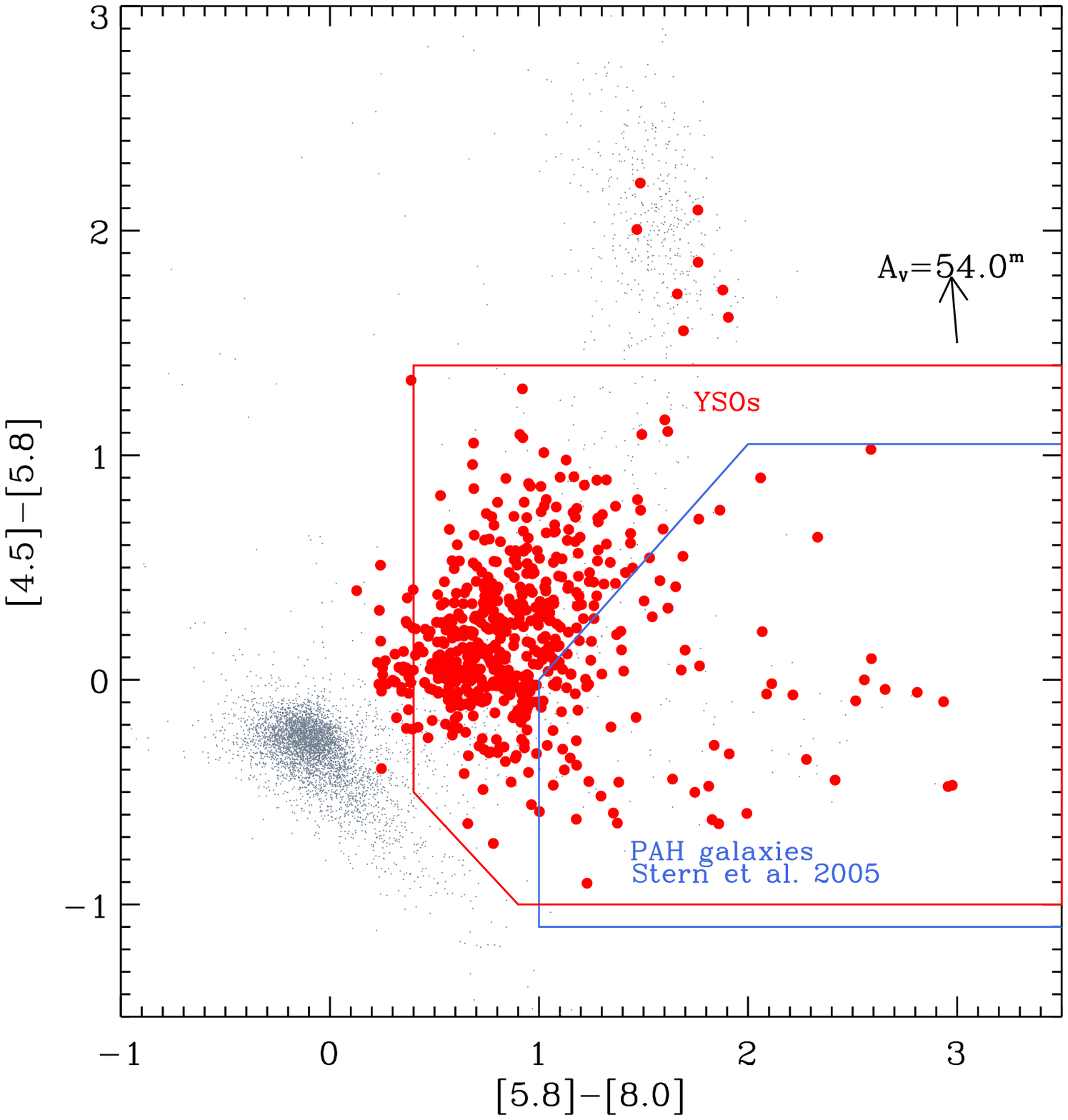}
	\includegraphics[width=9cm]{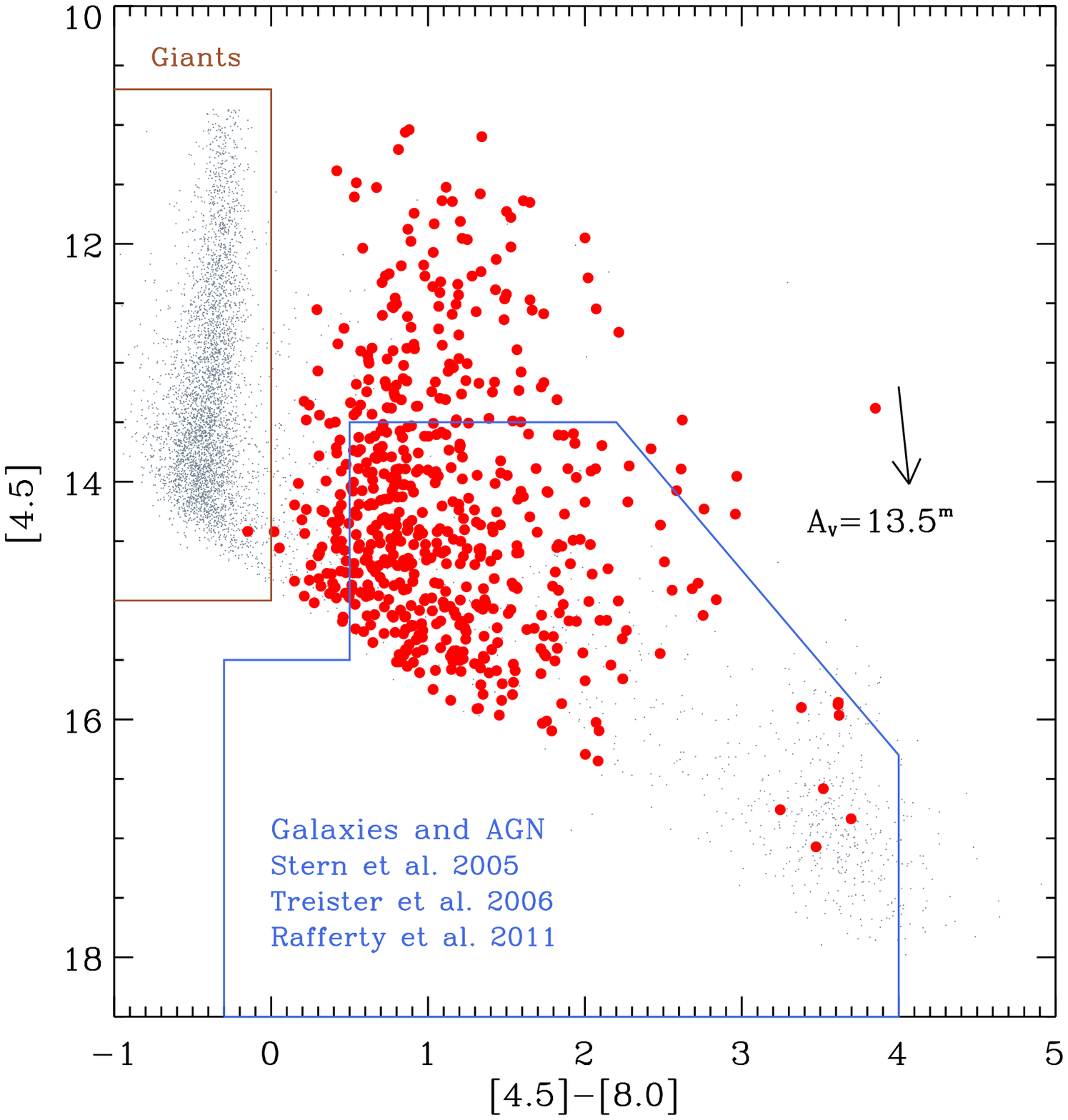}
	\includegraphics[width=9cm]{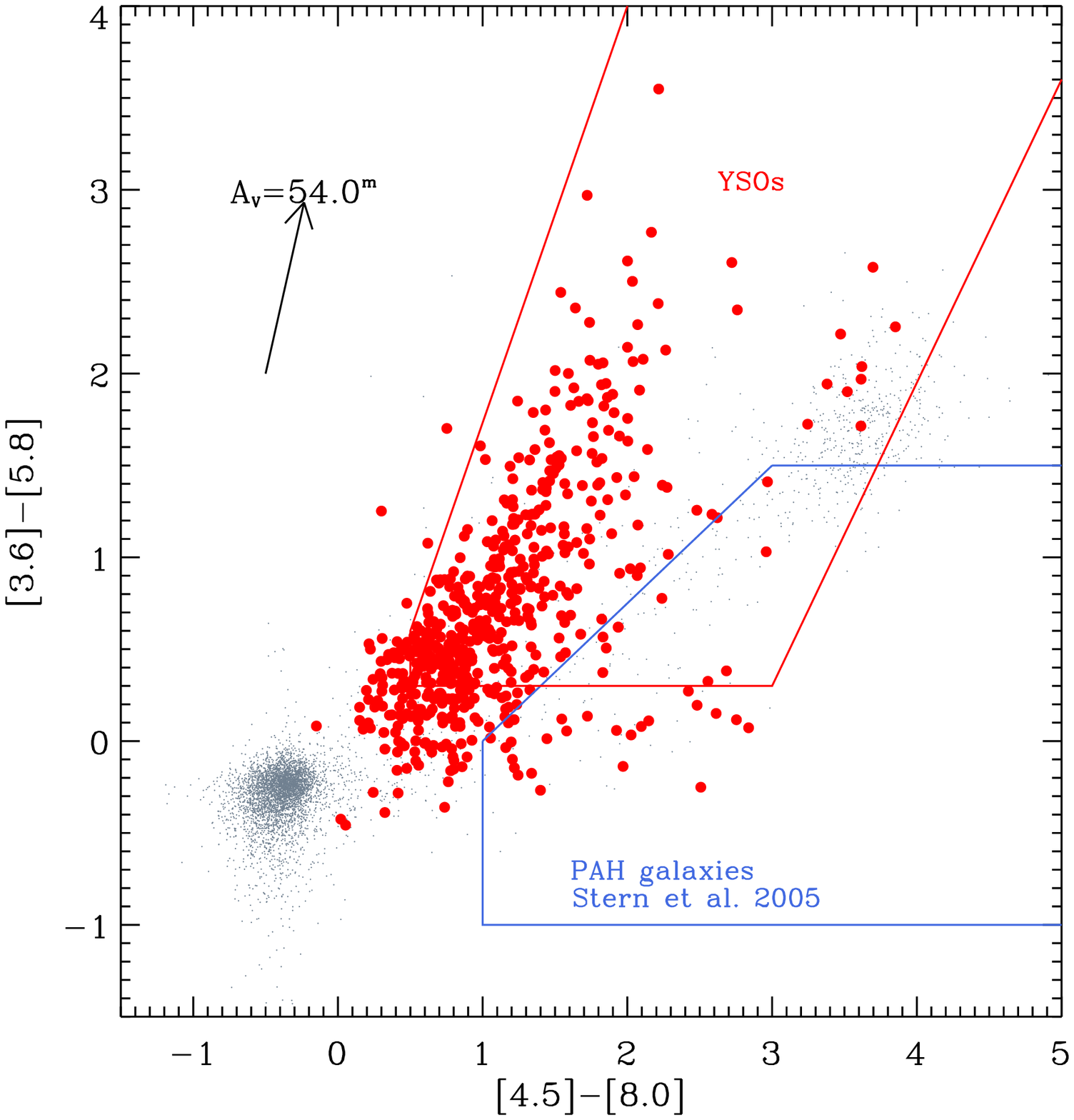}
	\includegraphics[width=9cm]{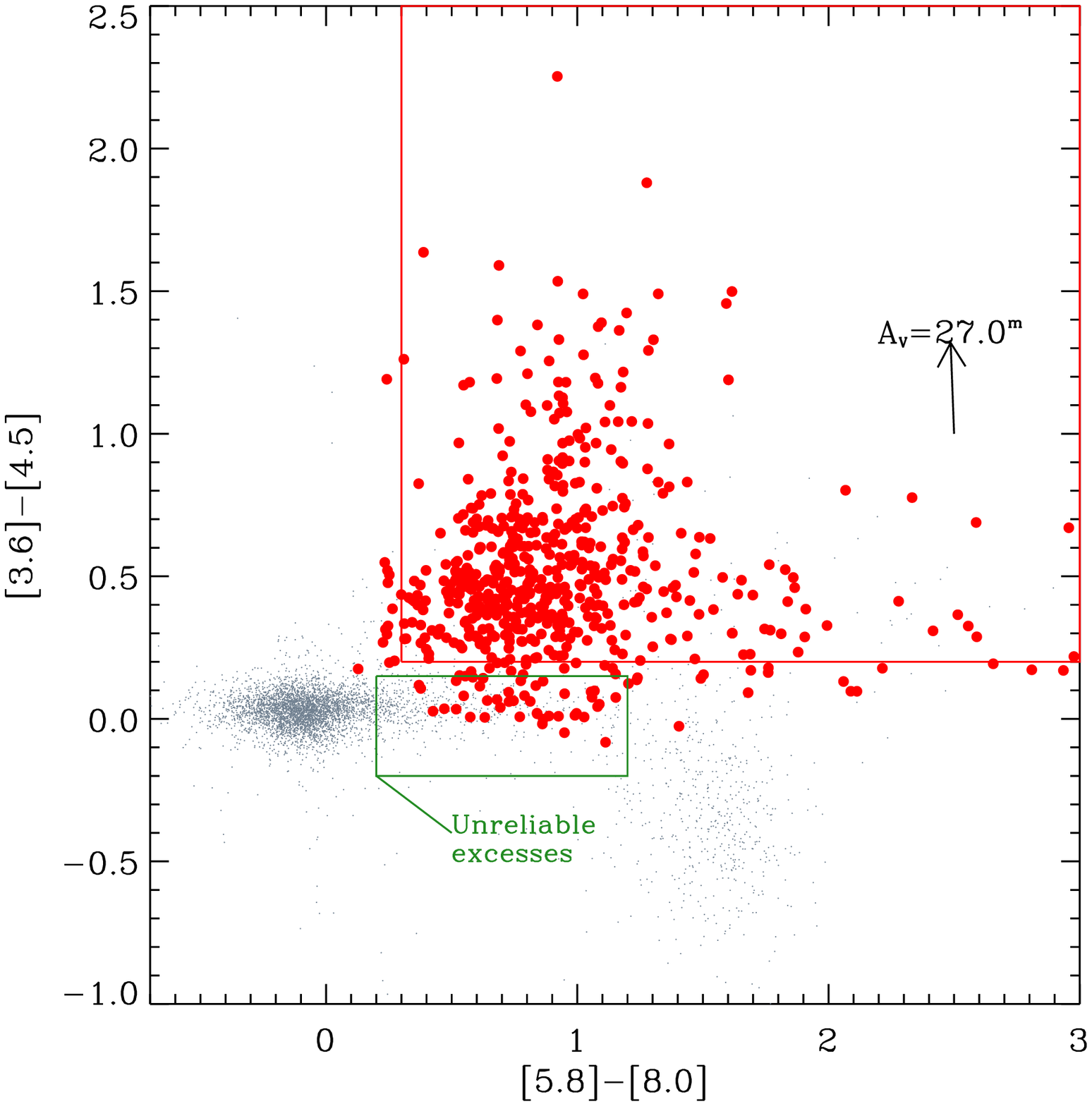}
	\caption{Infrared diagrams of all sources falling in the studied field meeting the criteria of good photometry. Figure layout and symbols as in Fig. \ref{diagrams3_img}. In these diagrams we show the loci expected to be populated by stars with disks, extragalactic sources, giants with circumstellar dust, and unreliable stars with excesses.}
	\label{diagrams4_img}
	\end{figure*}
		
$  $
\vspace{10cm}
\newpage
\section{Details on the estimate of individual masses and ages}
\label{AppC}

    Individual parameters of candidate members were estimated by interpolating their positions in selected derreddened color-magnitude diagrams on a grid computed using the 0.5-10$\,$Myrs low-metallicity PARSEC isochrones. This method allows to estimate easily the parameters of a large sample of cluster members, requiring only the use of good photometry. However, it relies on several assumptions, such as a good knowledge of individual stellar extinctions and the use of stellar models that properly describe the pre-main sequence phase. Because of the intrinsic uncertainties associated with this technique, results slightly change by adopting different color-magnitude diagrams, and no strong argument exists that allows us to prefer a priori one diagram over the others. We thus calculated stellar parameters from seven diagrams listed in the first columns of Table \ref{stelpar_table}. In all cases, we discarded stars with errors in the involved colors larger than 0.15$^m$ and 0.1$^m$ in magnitudes. In order to take into account for the photometric errors, for each star the interpolation in a given diagram was repeated 300 times, each time drawing the input values of magnitude and color from a normal distribution centered on the nominal values and with a $\sigma$ equal to the photometric errors. The values of mass and age associated with each star from the given diagram and their errors are thus set as the median value and the standard deviation, respectively, of the resulting distribution of results. \par
    
\begin{table}[!h]
\caption{Diagrams used to estimate stellar parameters}             
\label{stelpar_table}      
\centering                          
\begin{tabular}{c c c c c}        
\hline\hline                 
Diagram     & N(nodisk) & N(disks) & Log(median age)  & Completeness range   \\    
\hline                        
            &             &          & [Myrs]           & M$_{\odot}$          \\
\hline
$i$ vs. $i-z^*$      &  194 & 94   &  6.34 &  0.8-2.0 \\
$r$ vs. $g-r^*$      &  142 & 67   &  6.30 &  1.0-1.4 \\
$r$ vs. $r-i^*$      &  156 & 69   &  6.07 &  1.0-2.2 \\
$r$ vs. $r-z^*$      &  177 & 79   &  6.17 &  1.0-2.0 \\
$r$ vs. $r-y^*$      &  174 & 74   &  6.11 &  1.0-2.2 \\
$G$ vs. $Bp-Rp^{**}$ &  124 & 55   &  6.25 &  1.0-2.0 \\
$J$ vs. $J-K$        &  161 & 24   &  6.15 &  0.8-1.8 \\
Averaged values      &  226 & 111  &  6.18 &  0.8-2.0 \\
\hline
\hline
\multicolumn{5}{l}{$^*$: from Pan-STARRS; $^{**}$: from Gaia/DR2} \\ 
\end{tabular}
\end{table}

 The second and third columns of Table \ref{stelpar_table} show the number of members with and without disks for which the given diagram allowed to estimate mass and age. These numbers vary since stars may fall outside the isochrone grid in given derreddened diagrams (mainly because of photometric uncertainties, incorrect individual extinction, or blue and red excesses due to disks), making impossible to estimate their stellar parameters. This is shown in Fig. \ref{dereddiag_fig}. In these diagrams, we have corrected the low-metallicity PARSEC isochrones with age ranging from 0.5$\,$Myrs to 10$\,$Myrs for the factor -5.+5$\times$log(4500), with 4500$\,$pc being the assumed distance to Dolidze~25. Stellar magnitudes and colors are instead de-reddened by using the individual extinctions found in Sect. \ref{ext_sect} and adopting the extinction coefficients listed in the Appendix \ref{AppB}.  

The resulting mass distributions are instead shown in Fig. \ref{deredmass_fig}, where it is evident how the resulting distribution changes by adopting different diagrams to estimate stellar parameters. We also compared the observed slopes (calculated by adopting the limits listed in the fifth column of Table \ref{stelpar_table}) of the mass distribution with that of the normalized Salpeter-Kroupa IMF with $\alpha$=2.35 \citep{KroupaWeidner2003ApJ.598.1076K}. We do not see convincing evidence supporting a deviation of the mass function from the universal law. In order to obtain a unique mass distribution (the one shown in the bottom right panel in Fig. \ref{imf_figure}), for each star we calculated the average mass value from the individual estimates obtained from the various diagrams where the star fall inside the isochrones grid. In this distribution, the observed slope matches the normalized Salpeter-Kroupa IMF slope in the mass range 0.8-2$\,$M$_{\odot}$.  

Fig. \ref{deredage_fig} shows instead the resulting distributions of stellar ages as calculated from each adopted diagram and the resulting average values. None of the distributions obtained from the diagrams show a predominant age, which may be due to a real age spread or to uncertainties associated with the method. The resulting median values are listed in Table \ref{stelpar_table}. The ``averaged'' distribution shows instead a dominant peak at about 1$\,$Myr, with a median age equal to 1.6$\,$Myrs, which is the average cluster age adopted in this work. 	

\newpage

    \begin{figure}[!h]
	\centering	
	\includegraphics[width=16cm]{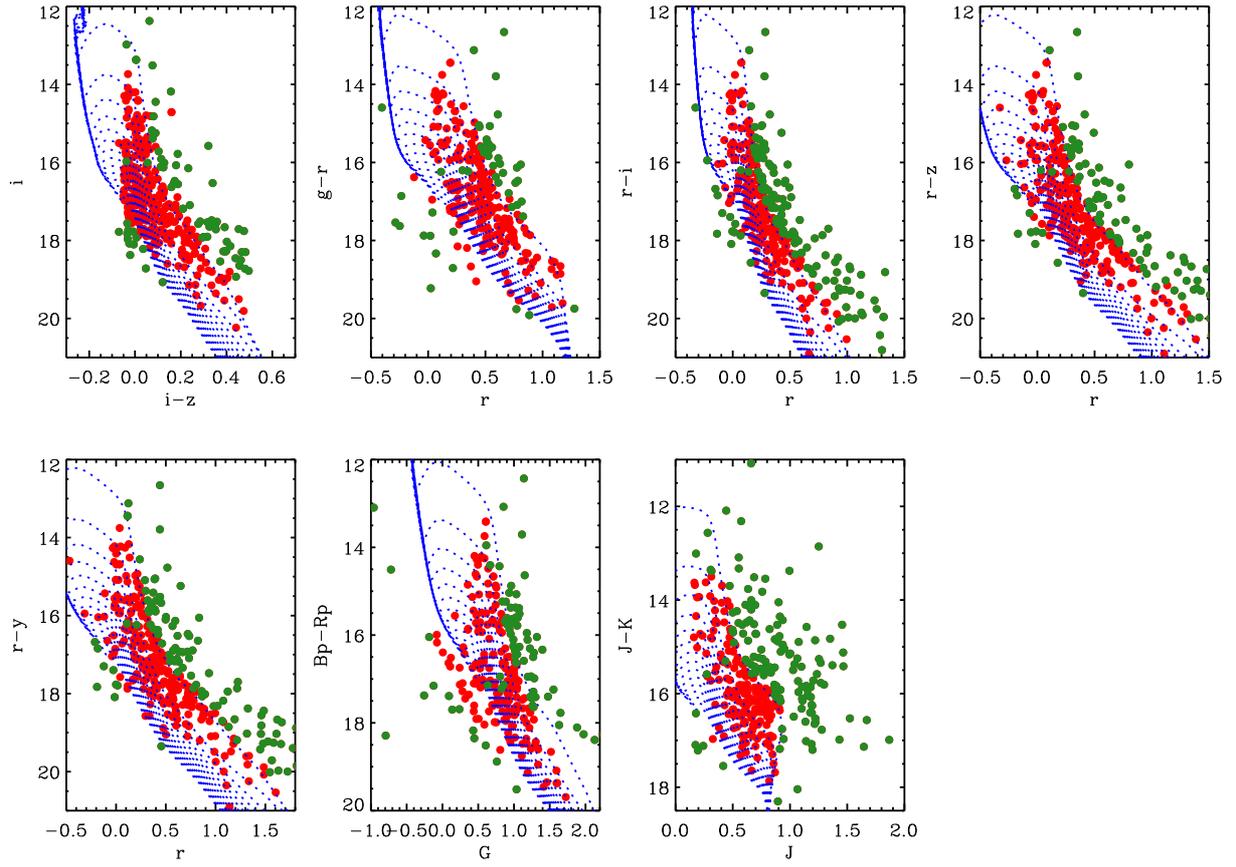}
	\caption{Derreddened diagrams used to estimate stellar parameters. Low-metallicity PARSEC isochrones (blue lines) were corrected for the distance (4500$\,$pc), while stellar magnitudes and colors were derreddened using the individual extinctions calculated in Sect. \ref{ext_sect}. Stars for which parameters were estimated from the given diagram are marked in red, while the remaining in green.}
	\label{dereddiag_fig}
	\end{figure}

\newpage
    
    \begin{figure}[!h]
	\centering	
	\includegraphics[width=18cm]{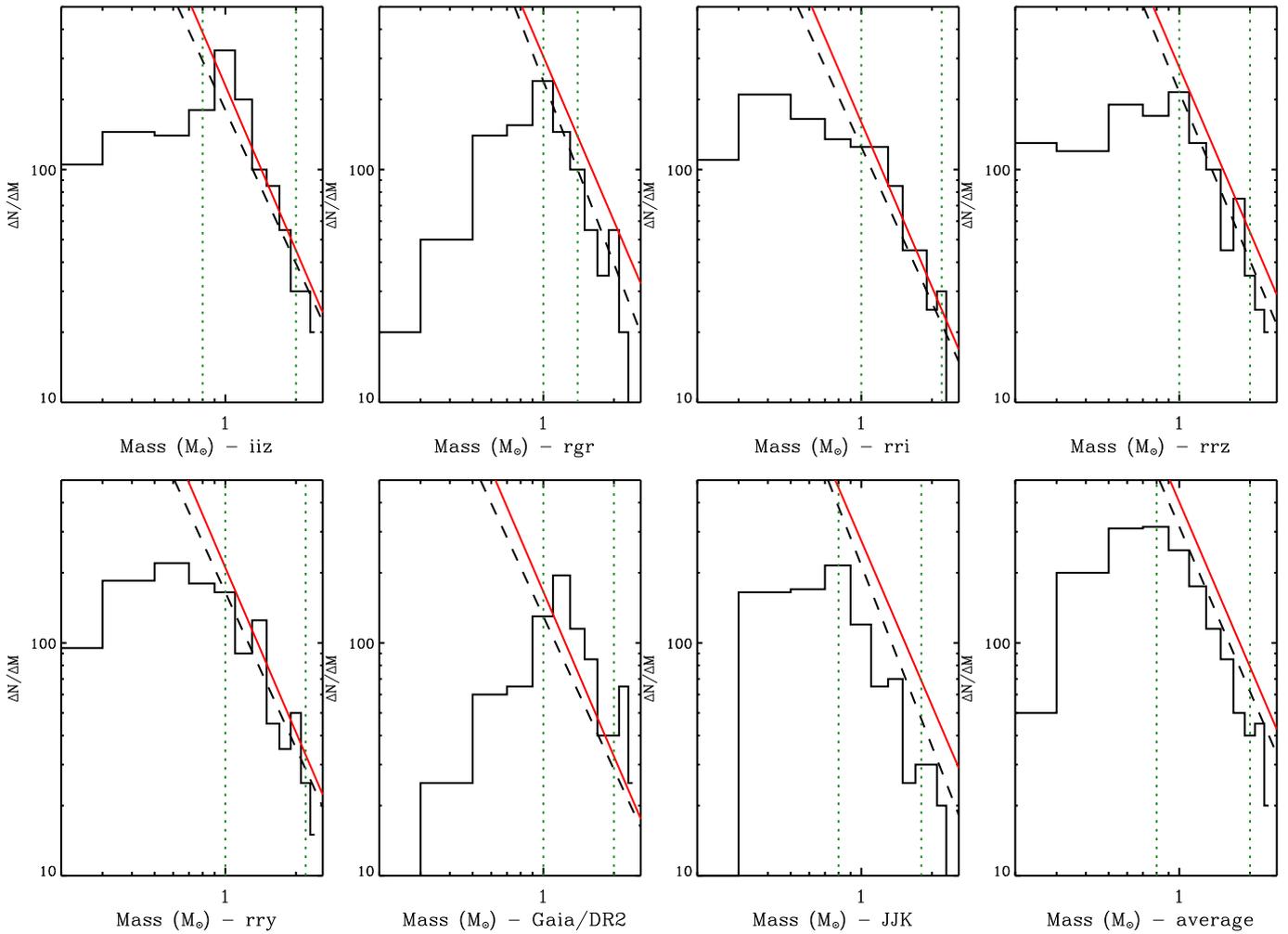}
	\caption{Mass function distributions of both disk-less and disk-bearing candidate members of Dolidze~25 (the latter considered only if inside the ACIS field) obtained from each of adopted diagrams. The solid black line is obtained from a linear fit in the log-log space on the mass distribution between the two limits marked with the vertical lines. The red line shows the normalized Salpeter-Kroupa IMF with $\alpha$=2.35 \citep{KroupaWeidner2003ApJ.598.1076K}. Both the range of completeness and the shape of the mass distribution slightly change if individual masses are estimated from different diagrams. The ``average'' distribution is obtained by averaging for each star the values obtained from the adopted diagrams.}
	\label{deredmass_fig}
	\end{figure}

\newpage


    \begin{figure}[!h]
	\centering	
	\includegraphics[width=18cm]{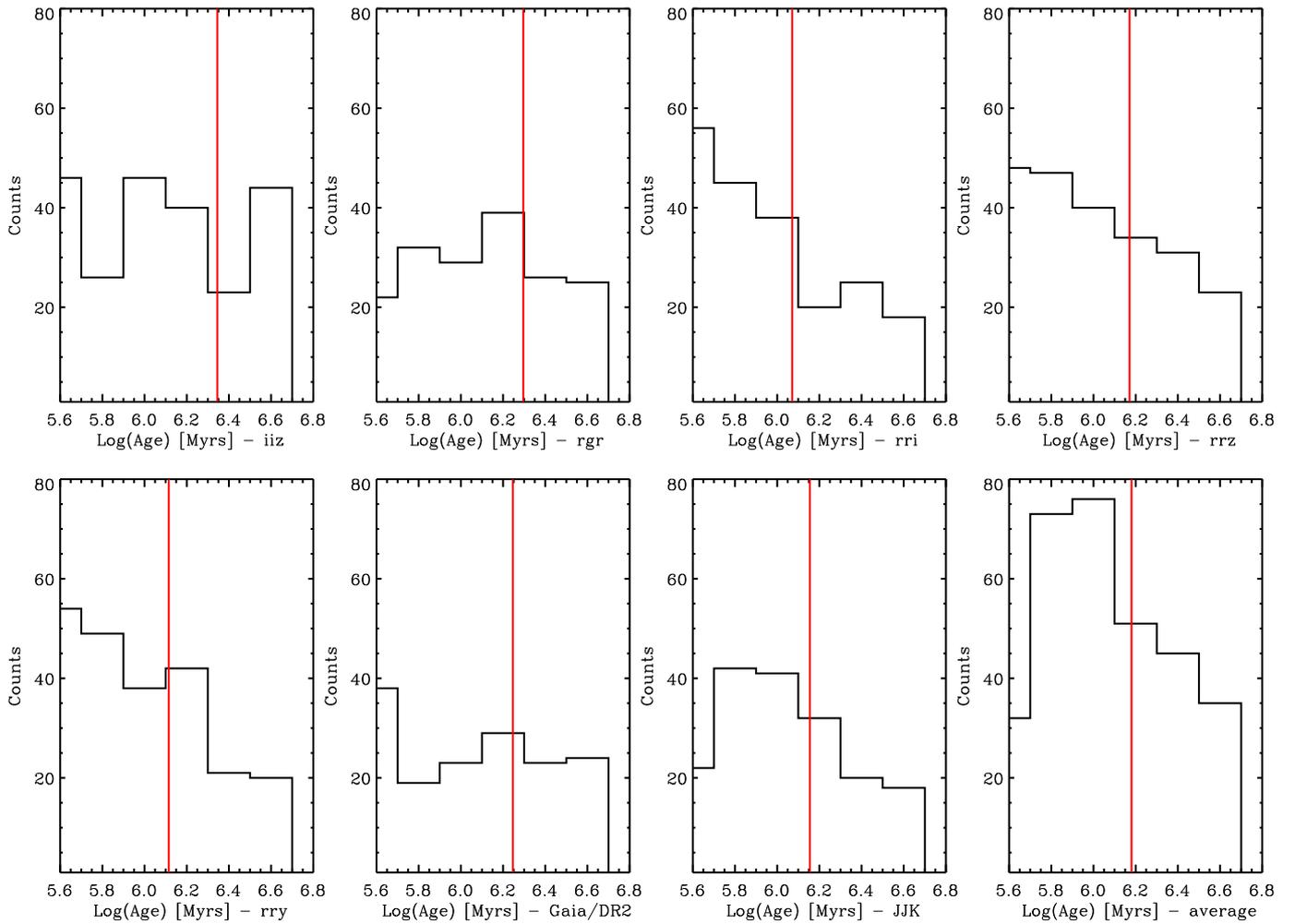}
	\caption{Distribution of stellar ages of both disk-less and disk-bearing candidate members of Dolidze~25 (the latter considered only if inside the ACIS field). The vertical red lines mark the median age values. The ``average'' distribution is obtained by averaging for each star the values obtained from the adopted diagrams.}
	\label{deredage_fig}
	\end{figure}
\pagebreak[4]
\newpage

\section{Catalog of the candidate young stars in Dolidze~25}
\label{app_members}

In this appendix we describe the catalog of the young stars in Dolidze~25 and the surrounding area. 

\begin{longtable}{cccc}
\caption{Catalog of the members of Dolidze~25 and Sh2-284} \\             
\hline\hline                 
Index & Field & Units & Description \\    
\hline                        
\endfirsthead
\caption{continued.} \\
\hline\hline
Index & Field & Units & Description \\    
\hline
\endhead
\hline
\endfoot
\hline
1   & ID            & ...   &   Star ID \\
2   & RA            & deg.  &   Star right ascension \\
3   & DEC           & deg.  &   Star declination    \\
4   & MAGU\_VP       & mag   &   VPHAS/DR2 $u$ band magnitude \\
5   & ERRMAGU\_VP    & mag   &   Error of the VPHAS/DR2 $u$ band magnitude \\
6   & MAGG\_VP       & mag   &   VPHAS/DR2 $g$ band magnitude \\
7   & ERRMAGG\_VP    & mag   &   Error of the VPHAS/DR2 $g$ band magnitude \\
8   & MAGR\_VP       & mag   &   VPHAS/DR2 $r$ band magnitude \\
9   & ERRMAGR\_VP    & mag   &   Error of the VPHAS/DR2 $r$ band magnitude \\
10  & MAGHA\_VP      & mag   &   VPHAS/DR2 H$\alpha$ band magnitude \\
11  & ERRMAGHA\_VP   & mag   &   Error of the VPHAS/DR2 H$\alpha$ band magnitude \\
12  & MAGI\_VP       & mag   &   VPHAS/DR2 $i$ band magnitude \\
13  & ERRMAGI\_VP    & mag   &   Error of the VPHAS/DR2 $i$ band magnitude \\
14  & MAGR\_IP       & mag   &   IPHAS/DR2 $r$ band magnitude \\
15  & ERRMAGR\_IP    & mag   &   Error of the IPHAS/DR2 $r$ band magnitude \\
16  & MAGI\_IP       & mag   &   IPHAS/DR2 $i$ band magnitude \\
17  & ERRMAGI\_IP    & mag   &   Error of the IPHAS/DR2 $i$ band magnitude \\
18  & MAGHA\_IP      & mag   &   IPHAS/DR2 H$\alpha$ band magnitude \\
19  & ERRMAGHA\_IP   & mag   &   Error of the IPHAS/DR2 H$\alpha$ band magnitude \\
20  & MAGG\_PAN      & mag   &   Pan-STARRS $g$ band magnitude \\
21  & ERRMAGG\_PAN   & mag   &   Error of the Pan-STARRS $g$ band magnitude \\
22  & MAGR\_PAN      & mag   &   Pan-STARRS $r$ band magnitude \\
23  & ERRMAGR\_PAN   & mag   &   Error of the Pan-STARRS $r$ band magnitude \\
24  & MAGI\_PAN      & mag   &   Pan-STARRS $i$ band magnitude \\
25  & ERRMAGI\_PAN   & mag   &   Error of the Pan-STARRS $i$ band magnitude \\
26  & MAGZ\_PAN      & mag   &   Pan-STARRS $z$ band magnitude \\
27  & ERRMAGZ\_PAN   & mag   &   Error of the Pan-STARRS $z$ band magnitude \\
28  & MAGY\_PAN      & mag   &   Pan-STARRS $y$ band magnitude \\
29  & ERRMAGY\_PAN   & mag   &   Error of the Pan-STARRS $y$ band magnitude \\
30  & MAGU\_DEL      & mag   &   $U$ band magnitude from Delgado et al. (2016)\\
31  & ERRMAGU\_DEL   & mag   &   Error of the $U$ band magnitude from Delgado et al. (2016) \\
32  & MAGB\_DEL      & mag   &   $U$ band magnitude from Delgado et al. (2016)\\
33  & ERRMAGB\_DEL   & mag   &   Error of the $U$ band magnitude from Delgado et al. (2016) \\
34  & MAGV\_DEL      & mag   &   $U$ band magnitude from Delgado et al. (2016)\\
35  & ERRMAGV\_DEL   & mag   &   Error of the $U$ band magnitude from Delgado et al. (2016) \\
36  & MAGR\_DEL      & mag   &   $U$ band magnitude from Delgado et al. (2016)\\
37  & ERRMAGR\_DEL   & mag   &   Error of the $U$ band magnitude from Delgado et al. (2016) \\
38  & MAGI\_DEL      & mag   &   $U$ band magnitude from Delgado et al. (2016)\\
39  & ERRMAGI\_DEL   & mag   &   Error of the $U$ band magnitude from Delgado et al. (2016) \\
40  & MAGJ          & mag   &   2MASS or UKIDSS/DR10 $J$ band magnitude \\
41  & ERRMAGJ       & mag   &   Error of the 2MASS or UKIDSS/DR10 $J$ band magnitude \\
42  & MAGH          & mag   &   2MASS or UKIDSS/DR10 $H$ band magnitude \\
43  & ERRMAGH       & mag   &   Error of the 2MASS or UKIDSS/DR10 $H$ band magnitude \\
44  & MAGK          & mag   &   2MASS or UKIDSS/DR10 $K$ band magnitude \\
45  & ERRMAGK       & mag   &   Error of the 2MASS or UKIDSS/DR10 $K$ band magnitude \\
46  & MAG1          & mag   &   Spitzer/IRAC [3.6] band magnitude \\
47  & ERRMAG1       & mag   &   Error of the Spitzer/IRAC [3.6] band magnitude \\
48  & MAG2          & mag   &   Spitzer/IRAC [4.5] band magnitude \\
49  & ERRMAG2       & mag   &   Error of the Spitzer/IRAC [4.5] band magnitude \\
50  & MAG3          & mag   &   Spitzer/IRAC [5.8] band magnitude \\
51  & ERRMAG3       & mag   &   Error of the Spitzer/IRAC [5.8] band magnitude \\
52  & MAG4          & mag   &   Spitzer/IRAC [8.0] band magnitude \\
53  & ERRMAG4       & mag   &   Error of the Spitzer/IRAC [8.0] band magnitude \\
54  & MAGW1         & mag   &   WISE [3.4] band magnitude \\
55  & ERRMAGW1      & mag   &   Error of the WISE [3.4] band magnitude \\
56  & MAGW2         & mag   &   WISE [4.6] band magnitude \\
57  & ERRMAGW2      & mag   &   Error of the WISE [4.6] band magnitude \\
58  & MAGW3         & mag   &   WISE [12] band magnitude \\
59  & ERRMAGW3      & mag   &   Error of the WISE [12] band magnitude \\
60  & MAGW4         & mag   &   WISE [22] band magnitude \\
61  & ERRMAGW4      & mag   &   Error of the WISE [22] band magnitude \\
62  & PARALLAX      & milliarcsec & Stellar parallaxes from Gaia/EDR3 \\
63  & ERR\_PARALLAX  & milliarcsec & Error of the stellar parallaxes from Gaia/EDR3 \\
63  & DISK          & ...   &   Equal to 1 for stars with disks \\
64  & XMEMBER       & ...   &   Equal to 1 for disk-less members \\
65  & SPECTR\_LITER  & ...   &   Equal to 1 for spectroscopic members \\
66  & AV            & mag   &   Individual stellar extinction \\
67  & ERRAV         & mag   &   Error of the individual stellar extinction \\
68  & AGE           & [Myrs]&   Log(10) of individual stellar age \\
69  & ERRAGE        & [Myrs]&   Error of the Log(10) of individual stellar age \\
70  & MASS          & M$_{\odot}$ & Individual stellar mass \\
71  & ERRMASS       & M$_{\odot}$ & Error of the individual stellar mass \\ 
72  & IPHAS\_DES     & ... & IPHAS/DR2 stellar designation \\
73  & VPHAS\_DES     & ... & VPHAS/DR2 stellar designation \\
74  & DELGADO\_DES   & ... & Delgado et al. (2016) stellar designation \\
75  & COROT\_DES     & ... & CoRoT stellar designation \\
76  & TWOMASS\_DES   & ... & 2MASS/PSC stellar designation \\
77  & UKIDSS\_DES    & ... & UKIDSS/DR10 stellar designation \\
78  & IRAC\_DES      & ... & Puga et al. (2009) stellar designation \\
79  & WISE\_DES      & ... & AllWISE Source Catalog stellar designation \\
80  & ACIS\_DES      & ... & Stellar designation in the X-ray sources catalog\\
81  & GAIADR2\_DES   & ... & GAIA/DR2 stellar designation \\
82  & GAIAEDR3\_DES  & ... & GAIA/EDR3 stellar designation \\
83  & PANSTARRS\_DES & ... & Pan-STARRS stellar designation \\
\hline
\hline
\end{longtable}

$  $    \par
\newpage
\section{Disk Fractions and ages of the clusters plotted in Fig. \ref{cluster_df}}
\label{AppD}

\end{onecolumn}

\begin{longtable}{ccccc}
\caption{Age and disk fraction of the clusters plotted in Fig. \ref{cluster_df}} \\             
\hline\hline                 
Cluster & Age & DF & dist. & References \\    
        & Myrs&    & pc    &            \\
\hline                        
\endfirsthead
\caption{continued.} \\
\hline\hline
Cluster & Age & DF & References \\    
        & Myrs&    &            \\
\hline
\endhead
\hline
\endfoot
\multicolumn{5}{c}{``Nearby'' and not massive clusters} \\ 
\hline
$\eta$Cha     &  $7_{4}^{10}$       &	$0.40\pm0.05$  & 94   &  \citet{Megeath2005ApJ.634L.113M},{\bf \citet{Rugel2018AA.609A.70R}}   \\
LowCent-Crux  &	 $10_{10}^{10}$     &	$0.09\pm0.03$  & 100  &  \citet{PecautMamajek2016MNRAS.461.794P}   \\
Taurus        &  $1.5_{1}^{2}$      &   $0.64\pm0.05$  & 130  &  \citet{LuhmanMamajek2012ApJ},{\bf \citet{Ribas2014AA.561A.54R}}   \\	
              &                     &                  &      &  \citet{Kraus2017ApJ.838.150K,Galli2019AA.630A.137G,ManzoMartinez2020ApJ.893.56M} \\
NGC1333       &  $0.5_{0}^{1}$      &	$0.66\pm0.06$  & 135  &  \citet{Ribas2014AA.561A.54R}   \\
Coronet		  &  $1.2_{0.5}^{1.9}$  &	$0.50\pm0.13$  & 138  &  \citet{Ribas2014AA.561A.54R}   \\
UpperSco      &  $9_{7}^{11}$       &	$0.16\pm0.06$  & 140  &  \citet{CarpenterMHM2006ApJ,SungSB2009},{\bf \citet{Ribas2014AA.561A.54R}}   \\
Lupus         &  $1.25_{1}^{1.5}$   &	$0.52\pm0.05$  & 140  &  {\bf \citet{Ribas2014AA.561A.54R}},\citet{Merin2008ApJS.177.551M}   \\
ChaI          &  $2.0_{1.5}^{2.5}$  &	$0.52\pm0.06$  & 160  &  \citet{Ribas2014AA.561A.54R}   \\
ChaII         &  $2.0_{0}^{4}$      &	$0.84\pm0.09$  & 178  &  \citet{Alcala2008ApJ.676.427A},{\bf \citet{Ribas2014AA.561A.54R}}   \\
IC348         &  $2.25_{2}^{2.5}$   &	$0.41\pm0.06$  & 300  &  \citet{LadaMLA2006,HernandezHMG2007,SungSB2009}   \\
              &                     &                  &      &  \citet{Ribas2014AA.561A.54R},{\bf \citet{Richert2018MNRAS.477.5191R}},\citep{ManzoMartinez2020ApJ.893.56M}   \\
25 Orionis    &  $8.5_{7.0}^{10 }$  &	$0.09\pm0.05$  & 330  &  \citet{Briceno2007ApJ.661.1119B,HernandezHMG2007}   \\
              &                     &                  &      &  \citet{SungSB2009},{\bf \citet{Ribas2014AA.561A.54R}}   \\
GammaVel      &  $7.5_{7}^{8}$      &	$0.06\pm0.01$  & 345  &  \citet{Hernandez2008ApJ.686.1195H,Jeffries2017MNRAS.464.1456J},{\bf \citet{ManzoMartinez2020ApJ.893.56M}}   \\
Berkeley59    &  $1.8_{1.6}^{2.0}$  &	$0.50\pm0.06$  & 400  &  \citet{Richert2018MNRAS.477.5191R}   \\
NGC2068/2071  &	 $2_{1}^{3}$        &	$0.54\pm0.13$  & 400  &  \citet{FlahertyMuzerolle2008AJ.135.966F},{\bf \citet{SungSB2009}}   \\
L1630N        &  $1.5_{1}^{2}$      &	$0.97\pm0.3	$  & 400  &  \citet{Spezzi2015AA.581A.140S}   \\
Lynds1641     &  $1.5_{1}^{2}$      &	$0.51\pm0.02$  & 400  &  \citet{Fang2013ApJS.207.5F}   \\
$\sigma$Ori   &	 $2.5_{2}^{3}$      &	$0.36\pm0.04$  & 414  &  \citet{HernandezHMG2007,Ribas2014AA.561A.54R},{\bf \citet{ManzoMartinez2020ApJ.893.56M}}   \\
$\lambda$Ori  &	 $5_{4}^{6}$        &	$0.19\pm0.04$  & 414  &  \citet{Hernandez2010ApJ.722.1226H,Kounkel2018AJ.156.84K,ManzoMartinez2020ApJ.893.56M}   \\
OriOB1b 	  &  $5_{4.5}^{5.5}$    &	$0.15\pm0.02$  & 414  &  \citet{HernandezHMG2007,Briceno2009RMxAC.35.27B},{\bf \citet{ManzoMartinez2020ApJ.893.56M}}   \\
Flame/NGC2023 &	 $0.8_{0.6}^{1}$    &	$0.71\pm0.08$  & 414  &  \citet{Richert2018MNRAS.477.5191R}   \\
ONC Flank	  &  $1.7_{1.5}^{1.9}$  &	$0.43\pm0.06$  & 414  &  \citet{Richert2018MNRAS.477.5191R}   \\
Serpens South &	 $1.8_{1}^{2.6}$    &	$0.58\pm0.19$  & 415  &  \citet{Richert2018MNRAS.477.5191R}   \\
Serpens       &  $0.5_{0}^{1}$      &	$0.75\pm0.16$  & 415  &  \citet{Winston2007ApJ.669.493W},{\bf \citet{SungSB2009}}   \\
W40           &  $0.8_{0.7}^{0.9}$  &	$0.79\pm0.07$  & 500  &  \citet{Richert2018MNRAS.477.5191R}   \\
LkH$\alpha$101&  $1.5_{1.2}^{1.8}$  &	$0.56\pm0.08$  & 510  &  \citet{Richert2018MNRAS.477.5191R}   \\
RCW36         &  $0.9_{0.8}^{1.0}$  &	$0.81\pm0.07$  & 700  &  \citet{Richert2018MNRAS.477.5191R}   \\
CepA          &  $1.4_{1.1}^{1.7}$  &	$0.65\pm0.1	$  & 700  &  \citet{Richert2018MNRAS.477.5191R}   \\
CepC          &  $2.2_{1.3}^{3.1}$  &	$0.44\pm0.12$  & 700  &  \citet{Richert2018MNRAS.477.5191R}   \\
CepOB3b-East  &	 $3.5_{3}^{4}$      &	$0.32\pm0.04$  & 700  &  \citet{Allen2012ApJ.750.125A}   \\
CepOB3b-West  &	 $3.5_{3}^{4}$      &	$0.50\pm0.06$  & 700  &  \citet{Allen2012ApJ.750.125A}   \\
MonR2		  &  $1.7_{1.5}^{1.9}$  &	$0.64\pm0.07$  & 830  &  \citet{Richert2018MNRAS.477.5191R}   \\
Trumpler37    &  $2.6_{2.3}^{2.9}$  &	$0.49\pm0.07$  & 900  &  \citet{Sicilia-Aguilar2006AJ.132.2135S},{\bf \citet{SungSB2009}}   \\
\hline
\multicolumn{5}{c}{Distant and not massive clusters} \\ 
\hline
NGC7129       &  $3_{2.5}^{3.5}$    &	$0.33\pm0.22$  & 1260 &  \citet{StelzerScholz2009AA.507.227S}   \\
Sh2-106       &  $0.8_{0.4}^{1.2}$  &	$0.53\pm0.1	$  & 1400 &  \citet{Richert2018MNRAS.477.5191R}   \\
NGC2282       &  $3.5_{2}^{5}$      &	$0.58\pm0.06$  & 1650 &  \citet{Dutta2015MNRAS.454.3597D}   \\
IC1795        &  $4_{3}^{5}$        &	$0.50\pm0.05$  & 2000 &  \citet{RoccatagliataBHG2011}   \\
AFGL333       &  $2.0_{1}^{3}$      &	$0.55\pm0.5	$  & 2000 &  \citet{Jose2016ApJ.822.49J}   \\
\hline
\multicolumn{5}{c}{Massive clusters*} \\ 
\hline
OMC           &  $1.5_{1.2}^{1.7}$  &	$0.45\pm0.07$  & 414  &  \citet{Richert2018MNRAS.477.5191R}   \\
NGC2264       &  $3.1_{2.8}^{3.4}$  &	$0.36\pm0.05$  & 751  &  \citet{ParkSung2002,BalogMRS2007}   \\
              &                     &                  &      &  \citet{Wang2008ApJ.675.464W},{\bf \citet{SungSB2009}}   \\
NGC2244       &  $1.7_{1.5}^{1.9}$  &	$0.36\pm0.05$  & 913  &  \citet{Marinas2013ApJ.772.81M,Rapson2014ApJ.794.124R}   \\
M8            &	 $2.3_{2.2}^{2.4}$  &	$0.41\pm0.06$  & 1300 &  \citet{Richert2018MNRAS.477.5191R}   \\
NGC6530 	  &  $0.7_{0.3}^{1.5}$  &	$0.20\pm0	$  & 1300 &  \citet{Damiani2006}   \\
CygnusOB2     &  $1.5_{0}^{3}$      &	$0.29\pm0.11$  & 1450 &  \citet{GuarcelloDWA2016arXiv}   \\
NGC2362       &  $2.5_{2.1}^{2.9}$  &	$0.12\pm0.04$  & 1480 &  \citet{DahmHillenbrand2007AJ.133.2072D,SungSB2009},{\bf \citet{Richert2018MNRAS.477.5191R}}   \\
NGC6231       &  $4_{3}^{5}$        &   $0.05\pm0.01$  & 1585 &  {\bf \citet{Damiani2016AA.596A.82D}},\citet{Baume1999AAS.137.233B,Sung2013AJ.145.37S} \\ 
Pismis24      &  $1.85_{1.0}^{2.7}$ &   $0.33\pm0.05$  & 1700 &  \citet{FangBKH2012} \\
NGC6611  	  &	 $1.2_{0.3}^{2.6}$  &	$0.34\pm0.03$  & 1750 &  \citet{GuarcelloMPP2010}   \\
W3Main        &  $3.0_{2.5}^{3.5}$  &	$0.07\pm0.02$  & 1950 &  \citet{Bik2014AA.561A.12B}   \\
M17 	      &  $1.1_{0.9}^{1.3}$  &	$0.60\pm0.15$  & 2000 &  \citet{Richert2018MNRAS.477.5191R}   \\
Trumpler14    &  $1.0_{0.5}^{1.5}$  &	$0.10\pm0.01$  & 2700 &  \citet{ReiterParker2019MNRAS.486.4354R}   \\
Trumpler16    &  $3.0_{2.5}^{3.5}$  &	$0.07\pm0.01$  & 2800 &  \citet{ReiterParker2019MNRAS.486.4354R}   \\
\hline
\multicolumn{5}{c}{Low-metallicity clusters} \\ 
\hline
NGC1893       &  $1.4_{0.8}^{1.8}$  &	$0.71\pm01	$  & 3600 &  \citet{Prisinzano2011AA.527A.77P}   \\
Sh2-207       &  $2.5_{2}^{3}$      &	$0.05\pm0.05$  & 9000 &  \citet{YasuiKTS2010}   \\
Sh2-208       &  $0.5_{0.5}^{0.5}$  &	$0.27\pm0.06$  & 9000 &  \citet{Yasui2016AJ.151.115Y}   \\
Sh2-209Main   &  $0.75_{0.5}^{1}$   &	$0.10\pm0.01$  &10000 &  \citet{YasuiKTS2010}   \\
Sh2-209Sub    &  $0.75_{0.5}^{1}$   &	$0.07\pm0.01$  &10000 &  \citet{YasuiKTS2010}   \\
Cloud2-N  	  &  $0.75_{0.5}^{1}$   &   $0.09\pm0.04$  &12000 &  \citet{YasuiKTS2010}   \\
Cloud2-S      &  $0.75_{0.5}^{1}$   &	$0.27\pm0.07$  &12000 &  \citet{YasuiKTS2010}   \\
\hline
\hline
\multicolumn{5}{l}{$^*$: Here we show an average DF value, which typically is 15\%-20\% higher than the values of the clusters core.} \\ 
\label{clusters_table}      
\end{longtable}

$  $    \par
\newpage
\section{Magnitude distributions}
\label{AppH}

In this section we show the magnitude distributions of the ``members'', ``control'', and ``ACIS FoV'' samples, as defined in Sect. \ref{complete_sect}. 

    \begin{figure}[!h]
	\centering	
	\includegraphics[width=18cm]{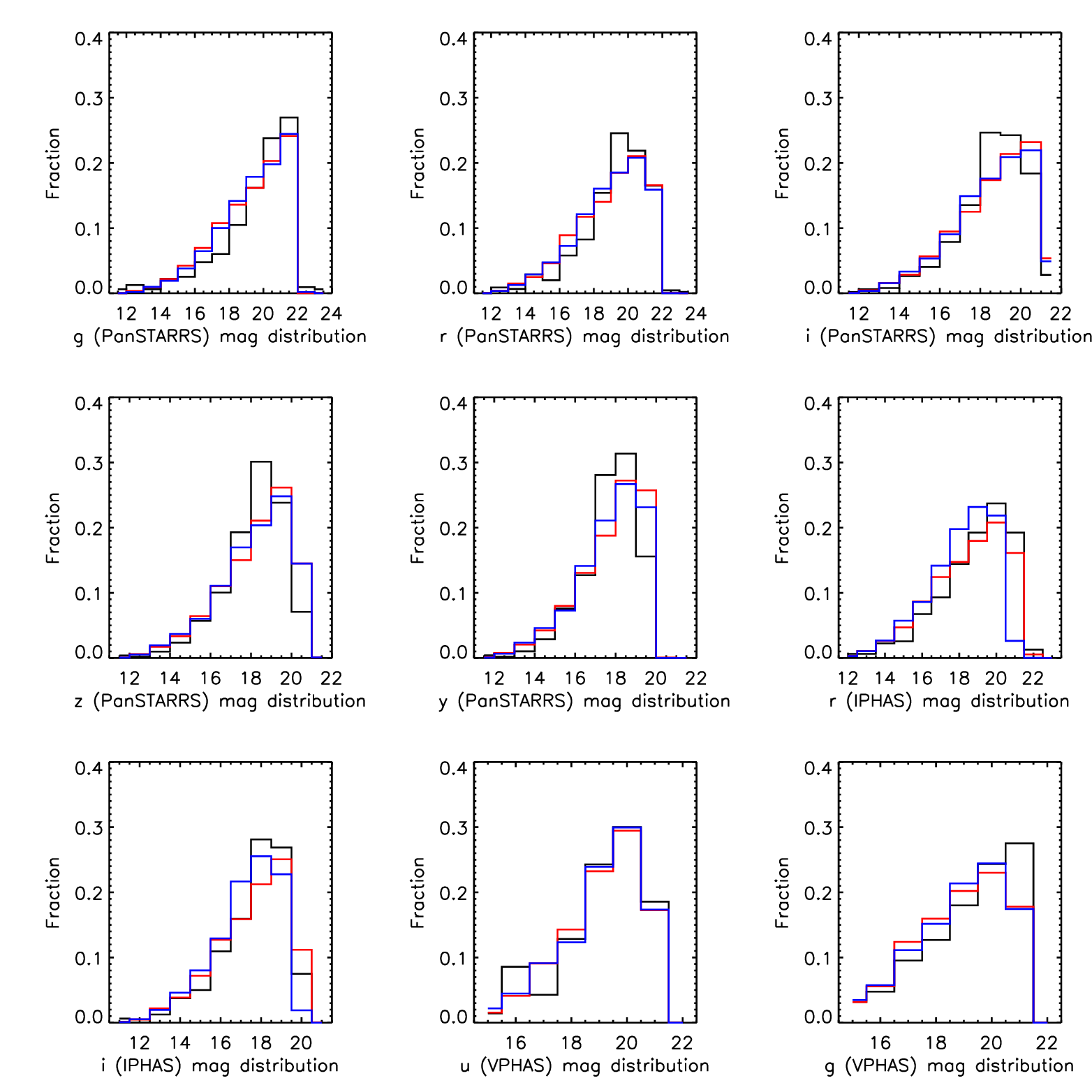}
	\caption{Magnitude distributions of the ``members'' (black), ``control'' (blue), and ``ACIS FoV'' (red) samples in given photometric bands.}
	\label{magdis1}
	\end{figure}

    \begin{figure}[!h]
	\centering	
	\includegraphics[width=18cm]{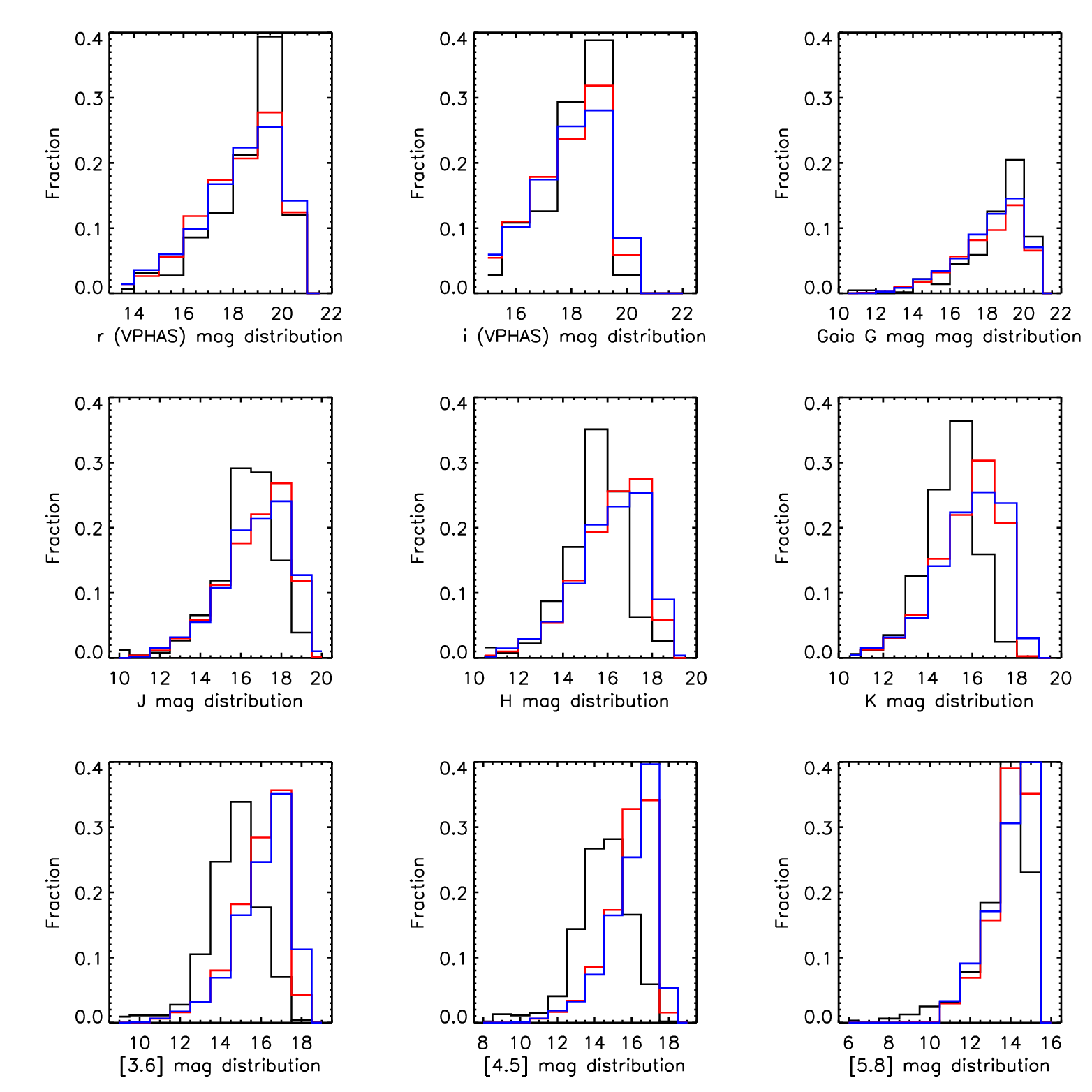}
	\caption{Magnitude distributions of the ``members'' (black), ``control'' (blue), and ``ACIS FoV'' (red) samples in given photometric bands.}
	\label{magdis1}
	\end{figure}

    \begin{figure}[!h]
	\centering	
	\includegraphics[width=18cm]{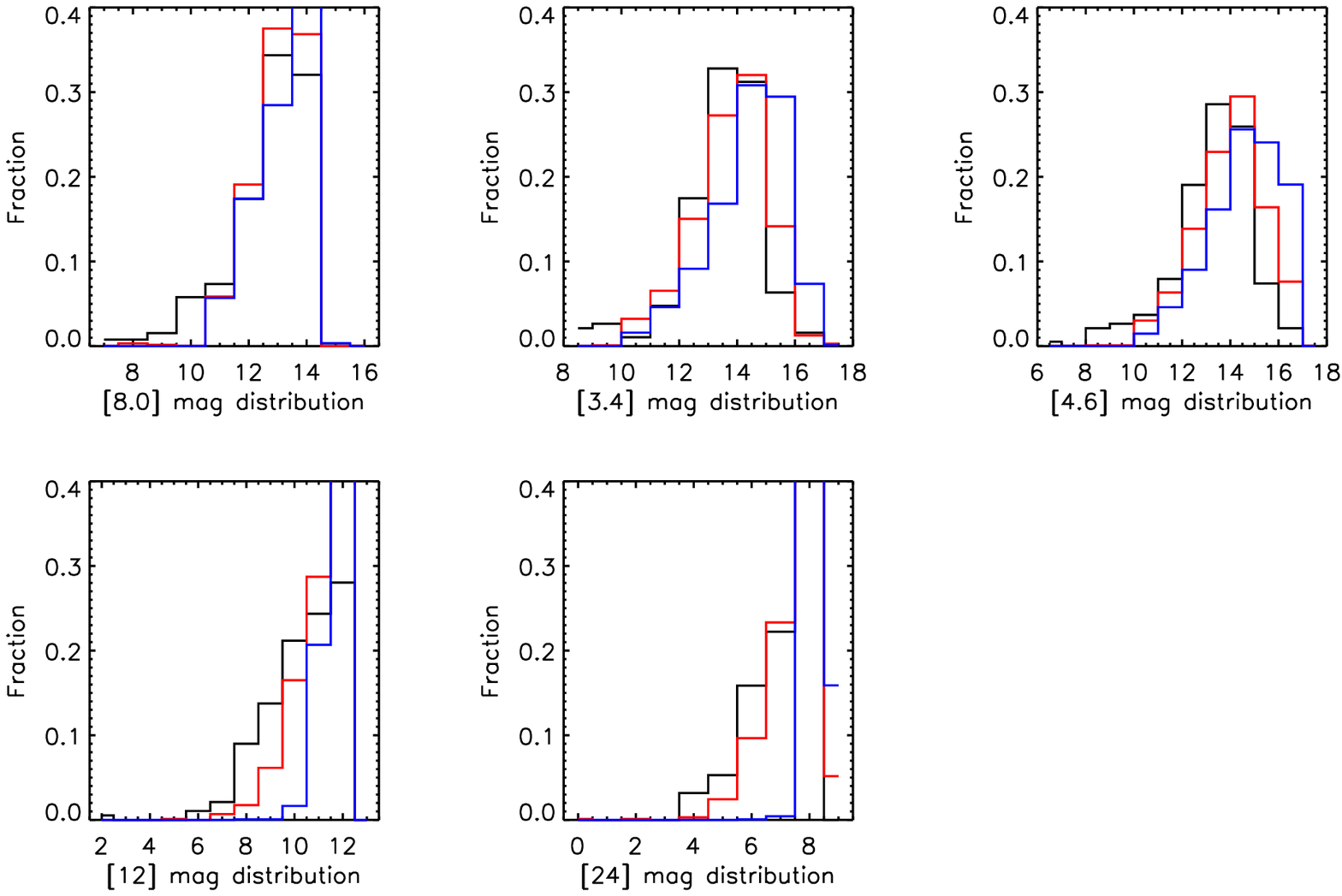}
	\caption{Magnitude distributions of the ``members'' (black), ``control'' (blue), and ``ACIS FoV'' (red) samples in given photometric bands.}
	\label{magdis1}
	\end{figure}

\end{appendix}

\end{document}